\documentclass[a4paper,twocolumn,aps,pre,showpacs,nofootinbib]{revtex4-1}

\usepackage{amsmath,amssymb,graphicx,color,times}

\def\<{\langle}
\def\>{\rangle}
\newcommand{\cp}{{c\ '}}
\newcommand{\PA}{{Physica~}}
\newcommand{\NAT}{{Nature}}
\newcommand{\EPLOLD}{{Europhys.~Lett.~}}
\newcommand{\EPL}{{EPL~}}
\newcommand{\JSTAT}{{J.~Stat.~Mech.~}}
\newcommand{\PR}{Phys.~Rev.~}
\newcommand{\PRL}{Phys.~Rev.~Lett.~}
\newcommand{\PRB}{Phys.~Rev.~B~}
\newcommand{\PRE}{Phys.~Rev.~E~}
\newcommand{\JCP}{J.~Chem.~Phys.~}
\newcommand{\JPCM}{J.~Phys.~: Condens.~Matter~}

\newcommand{\upd}{{\ensuremath{\textrm{d}}}}
\newcommand{\eref}[1]{Eq.~\eqref{#1}}
\newcommand{\fref}[1]{Fig.~\ref{#1}}
\usepackage{upgreek}
\newcommand{\BC}{BC}
\newcommand{\MC}{MC}
\newcommand{\po}{\ensuremath{{(+,o)}}}
\newcommand{\pp}{\ensuremath{{(+,+)}}}
\newcommand{\oo}{\ensuremath{{(o,o)}}}
\newcommand{\mo}{\ensuremath{{(-,o)}}}
\newcommand{\pmbc}{\ensuremath{{(+,-)}}}
\renewcommand*{\vec}[1]{\mathbf{#1}}

\makeatletter
\newcommand{\raisemath}[1]{\mathpalette{\raisem@th{#1}}}
\newcommand{\raisem@th}[3]{\raisebox{#1}{$#2#3$}}
\makeatother

\begin{document}
\title{Critical Casimir forces between homogeneous and chemically striped surfaces}
\author{\firstname{Francesco} \surname{Parisen Toldin}}
\email{francesco.parisentoldin@physik.uni-wuerzburg.de}
\affiliation{Institut f\"ur Theoretische Physik und Astrophysik, Universit\"at W\"urzburg, Am Hubland, D-97074 W\"urzburg, Germany}
\affiliation{Max-Planck-Institut f\"ur Physik komplexer Systeme, N\"othnitzer Str.~38, D-01187 Dresden, Germany}
\author{\firstname{Matthias} \surname{Tr\"ondle}}
\email{troendle@is.mpg.de}
\author{\firstname{S.} \surname{Dietrich}}
\email{dietrich@is.mpg.de}
\affiliation{Max-Planck-Institut f\"ur Intelligente Systeme, Heisenbergstr.~3, D-70569 Stuttgart, Germany}
\affiliation{IV. Institut f\"ur Theoretische Physik, Universit\"at Stuttgart, Pfaffenwaldring 57, D-70569 Stuttgart, Germany}


\begin{abstract}

  Recent experiments have measured the critical Casimir force acting on a colloid immersed in a binary liquid mixture near its continuous demixing  phase transition and exposed to a chemically structured substrate. Motivated by these experiments, we study the critical behavior of a system, which belongs to the Ising universality class, for the film geometry with one planar wall chemically striped, such that there is a laterally alternating adsorption preference for the two species of the binary liquid mixture, which is implemented by surface fields. For the opposite wall we employ alternatively a homogeneous adsorption preference or homogeneous Dirichlet boundary conditions, which within a lattice model are realized by open boundary conditions. By means of mean-field theory, Monte Carlo simulations, and finite-size scaling analysis we determine the critical Casimir force acting on the two parallel walls and its corresponding universal scaling function. We show that in the limit of stripe widths small compared with the film thickness, on the striped surface the system effectively realizes Dirichlet boundary conditions, which generically do not hold for actual fluids. Moreover, the critical Casimir force is found to be attractive or repulsive, depending on the width of the stripes of the chemically patterned surface and on the boundary condition applied to the opposing surface.

\end{abstract}
\maketitle

\section{Introduction}
\label{sec:intro}

As an intriguing consequence of their presence, fluctuations of an embedding medium may manifest themselves in terms of effective forces
acting on its confining boundaries. %
The critical Casimir force is  such a fluctuation-induced force which arises due to the emergence of long-ranged thermal fluctuations if 
a fluid close to a second-order phase transition is confined between surfaces.
This phenomenon, first predicted by Fisher and de~Gennes \cite{FG-78} is the analog of the Casimir effect in quantum electrodynamics \cite{Casimir-48}. 
Reference \cite{Gambassi-09} provides a recent review which illustrates analogies as well as differences between these two effects and guides the 
reader towards further reviews of the subject and the pertinent original literature.

The dependence of the critical Casimir force on the distance between the confinements and on temperature is characterized by a universal scaling function, which is determined by the bulk and surface universality classes (UC) \cite{Binder-83,Diehl-86} 
of the confined system. It is independent of microscopic details of the system, and its form depends only on a few global and general properties, such as the spatial dimension $d$, the number of components of the order parameter, the shape of the confinement, and the type of boundary conditions ({\BC}) \cite{Krech-94,Krech-99,BTD-00}.

In recent years the critical Casimir effect has attracted numerous experimental \cite{he4,cwetting,qwetting,HHGDB-08,GMHNHBD-09,SZHHB-08,TZGVHBD-11,NHB-09,NDHCNVB-11,BOSGWS-09,Schall-aggregation} and even more theoretical investigations. Critical Casimir forces can be  inferred indirectly by studying wetting  films of fluids close to a critical end point \cite{NI-85,KD-92b}. In this context, $^4$He wetting films close to the onset of superfluidity \cite{he4} and wetting  films of classical \cite{cwetting} and quantum \cite{qwetting} binary liquid mixtures have been studied experimentally.
Only recently direct measurements of the critical Casimir force have been reported \cite{HHGDB-08,GMHNHBD-09,SZHHB-08,NHB-09,TZGVHBD-11,NDHCNVB-11} by monitoring individual colloidal particles immersed into a binary liquid mixture close to its critical demixing point and exposed to a planar wall. The critical Casimir effect has also been studied via its influence on aggregation phenomena \cite{BOSGWS-09,Schall-aggregation}.

Not only the strength of critical Casimir forces can be tuned by small temperature changes but even their sign 
depends on the {\BC} of the confining boundaries.
The two interfaces of a $^4$He film impose a symmetry-preserving Dirichlet {\BC} [denoted by $(o)$] on 
the superfluid order-parameter at both sides of the film, which causes attractive critical Casimir forces leading to a thinning of the 
film near the $\lambda$ transition \cite{he4,NI-85,KD-92b}.
However, for classical binary liquid mixtures (or simple fluids), surfaces preferentially adsorb one of the two species of
the mixture (or the gaseous or the liquid phase of a simple fluid, respectively).
This corresponds to symmetry-breaking {\BC} (denoted as $(+)$ or $(-)$ {\BC}) acting on the order parameter which is, e.g., the concentration
difference in a binary liquid.
Within the theoretical description $(\pm)$ {\BC} are realized by surface fields and the $(o)$ {\BC} by their absence.
\par
The emergence of long-ranged thermal fluctuations close to a second-order phase transition leads to a mesoscopic extent of 
the adsorption layer close to surfaces with $(\pm)$ {\BC}.
Depending on whether the adsorption preferences of the confining surfaces of the fluid are the same $(\pm,\pm)$ or different $(+,-)$,
critical Casimir forces acting on them are either attractive $(\pm,\pm)$ or repulsive $(+,-)$ \cite{cwetting,HHGDB-08,GMHNHBD-09,NHB-09}.
The critical Casimir force between walls with $(\pm)$ {\BC} is the combined effect of the change
of the fluctuation spectrum due to the confinement and the interference of the adsorption layers, which are present even within
mean-field theory.
The shapes of the adsorption layers themselves are strongly influenced by non-Gaussian fluctuations, i.e., they differ from
mean-field predictions.
In this sense, the effective forces acting on surfaces which confine a (near-) critical fluid 
provide a classical analog of the Casimir effect both in the case of symmetry-breaking and in the case of symmetry-preserving {\BC}.

 Early theoretical investigations of the  critical Casimir force used, to a large extent, field-theoretical methods 
(see, e.g., Ref.~\cite{PTD-10} for a list of references). 
Only recently have Monte Carlo (\MC) simulations allowed for  their quantitatively  reliable computation. Early numerical simulations for the critical Casimir force have been employed in Ref.~\cite{Krech-97} for the film geometry with laterally homogeneous {\BC}. More recently the critical Casimir force has been determined by numerical simulations for the $XY$ UC \cite{DK-04,Hucht-07,VGMD-07,VGMD-08,Hasenbusch-09b,Hasenbusch-09c,Hasenbusch-09d}, which describes the critical properties of the superfluid phase transition in $^4$He, as well as the Ising UC \cite{DK-04,VGMD-07,VGMD-08,Hasenbusch-10c,PTD-10,HGS-11,Hasenbusch-11,VMD-11,Hasenbusch-12,Hasenbusch-12b} which describes, {\it inter alia}, the  experimentally relevant demixing transition in a binary liquid mixture.

Since Casimir forces may affect or empower future devices on the micro- and nanoscales, their modifications due to the presence of nano- or microstructures on the 
substrates has been a topic of intense research during the past decade.
Recent theoretical and experimental studies of QED Casimir forces (see, e.g., Ref.~\cite{casimirtopo} and references therein) 
as well as critical Casimir forces \cite{troendle:2008} for \emph{topologically} structured
substrates exhibit remarkable deviations from the corresponding ones for planar walls as well as the occurrence of lateral forces.
However, only \emph{chemically} patterned substrates allow for interesting combinations of attractive and repulsive critical Casimir forces
so that, among the various realizations of the critical Casimir  effect, the force in the presence of a  chemically patterned substrate has 
recently attracted particular interest \cite{GD-11,TZGVHBD-11}. 

Experiments with binary liquid mixtures  as solvents have been used to study critical Casimir forces acting on  dissolved colloids close to a chemically structured substrate 
\cite{TZGVHBD-11,SZHHB-08,NHB-09},  which creates a laterally varying adsorption preference  for both components of the solvent. Such  kind of systems have been investigated theoretically for the film geometry within mean-field theory \cite{SSD-06}, within Gaussian approximation \cite{KPHSD-04}, and with {\MC} simulations in a three-dimensional film geometry in the presence of a single chemical step \cite{PTD-10}. The critical Casimir force in the presence of a patterned substrate has also been studied in the case of a sphere near a planar wall within the Derjaguin approximation \cite{TKGHD-09,TKGHD-10}.  
If the lateral chemical patterns do not consist of stripes with sharp chemical steps between areas of strong but opposite adsorption preferences, one faces spatial regions
characterized by surface fields of medium strength.
This case has been studied so far only for laterally homogeneous {\BC} in the presence of variable boundary fields.
This case already gives rise to interesting crossover phenomena, which have been studied within mean-field theory \cite{MMD-10},  by exact calculations in  two spatial dimensions \cite{AM-10,Borjan-12}, and with {\MC} simulations \cite{Hasenbusch-11,VMD-11}.

Motivated by the aforementioned experimental results, and based on previous investigations  by two of the authors \cite{PTD-10}, 
here we present a {\MC} study of a three-dimensional lattice model in the film geometry, representing the Ising UC in the presence of a chemically  striped substrate. 
Moreover, we compare the universal scaling functions of the critical Casimir forces obtained from these {\MC} results with the corresponding mean-field results,
which we obtain by generalizing a previous study \cite{SSD-06} and which are valid in $d=4$ spatial dimensions.
{
}%
We employ periodic boundary conditions in the lateral directions and different {\BC} for the two surfaces  confining the slab.
To this end, we consider a film of thickness $L$ confined along the normal $z$ direction on one side by a surface at which the order parameter 
of the fluid exhibits a laterally homogeneous {\BC} which corresponds either to strong adsorption $(+)$ or to the so-called ordinary surface transition
$(o)$ \cite{Binder-83,Diehl-86}.
The other side of the film is confined by a surface which is periodically patterned
by stripes  leading to strong, alternating adsorption preferences  corresponding to $(+)$ or $(-)$ {\BC}, respectively, varying along the 
lateral $x$-direction.
\begin{figure}[t]
\begin{center}
\includegraphics[width=\linewidth,keepaspectratio]{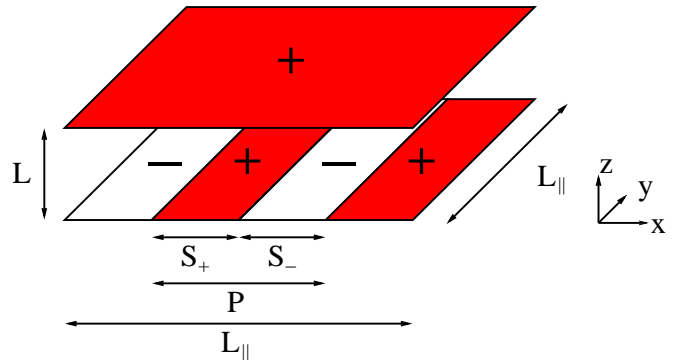}
\end{center}
\caption{(Color online) Film geometry confined by a  laterally homogeneous upper surface and by a lower surface with alternating stripes  of equal width. At both surfaces the spins are fixed. We choose $S_+=S_-$ so that the period $P=S_++S_-=2S_+$.}
\label{bcstripes}
\end{figure}

\begin{figure}[b]
\begin{center}
\includegraphics[width=\linewidth,keepaspectratio]{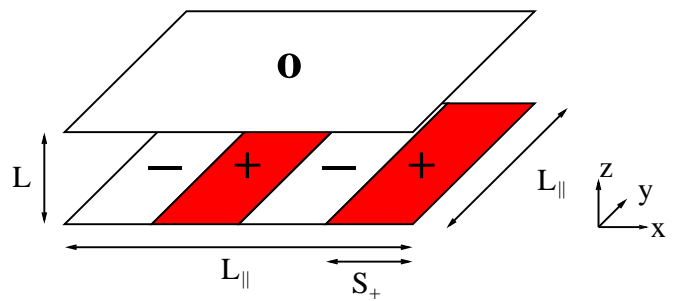}
\end{center}
\caption{(Color online) Film geometry confined by  an upper surface with open {\BC} and by  a lower surface with alternating stripes of equal width with fixed spins.}
\label{bcopenstripes}
\end{figure}

Here we focus on stripes of equal width $S_+=S_-=P/2$ corresponding to half of the period $P$ along the $x$-direction, so that
the important geometrical parameter is given by $\kappa\equiv S_+/L$, which relates the width of the stripes to the film thickness 
(see Figs.~\ref{bcstripes} and \ref{bcopenstripes}).
Within the lattice model this system is realized by either fixing
the Ising spins in the upper surface to $+1$  or imposing an open boundary by not fixing them,
whereas the lower surface consists of alternating stripes  of equal width, where the spins are fixed to $+1$ and $-1$. 
The chemical steps separating the stripes are taken to be sharp.

\par
\begin{figure}[t]
\begin{center}
\includegraphics[width=\linewidth,keepaspectratio]{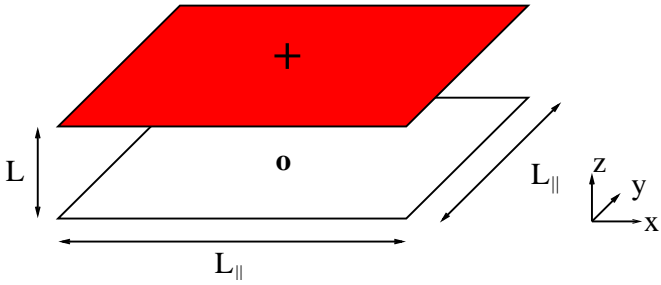}
\end{center}
\caption{(Color online) Film geometry confined by a laterally homogeneous upper surface with fixed spins and by a lower surface with open {\BC}.}
\label{bcplusopen}
\end{figure}
\begin{figure}[b]
\begin{center}
\includegraphics[width=\linewidth,keepaspectratio]{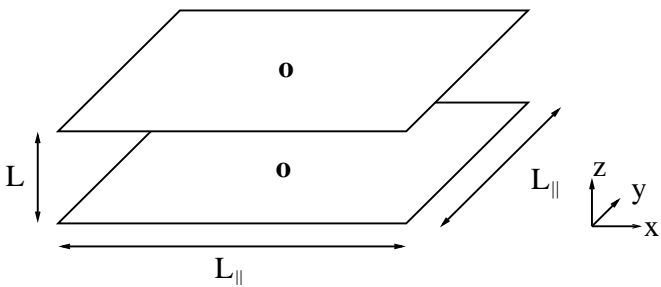}
\end{center}
\caption{Film geometry confined by a lower and an upper surface both with open {\BC}.}
\label{bcopenopen}
\end{figure}

Our results show that, in the limit of  stripe widths small compared to the film thickness, the lower surface effectively realizes Dirichlet {\BC}.
Such {\BC} can also be obtained in the presence of a surface characterized by a locally random adsorption preference, such that on average there is no preferential adsorption for one of the two species \cite{PT-13}.
Thus the system reduces for $\kappa\to0$ to $(+,o)$ or $(o,o)$ {\BC}, and, in order to  be able to compare with this limiting case, 
here we also consider a  film in which both surfaces have a laterally homogeneous {\BC}  from the outset (see Figs.~\ref{bcplusopen} and \ref{bcopenopen}).
This may provide a novel possibility of studying also symmetry-preserving {\BC} for simple fluids and binary liquid mixtures which are
difficult to establish experimentally otherwise \cite{NHB-09}.
{
}
\par

In order to extract universal quantities from {\MC} simulations,  it is important to take corrections to scaling into account in order to be able to extrapolate data for systems of finite size $L$ to the thermodynamic limit $L\rightarrow\infty$. In particular, in the standard three-dimensional Ising model, scaling corrections are proportional to $L^{-\omega}$, with $\omega=0.832(6)$ \cite{Hasenbusch-10}. The presence of nonperiodic boundary conditions, such as in the  direction normal to the film, gives rise to additional scaling corrections, the leading one being proportional to $L^{-1}$, which is numerically difficult to disentangle from the previous one. Following Refs.~\cite{Hasenbusch-09b,Hasenbusch-09c,Hasenbusch-10c,PTD-10,Hasenbusch-11}, in order to avoid the simultaneous presence of these competing corrections, we have studied a so-called improved model \cite{PV-02}, for which the leading scaling corrections $\propto L^{-\omega}$ are suppressed for all observables so that the correction $\propto L^{-1}$ becomes the leading one.
\par
This paper is organized  such that in  
Sec.~\ref{sec:fss} the finite-size scaling behavior, as expected for the system under study, is established. 
In Sec.~\ref{sec:model} we introduce the  lattice model studied here. 
In Secs.~\ref{sec:critical} and \ref{sec:off_criticality} we present our  MC results for the critical Casimir force at $T=T_c$ 
and for the universal scaling function of the critical Casimir force at $T\neq T_c$,  respectively. 
The corresponding results obtained within mean-field theory ($d=4$) are presented in Sec.~\ref{sec:mft} and compared
with the actual behavior in $d=3$ in Sec.~\ref{sec:comparison}.
We summarize our  main findings in Sec.~\ref{sec:summary}. In Appendix \ref{sec:mc} we  provide certain important technical details of the MC simulations. 
In Appendix \ref{sec:bulk} we report details of the determination of the bulk free-energy density which is needed in order to compute the critical Casimir force.

\section{Finite-size scaling and critical Casimir force}
\label{sec:fss}
In this section we recall the finite-size scaling (FSS) behavior of a system in the film geometry 
$L\times  L_\parallel^{d-1}$  in $d$ spatial dimensions, 
which in the thermodynamic limit exhibits a second-order phase transition at the temperature $T=T_c$. Here, we restrict ourselves to the {\BC} described  above; 
a broader discussion of finite-size scaling for nonperiodic {\BC}  can be found in Ref.~\cite{PTD-10}. 
In the following, for the sake of  brevity, we do not analyze separately the FSS behavior of the {\BC} illustrated in Figs.~\ref{bcplusopen} and \ref{bcopenopen}, where there are no stripes.  These two cases can be obtained by taking the limit $\kappa=S_+/L\to 0$ in the {\BC} of Figs.~\ref{bcstripes} and \ref{bcopenstripes}, respectively. 

In the critical region and in the absence of an external  bulk field, the free-energy density $\cal F$ per $k_BT$ of the system (i.e., the free energy divided by $LL_\parallel^{d-1} k_BT$) can be decomposed into a \textit{s}ingular contribution and a \textit{n}on-\textit{s}ingular background term:
\begin{equation}
{\cal F}(t,L,L_\parallel,S_+) = {\cal F}^{\rm (s)}(t,L,L_\parallel,S_+) + {\cal F}^{\rm (ns)}(t,L,L_\parallel,S_+),
\label{free_sns}
\end{equation}
where $t\equiv (T-T_c)/T_c$ is the reduced temperature. The nonsingular background ${\cal F}^{\rm (ns)}$ can be further decomposed into specific geometric  contributions, corresponding to bulk, surface, and line contributions, which are analytic functions of $t$. The singular part of the free-energy density is instead a nonanalytic function of at least one of its variables. According to renormalization-group (RG) theory \cite{Wegner-76} and neglecting corrections to scaling, in spatial dimension $d$ the singular part of the free-energy density obeys the following scaling property:
\begin{eqnarray}
\label{free_full_fss}
&{\cal F}^{\rm (s)}(t,L,L_\parallel,S_+) = \frac{1}{L^d}f\left(\tau,\kappa,\rho\right), \nonumber \\
&\tau \equiv t\left(L/\xi_0^+\right)^{1/\nu}, \nonumber \\
&\kappa\equiv S_+/L, \nonumber\\
&\rho \equiv L/L_\parallel,
\end{eqnarray}
where $\nu$ is the  critical exponent of the bulk correlation length and $\xi_0^+$ is its nonuniversal amplitude,
\begin{equation}
\label{xi_crit}
\xi(t\rightarrow 0^\pm)=\xi_0^\pm|t|^{-\nu}.
\end{equation}
The function $f(\tau,\kappa,\rho)$ is a universal scaling function, i.e., it depends only on the bulk universality class and on the {\BC} applied at the two surfaces. As in Ref.~\cite{SSD-06},  the scaling ansatz in \eref{free_full_fss} generalizes the one for laterally homogeneous {\BC} by  an additional dependence on the scaling variable $\kappa$.  In the following we neglect the dependence on the aspect ratio $\rho\equiv L/L_\parallel$  because here we are interested in the film geometry with $L_\parallel\gg L$. In this limit and for the {\BC} considered here, the dependence on the aspect ratio is expected to be negligible. Our MC data support this observation (see also the discussion in Sec.~\ref{sec:critical} below). The bulk free-energy density $f_{\rm bulk}(t)$ is defined as
\begin{equation}
f_{\rm bulk}(t)\equiv \lim_{L,L_\parallel\rightarrow\infty}{\cal F}(t,L,L_\parallel,S_+)
\end{equation}
and it is independent of the {\BC}. Analogously to \eref{free_sns}, $f_{\rm bulk}(t)$ can also be decomposed into a singular contribution and a nonsingular background,
\begin{equation}
f_{\rm bulk}(t)=f^{(s)}_{\rm bulk}(t)+f^{(ns)}_{\rm bulk}(t)
\end{equation}
with $f^{(s)}_{\rm bulk}(t\to0)\propto |t|^{d\nu}=|t|^{2-\alpha}$, where $\alpha$ is a standard bulk critical exponent. The excess free energy $f_{\rm ex}^{\rm (s)}$ is defined as the remainder of the free-energy density ${\cal F}^{\rm (s)}$ after subtraction of the bulk contribution,
\begin{equation}
\label{free_ex_def}
f_{\rm ex}^{\rm (s)}(t,L,L_\parallel,S_+)\equiv {\cal F}^{\rm (s)}(t,L,L_\parallel,S_+)-f^{\rm (s)}_{\rm bulk}(t).
\end{equation}
According to \eref{free_full_fss} it  exhibits the following scaling  behavior:
\begin{equation}
\label{scaling_excess}
f_{\rm ex}^{\rm (s)}(t,L,L_\parallel,S_+) = \frac{1}{L^d}\Delta\left(\tau=t\left(L/\xi_0^+\right)^{1/\nu}\!,\kappa=S_+/L\right).
\end{equation}
\par
The critical Casimir force $F_C$ per area $L_\parallel^{(d-1)}$ and per $k_BT$ is defined as
\begin{equation}
\label{casimir_def}
F_C\equiv-\frac{\partial \left(Lf^{\rm (s)}_{\rm ex}\right)}{\partial L}\Big|_{t,L_\parallel,S_+}.
\end{equation}
Due to Eqs.~\eqref{free_full_fss}--\eqref{casimir_def}, the critical Casimir force exhibits the following scaling behavior:
\begin{equation}
\label{casimir_fss_leading}
F_C\left(t,L,L_\parallel,S_+\right)=\frac{1}{L^d}\theta\left(\tau= t\left(L/\xi_0^+\right)^{1/\nu},\kappa=S_+/L\right),
\end{equation}
where $\theta(\tau,\kappa)$ is a universal scaling function. At the critical point  one has $\tau=0$, so that  at criticality the force is given by
\begin{equation}
F_C\left(t=0,L,L_\parallel,S_+\right)=\frac{1}{L^d}\Theta\left(\kappa\right),
\end{equation}
with
\begin{equation}
\label{theta_Delta}
\Theta(\kappa)\equiv\theta(0,\kappa).
\end{equation}
\par
In the limit of very narrow stripes, i.e., $\kappa\to0$, the character of a striped surface effectively 
approaches the one for a homogeneous one with $(o)$ {\BC}.
Dirichlet {\BC} are also obtained with an inhomogeneous surface characterized by a locally random adsorption preference, such that on average the fraction of the surface which prefers one component is equal to the fraction which prefers the other one \cite{PT-13}.
Thus, the scaling functions of the critical Casimir force
approach the ones for the critical Casimir force acting on two homogeneous surfaces with $(+,o)$ or $(o,o)$ {\BC}, respectively, i.e.,
\begin{equation} 
  \label{eq:limit-1}
  \theta_{+/o}(\tau,\kappa)\xrightarrow{\kappa\to0}
  \left\{\begin{aligned}
    \theta_{\po}(\tau), &\;\;(+) \ \text{vs stripes for $L\gg S_+$},\\
    \theta_{\oo}(\tau), &\;\;(o) \ \text{vs stripes for $L\gg S_+$,}
  \end{aligned}\right.
\end{equation} 
where the subscript $+/o$ indicates the corresponding type of {\BC} at the homogeneous surface.
\par
On the other hand, for very broad stripes, i.e., $\kappa\to\infty$, the limiting behavior for the case of a homogeneous
$(+)$ wall opposite to a striped surface (\fref{bcstripes}) is given by the average of the two homogeneous cases
for $(+,+)$ and $(+,-)$ {\BC}, respectively.
In this case, i.e.,  for $\kappa\gg1$ the system effectively corresponds to the one for isolated chemical steps opposite to a homogeneous wall,
connecting regions which are almost laterally homogeneous and correspond to $(+,-)$ or $(+,+)$ {\BC}.
As discussed in detail in Ref.~\cite{PTD-10}, every isolated chemical step represents a line defect which gives rise to a contribution to the scaling function of the critical Casimir force proportional to $\rho=L/L_\parallel$. In the present case we have $N_{\rm steps}=L_\parallel/S_+$ of such steps. Thus, assuming additivity, which holds for well separated chemical steps, i.e., for $S_+\gg L$,
the contributions from the nearly isolated chemical steps to the scaling function of the critical Casimir force per unit area vanish $\propto N_{\rm steps}\rho=\kappa^{-1}$.
The asymptotic behavior for $L\ll S_+$ of the universal scaling function for the critical
Casimir force for a $(+)$ wall vs a striped surface is therefore given by
\begin{equation}
  \label{eq:limit-2}
  \theta_+(\tau,\kappa\gg1) = \frac{1}{2}\left(\theta_{\pp}(\tau)+\theta_{\pmbc}(\tau)\right) +\frac{E(\tau)}{2\kappa},
\end{equation} 
where $E(\tau)$ represents the universal contribution of a pair of individual chemical steps, which has been determined in Ref.~\cite{PTD-10}; the factor $2$ in the denominator of the last term of \eref{eq:limit-2} has been chosen as to match with the notation of Ref.~\cite{PTD-10}.
\par
Similarly to \eref{eq:limit-2}, for the case of a $(o)$ wall vs a striped surface (\fref{bcopenstripes})
$\theta(\tau,\kappa)$ approaches
\begin{equation} 
  \label{eq:limit-3}
  \theta_o(\tau,\kappa\gg1) - \theta_{\po}(\tau)\propto \kappa^{-1},
\end{equation} 
because $\theta_{\po}(\tau)= \theta_{\mo}(\tau)$.
\par
\begin{figure}[t]
\begin{center}
\includegraphics[width=\linewidth,keepaspectratio]{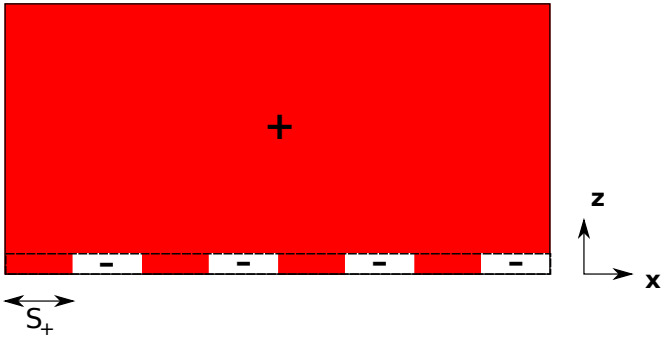}
\end{center}
\caption{(Color online) A section of the ground-state configuration at $y=const$ for the {\BC} of Fig.~\protect\ref{bcstripes} and for the {\BC} of Fig.~\protect\ref{bcopenstripes} with $\kappa < 2$; the ground-state configuration is translationally invariant along the $y$ direction. The dashed line at the alternating bottom denotes the layer of fixed spins. An equivalent configuration is obtained by fixing the spins to $S=-1$ in the region above the alternating bottom layer of fixed spins.}
\label{homogeneousgs}
\end{figure}
\begin{figure}[b]
\begin{center}
\includegraphics[width=\linewidth,keepaspectratio]{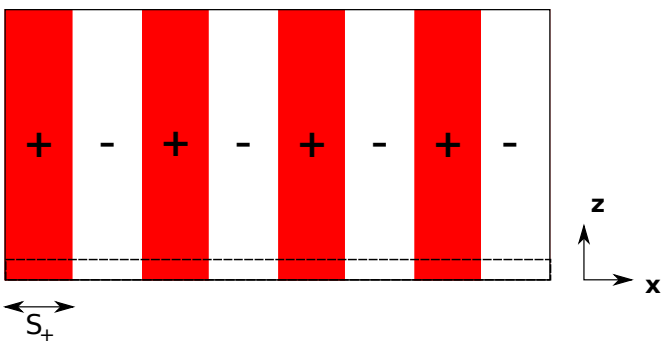}
\end{center}
\caption{(Color online) Same as Fig.~\protect\ref{homogeneousgs} for the {\BC} of Fig.~\protect\ref{bcopenstripes} and for $\kappa>2$.}
\label{stripesgs}
\end{figure}
For $\tau<0$,
due to the presence of the chemical steps between the stripes,
interfaces form, which separate the domains of positive and negative order parameter.
As will be discussed below, for the case of a $(+)$ wall opposite to a striped surface as well as
for a $(o)$ wall opposite to a striped surface and $\kappa<2$, these interfaces align on average
\emph{parallel} to the film surfaces.
In \fref{homogeneousgs} we illustrate the ground-state configuration corresponding to these {\BC}.
By contrast, for a $(o)$ wall opposite to a striped surface and $\kappa>2$ the emerging interfaces
for $\tau<0$ preferentially align \emph{perpendicularly} to the film surfaces in order to minimize the
interface area.
The corresponding ground-state configuration is illustrated in \fref{stripesgs}.
As discussed in Sec.~\ref{sec:mft} below, for the latter case the proportionality
constant in \eref{eq:limit-3} is determined by contributions from these interfaces
and is given by $-R_\sigma|\tau|^\mu$, where $R_\sigma= \sigma_0(\xi_0^+)^{d-1}/(k_BT_c)$ is the universal amplitude ratio
for the interfacial tension $\sigma=\sigma_0|t|^\mu$ associated with the spatially coexisting bulk phases 
and $\mu=(d-1)\nu$ is its critical exponent.
Thus, for the limit $\tau\ll-1$ and $\kappa>2$ the scaling function of the critical Casimir force 
between a $(o)$ wall and a striped surface approaches 
\begin{equation} 
  \label{eq:sum-1}
  \theta_o(\tau\ll-1,\kappa>2)\simeq\theta_{\po}(\tau)-\frac{R_\sigma}{\kappa}|\tau|^{\mu}.
\end{equation} 
Accordingly, the limits for  $\tau\to-\infty$ and $\kappa\to\infty$ do not commute.

\section{Lattice model and observables}
\label{sec:model}
In order to compute the critical Casimir force for a binary liquid mixture close to its critical  demixing point,  as in Ref.~\cite{PTD-10} we study the so-called improved Blume-Capel model \cite{Blume-66,Capel-66} as a representative of the 3D Ising universality class. It is defined on a three-dimensional simple cubic lattice, with a spin variable $S_i$ on each site $i$ which can take the values $S_i=-1$, $0$, $1$. The reduced, dimensionless Hamiltonian for nearest-neighbor interactions is
\begin{equation}
\label{bc}
{\cal H}=-\beta\sum_{<i j>}S_i S_j + D\sum_i S_i^2,\qquad S_i=-1,0,1,
\end{equation}
so that the Gibbs weight is $\exp(-\cal H)$ and the partition function is
\begin{equation}
\label{Z}
Z(\beta,L,L_\parallel)\equiv\sum_{\{\cal C\}} \exp(-{\cal H}),
\end{equation}
where $\{\cal C\}$ is the configuration space of the Hamiltonian  given in \eref{bc}.  We note that the partition function in \eref{Z} depends implicitly also on the {\BC} (see the discussion below). In line with the convention used in Refs.~\cite{Hasenbusch-10c,Hasenbusch-01,Hasenbusch-10,PTD-10}, in the following we shall keep $D$ constant, considering it as a part of the integration measure over $\{S_i\}$, while we vary the coupling parameter $\beta$, which is proportional to the inverse temperature, $\beta\sim 1/T$. In the limit $D\rightarrow -\infty$, one recovers the usual Ising model, because in this limit any state for which there is an $i_0$ such that $S_{i_0}=0$ is suppressed relative to the states $\{S_i=\pm 1\}$. For $d\ge 2$, the model exhibits a phase transition at $\beta_c=\beta_c(D)$ which is second order for $D\le D_{\rm tri}$ and first order for $D>D_{\rm tri}$. The value of $D_{\rm tri}$ in $d=3$ has been determined as $D_{\rm tri}=2.006(8)$ in Ref.~\cite{Deserno-97}, as $D_{\rm tri}\simeq 2.05$ in Ref.~\cite{HB-98}, and more recently as $D_{\rm tri}=2.0313(4)$ in Ref.~\cite{DB-04}.

We consider a three-dimensional  simple cubic lattice $L_z\times L_x\times L_y$, with $L_y=L_x$ and periodic {\BC} in the lateral directions $x$ and $y$. For the two confining surfaces we employ the {\BC} shown in Figs.~\ref{bcstripes}--\ref{bcopenopen}. The {\BC} illustrated in \fref{bcstripes} are realized by fixing the spins at the two surfaces $z=0$ and $z=L_z-1$, so that there are $L_z-2$ layers of fluctuating spins. The spins at the upper surface $z=L_z-1$ are fixed to $+1$, and the lower surface $z=0$  mimics a patterned substrate, so that the surface is divided into stripes of equal width $s_+$ and alternating {\BC} with the spins fixed to $+1$ or $-1$, respectively.

Here and in the following all lengths are measured in units of the lattice constant $a$. The size $L_z$ indicates the total number of lattice layers, including eventually the layers of fixed spins. Therefore the thickness $L$, the lateral size $L_\parallel$, and stripe width $S_+$ are related to the dimensionless lattice lengths $L_z$, $L_x$, and $s_+$ according to $L=(L_z-1)a$, $L_\parallel=L_xa$, and $S_+ = s_+a$, respectively. For the sake of simplicity, here and in the following sections (\ref{sec:critical} and \ref{sec:off_criticality}), we employ a slightly different definition of the scaling variables $\tau$ and $\kappa$. We consider $\tau_l \equiv t (L_z/\xi_{0l}^+)^{1/\nu}$ and $\kappa_l\equiv s_+/L_z$, where $\xi_{0l}^+=\xi_0^+/a$ is the dimensionless nonuniversal amplitude of the correlation length on the {\it l}attice, measured in units of the lattice constant. Accordingly, we also redefine the aspect ratio as $\rho_l\equiv L_z/L_x$. By comparing these new definitions with the previous ones introduced in \eref{free_full_fss}, we observe that, for $L\rightarrow\infty$, $t (L_z/\xi_{0l}^+)^{1/\nu} = t (L/\xi_0^+)^{1/\nu} + O(1/L)$, $s_+/L_z = S_+/L + O(1/L)$, and $L_z / L_x = L/L_\parallel + O(1/L)$. Therefore, the FSS limit, i.e., the limit $L_z\rightarrow\infty$ at fixed $\tau_l$, $\kappa_l$, as well as the limit of vanishing aspect ratio $\rho_l\rightarrow 0$, are unaltered by these new definitions. In order to avoid a clumsy notation, in the following we omit the index $l$.

Here we consider the limit of a vanishing aspect ratio $\rho=L_z/L_x\rightarrow 0$, which is obtained via extrapolation by computing the critical Casimir force for three different aspect ratios $\rho<1$ (see the discussion in the following sections). As discussed at the end of Sec.~\ref{sec:fss}, for the {\BC} illustrated in \fref{bcstripes}, in the limit $\rho\rightarrow 0$ the  subsequent limit $\kappa\equiv s_+/L_z\rightarrow\infty$ corresponds to the presence of an isolated chemical step. In such a geometry, the isolated chemical step gives rise to a line defect which, in turn, results into a linear aspect ratio dependence of the critical Casimir force. In the limit of vanishing aspect ratio the force reduces to the mean value of the force for homogeneous $(+,+)$ and $(+,-)$ {\BC},  for which the two surfaces display the same (respectively, opposite) adsorption preference \cite{PTD-10} [compare with \eref{eq:limit-2}]. In the opposite limit $\kappa\rightarrow 0$, the lower surface is expected to effectively realize Dirichlet {\BC} [compare the upper part of \eref{eq:limit-1}].
Such {\BC} can also be obtained by considering a surface at which the spins are randomly fixed to $+1$ or $-1$ with equal probability; this mimics a surface with a random local adsorption preference, with on average no preferential adsorption for one of the two species \cite{PT-13}.
In order to analyze the limit $\kappa\rightarrow 0$, as a reference system we study a film geometry $L_z\times L_x\times L_x$ with periodic {\BC} in the lateral directions $x$ and $y$, fixed spins at the surface $z=L_z-1$, and open {\BC} on the lower surface, so that there are $L_z-1$ layers of fluctuating spins. This geometry is illustrated in \fref{bcplusopen}. In the following, we shall denote this {\BC} as $(+,o)$.

In addition, we consider the three-dimensional film geometry $L_z\times L_x\times L_x$ with periodic {\BC} in the lateral directions $x$ and $y$, with fixed spins at the lower surface $z=0$ and open {\BC} at the upper surface, so that there are $L_z-1$ layers of fluctuating spins. For the lower surface $z=0$ we employ a  pattern such that the surface is divided  into alternating stripes of equal width $s_+$ with the spins fixed to either $+1$ or $-1$. This geometry is illustrated in \fref{bcopenstripes}. Two interesting limiting cases arise from this geometry. In the limit of large stripes, i.e., for $\kappa=s_+/L_z\rightarrow\infty$ and for vanishing aspect ratio, the lower surface effectively realizes an isolated chemical step. In analogy with the results of Ref.~\cite{PTD-10}, in this limiting case the critical Casimir force is the mean value of the force for $(+,o)$ and {\mo} {\BC}, which corresponds to a film geometry where one of the confining surface implements Dirichlet {\BC}, and the other surface exhibits a homogeneous adsorption preference for one of two components  of the fluid. In the absence of an external bulk magnetic field these two {\BC} are equivalent. Therefore we conclude that in the limit $\kappa=s_+/L_z\rightarrow\infty$ and for vanishing aspect ratio, the critical Casimir force for the {\BC} of \fref{bcopenstripes} reduces to the force for the $(+,o)$ {\BC} illustrated in \fref{bcplusopen} [compare with \eref{eq:limit-3}].

In the opposite limit $\kappa\rightarrow 0$, the lower surface effectively realizes Dirichlet {\BC}, so that the system reduces to a film geometry with Dirichlet {\BC} on both surfaces [compare with the lower part of \eref{eq:limit-1}]. In order to analyze this limit, as a reference system we consider here a three-dimensional film geometry $L_z\times L_x\times L_x$ with periodic {\BC} in the lateral directions $x$ and $y$ and open {\BC} at both surfaces, so that there are $L_z$ layers of fluctuating spins (see \fref{bcopenopen}). In the following we shall denote this film {\BC} as $(o,o)$.

 For the lattice model corresponding to \eref{bc}, the scaling behavior discussed in Eqs.~(\ref{free_full_fss}), (\ref{scaling_excess}), and (\ref{casimir_fss_leading}) is valid only up to  contributions due to corrections to scaling. We distinguish two types of scaling corrections: nonanalytic and analytic ones. The nonanalytic corrections are due to the presence of irrelevant operators. In this case, in \eref{free_full_fss}, additional  scaling field contributions arise, which are characterized by negative RG dimensions. In the FSS limit, i.e., for $L_z\rightarrow\infty$, $t\rightarrow 0$ at fixed $\xi/L_z$, this results in the following expression for the singular part of the free-energy density ${\cal F}^{\rm (s)}$ in the absence of external  bulk fields:
\begin{equation}
\begin{split}
{\cal F}^{\rm (s)}(t,L=a(L_z-1),L_\parallel=aL_x,S_+=as_+) \\
= \frac{1}{L_z^d}\left[f\left(\tau,\kappa,\rho\right) + \sum_{i, k\ge 1} L_z^{ky_i}g_i\left(\tau,\kappa,\rho\right)\right],
\end{split}
\label{free_full_fss_corrections}
\end{equation}
where $y_i<0$, $i\ge 1$, are the RG dimensions of the irrelevant operators and $g_i$ are smooth functions which are universal up to a normalization constant. The leading correction is given by the operator that has the least negative dimension. This is usually denoted by $\omega$, so that the leading scaling corrections are $\propto L_z^{-\omega}$. For the standard three-dimensional Ising model one has $\omega=0.832(6)$ \cite{Hasenbusch-10}. In a family of models characterized by an irrelevant parameter $\lambda$, it can occur that for a certain choice of $\lambda$ the  amplitude of the leading correction-to-scaling term $\propto L_z^{-\omega}$ vanishes.  In these so-called \emph{improved} models, the observed scaling corrections usually decay much more rapidly, i.e., as $L_z^{-\omega_2}$  with $\omega_2=1.67(11)$ according to Ref.~\cite{NR-84} and $\omega_2\simeq 1.89$ according to Ref.~\cite{BJL-07} for the three-dimensional Ising universality class.  This scenario holds for the Blume-Capel model described by \eref{bc}, where $D$ is an irrelevant parameter for $D<D_{\rm tri}$. At $D=0.656(20)$ \cite{Hasenbusch-10} the model is improved. In the present work we fix $D=0.655$, which is the value of $D$ used in most of the recent simulations of the improved Blume-Capel model \cite{Hasenbusch-10c,Hasenbusch-11,Hasenbusch-12b,Hasenbusch-10}. For this value of the reduced coupling $D$ the model is critical for $\beta=\beta_c=0.387721735(25)$ \cite{Hasenbusch-10}.
The presence of two confining surfaces can in general give rise to additional nonanalytic scaling corrections due to the presence of surface irrelevant operators. In particular, the symmetry-breaking {\BC} considered here generate odd-parity irrelevant surface operators, the leading one being the cubic operator; in a field-theoretic approach, such an irrelevant perturbation corresponds to a surface $\phi^3$ term \cite{CD-90}. According to the results of Ref.~\cite{CD-90}, the correction-to-scaling exponent due to this surface operator is $\omega_w=\varepsilon+O(\varepsilon^2)$, in $4-\varepsilon$ spatial dimensions. We are not aware of a quantitatively reliable determination of the RG dimension of such an irrelevant operator. Previous numerical studies \cite{Hasenbusch-10c,Hasenbusch-11,Hasenbusch-12b,PTD-10}, as well as the results which we present here, have not detected the presence of such scaling corrections.

Another type of scaling corrections is provided by so-called analytic scaling corrections, which can stem from various sources. Nonlinear terms in the expansion of the scaling field $\tau$ \cite{AF-83} result in scaling corrections $\propto L_z^{-1/\nu}$. Analytic corrections can also be due to the boundary conditions: {\BC} which are not periodic in all directions induce additional corrections, which are proportional to $L_z^{-1}$. It was first proposed in Ref.~\cite{CF-76}, in the context of studying surface susceptibilities, that such scaling corrections can be absorbed by the substitution $L_z\rightarrow L_z+c$, where $c$ is a nonuniversal, temperature--independent length.  Recently, this property has been checked numerically in Refs.~\cite{Hasenbusch-08,Hasenbusch-09,Hasenbusch-09b} for the $XY$ model with free surfaces, in Ref.~\cite{Hasenbusch-10c} for the Ising model with homogeneously fixed surface spins, and in Refs.~\cite{PTD-10,Hasenbusch-11} for the Ising model with  laterally inhomogeneous surfaces.

Here we study the critical Casimir force using  the improved Blume-Capel model according to \eref{bc}. On the basis of the above discussion, for such a model the leading scaling corrections are expected to be proportional to $L_z^{-1}$. Furthermore, assuming that also in this case in leading order such a scaling correction  can be absorbed by the substitution $L_z\rightarrow L_z+c$, \eref{casimir_fss_leading} is replaced by
\begin{equation}
\begin{split}
F_C\left(t,L=a(L_z-1),L_\parallel=aL_x,S_+=as_+\right)\\
=\frac{1}{(L_z+c)^3}\theta\left(t\left(\frac{L_z+c}{\xi_{0l}^+}\right)^{1/\nu},\frac{s_+}{L_z+c}\right).
\end{split}
\label{casimir_fss}
\end{equation}
In the case of laterally homogeneous {\BC} in Figs.~\ref{bcplusopen} and \ref{bcopenopen}, the dimensionless quantity $c$ (such that $ca$ is a length) enters only via the volume factor and via the scaling variable $\tau$. Scaling corrections to \eref{casimir_fss} are expected to decay as $\propto L_z^{-\omega_2}$ (with $\omega_2=1.67(11)$ \cite{NR-84} or $\omega_2\simeq 1.89$ \cite{BJL-07}, see above).

We introduce the reduced energy density $E(\beta,L_z,L_x,s_+)$ in units of $-k_BT$, which is used in order to compute the critical Casimir force,
\begin{equation}
\label{Edef}
E(\beta,L_z,L_x,s_+) \equiv \frac{1}{V}\Bigg\<\sum_{<i j>}S_i S_j\Bigg\>,
\end{equation}
where $V\equiv L_zL_x^2$ is the total number of spins and $\<\dots\>$ denotes the thermal average. (Note that, according to \eref{bc}, $-\frac{\partial \cal H}{\partial \beta}$ has no contribution $\sim \sum_i S_i^2$.) The reduced free-energy density $F(\beta,L_z,L_x,s_+)$ is defined as
\begin{multline}
\label{Fdef}
F(\beta,L_z,L_x,s_+)\\
\equiv \frac{1}{V}\ln\left(\frac{Z(\beta,L=a(L_z-1),L_\parallel=aL_x)}{Z(0,L=a(L_z-1),L_\parallel=aL_x)}\right).
\end{multline}
 Thus $F(\beta,L_z,L_x,s_+)$ is the free energy per spin and in units of $-k_BT$. It is normalized such that $F(\beta=0,L_z,L_x,s_+)=0$. With this normalization one has
\begin{equation}
\label{F_from_E}
F(\beta,L_z,L_x,s_+) = \int_0^\beta d\beta' E(\beta',L_z,L_x,s_+).
\end{equation}
The relation between ${\cal F}(t,L_z,L_x,s_+)$ and the reduced free-energy density $F(\beta,L_z,L_x,s_+)$ defined in \eref{Fdef} is given by
\begin{multline}
F(\beta,L_z,L_x,s_+) \\= -{\cal F}(t,L=a(L_z-1),L_\parallel=aL_x,S_+=as_+) \\+ {\cal F}(t\rightarrow\infty,L=a(L_z-1),L_\parallel=aL_x,S_+=as_+).
\label{F_from_free_sns}
\end{multline}
Finally, the reduced bulk free-energy density $F_{\rm bulk}(\beta)$ is defined by taking the thermodynamic limit of \eref{Fdef},
\begin{equation}
F_{\rm bulk}(\beta) = \lim_{L_z,L_x\rightarrow\infty}F(\beta,L_z,L_x,s_+).
\label{Fbulkdef}
\end{equation}

\section{Critical Casimir amplitude at $T_c$}
\label{sec:critical}
In order to determine the critical Casimir force at $T_c$, we follow the approach introduced in Ref.~\cite{VGMD-07} and also used in Refs.~\cite{VGMD-08,PTD-10,VMD-11}, which we briefly describe here. For two reduced Hamiltonians ${\cal H}_1$ and ${\cal H}_2$ associated with the same configuration space $\{C\}$ we construct the convex combination ${\cal H}(\lambda)$
\begin{equation}
\label{crossover_H}
{\cal H}(\lambda) \equiv \left(1-\lambda\right){\cal H}_1 + \lambda{\cal H}_2,\qquad \lambda\in \left[0,1\right].
\end{equation}
This Hamiltonian ${\cal H}(\lambda)$ leads to a free energy ${\mathrm F}(\lambda)$ in units of $k_BT$. \footnote{Note that the free energy ${\mathrm F}(\lambda)$ in units of $k_BT$ differs from the reduced free-energy density $F(\beta,L_z,L_x,s_+)$ defined in \eref{Fdef}.} Its derivative is
\begin{equation}
\label{free_derivative}
\frac{\partial {\mathrm F}(\lambda)}{\partial\lambda}=\frac{\sum_{\{C\}}\frac{\partial {\cal H}(\lambda)}{\partial\lambda}e^{-{\cal H}(\lambda)}}{\sum_{\{C\}}e^{-{\cal H}(\lambda)}}.
\end{equation}
Combining Eqs.~(\ref{crossover_H}) and (\ref{free_derivative}) we can determine the free-energy difference as
\begin{equation}
\label{free_diff}
{\mathrm F}(1)-{\mathrm F}(0)=\int_0^1 d\lambda \frac{\partial {\mathrm F}(\lambda)}{\partial\lambda}=\int_0^1 d\lambda \<{\cal H}_2-{\cal H}_1\>_\lambda,
\end{equation}
where $\<{\cal H}_2-{\cal H}_1\>_\lambda$ is the thermal average of the observable ${\cal H}_2-{\cal H}_1$ with the statistical weight $\exp(-{\cal H}(\lambda))$. For every $\lambda$ this average is accessible to standard {\MC} simulations. Finally, the integral appearing in \eref{free_diff} is performed numerically, yielding the free-energy difference between the systems governed by the Hamiltonians ${\cal H}_2$ and ${\cal H}_1$, respectively.

We apply \eref{free_diff} with ${\cal H}_1$ as the Hamiltonian of the lattice $L_z\times L_x\times L_x$ with the {\BC} illustrated in Figs.~\ref{bcstripes}--\ref{bcopenopen}, and ${\cal H}_2$ as the Hamiltonian of the lattice $(L_z-1)\times L_x\times L_x$ plus a  completely separated two-dimensional layer of  noninteracting spins  governed by the reduced Hamiltonian of \eref{bc} with $\beta=0$, so that both Hamiltonians share the same configuration space. This layer can be inserted into the film by varying the coupling $(1-\lambda)\beta$ with its neighboring planes between $0$ and $\beta$. With this we evaluate the following quantity:
\begin{equation}
\label{I_def}
I\left(\beta, L_z, L_x, s_+\right) \equiv \frac{1}{L_x^2}\int_0^1 d\lambda \<{\cal H}_2-{\cal H}_1\>_\lambda.
\end{equation}
By using the definitions of the excess free energy [\eref{free_ex_def}] and of the critical Casimir force [\eref{casimir_def}] one finds \cite{PTD-10}
\begin{multline}
\label{I_casimir}
I\left(\beta, L_z, L_x, s_+\right) = F_{\rm bulk}(\beta) \\
+ F_C\left(t, L=a\left(L_z-\frac{3}{2}\right), L_\parallel=aL_x, S_+=as_+\right),
\end{multline}
where corrections $\propto L_z^{-2}$ have been neglected. In computing the critical Casimir force, the derivative in \eref{casimir_def} is implemented by a finite difference between the free energies of a film of thickness $L=a(L_z-1)$ and of a film of thickness $L-a=a(L_z-2)$, so that the resulting critical Casimir force corresponds to the intermediate thickness $a(L_z-3/2)$. This choice ensures that in the FSS limit no additional scaling corrections $\propto L_z^{-1}$ are generated \cite{PTD-10}. By inserting \eref{casimir_fss} into \eref{I_casimir} we obtain the following scaling form for $I\left(\beta, L_z, L_x, s_+\right)$:
\begin{equation}
\label{I_casimir_scaling}
\begin{split}
&I\left(\beta, L_z, L_x, s_+\right) = F_{\rm bulk}(\beta) \\
&+\frac{1}{\left(L_z-\tfrac{1}{2}+c\right)^3}\theta\left(t\left(\frac{L_z-\tfrac{1}{2}+c}{\xi_0^+}\right)^{1/\nu},\frac{s_+}{L_z-\tfrac{1}{2}+c}\right).
\end{split}
\end{equation}
At the bulk critical temperature \eref{I_casimir_scaling}  turns into
\begin{equation}
\label{I_critical}
\begin{split}
&I\left(\beta_c, L_z, L_x, s_+\right) = F_{\rm bulk}(\beta_c) \\
&+\frac{1}{(L_z-1/2+c)^3}\Theta\left(\frac{s_+}{L_z-1/2+c}\right).
\end{split}
\end{equation}
Equation~(\ref{I_critical}) can be rewritten as
\begin{equation}
\label{I_critical2}
\begin{split}
&I\left(\beta_c, L_z, L_x, s_+\right) = F_{\rm bulk}(\beta_c)\\
& + \frac{1}{(L_z-1/2+\cp)^3}\Theta\left(\kappa=\frac{s_+}{L_z}\right) + O\left(L_z^{-3}\left(\frac{c}{L_z}\right)^2\right),
\end{split}
\end{equation}
with $\cp$ given by
\begin{equation}
\cp=c +\left(c-\frac{1}{2}\right)\frac{\kappa}{3\Theta(\kappa)}\frac{\partial\Theta(\kappa)}{\partial\kappa}.
\label{cprime}
\end{equation}
In a series of {\MC} simulations, we have evaluated the quantity $I\left(\beta_c, L_z, L_x, s_+\right)$ for lattice sizes $L_z=8$, $12$, $16$, $24$, $32$, and $48$ with the {\BC} illustrated in \fref{bcstripes} for $\kappa=1/4$, $1/2$, $1$,  $2$, and $3$ as well as with the {\BC} of \fref{bcplusopen}, which corresponds to the limit $\kappa\rightarrow 0$.
We have also computed $I\left(\beta_c, L_z, L_x, s_+\right)$ for lattice sizes $L=8$, $12$, $16$, $24$, and $32$ with the {\BC} illustrated in \fref{bcopenstripes} for $\kappa=1/4$, $1/2$, $3/4$, $1$, $2$, and $3$ as well as with {\BC} of \fref{bcopenopen}, which corresponds to the limit $\kappa\rightarrow 0$. Certain important details of the simulations are reported in Appendix \ref{sec:mc}. Since we are interested in the film geometry, which corresponds to the limit of a vanishing aspect ratio $\rho=L_z/L_x$, we have simulated every {\BC} for three aspect ratios $\rho\le 1/8$, such that there is always an even number of stripes in the lower confining surface.  An odd or noninteger number of stripes would give rise to a line defect which in turn,  for $\rho\rightarrow 0$, would result into an unwelcome linear aspect-ratio dependence \cite{PTD-10}. Within the present numerical accuracy, for $\rho\le 1/8$ the MC data do not  show a visible dependence on $\rho$. Thus we consider our results obtained for nonvanishing $\rho\le 1/8$ as a  reliable extrapolation to the limit $\rho\rightarrow 0$.  {\it A posteriori}, this also justifies the scaling ansatz in Eqs.~(\ref{scaling_excess})--(\ref{theta_Delta}), in which the dependence on $\rho$ has been neglected.
We have simulated the Blume-Capel  model with the Hamiltonian  given in \eref{bc}, choosing the values of the reduced couplings as $D=0.655$ and $\beta_c=0.387721735$.  This corresponds to the critical point of the improved model \cite{Hasenbusch-10},  for which the \eref{I_critical2}  is expected to describe correctly the corrections to scaling. We  have fitted our {\MC} data  directly to the quantity $I\left(\beta_c=0.387721735, L_z, L_x,s_+\right)$ in \eref{I_critical2}, leaving $F_{\rm bulk}(\beta_c)$, $\Theta$, and $\cp$ as free parameters. In order to control a possible systematic error due to subleading scaling corrections, we have repeated the fits discarding the smallest lattices.  For the {\BC} of Figs.~\ref{bcstripes} and \ref{bcplusopen}, and for various  values of ratio $\kappa$, in Tables \ref{fit_critical} and \ref{fit_critical2} we report the fit results as a function of the smallest lattice size $L_{\rm min}$ taken into account for the fit.  In Tables \ref{fit_open_critical} and \ref{fit_open_critical2} we report the corresponding fit results for the {\BC} of Figs.~\ref{bcopenstripes} and \ref{bcopenopen}.

\begin{table*}
\begin{tabular}{l@{\hspace{2em}}c@{\hspace{4em}}c@{\hspace{4em}}c}
\hline
\hline
$L_{\rm min}$ & $\kappa\rightarrow 0:(+,o)$                        & $\kappa=1/4$                   & $\kappa=1/2$ \\
\hline
$8$          & $\chi^2/DOF=8.7/15$            & $\chi^2/DOF=12.3/15$           & $\chi^2/DOF=16.1/15$       \\
             & $F_{\rm bulk}(\beta_c)=0.0757369(2)$ & $F_{\rm bulk}(\beta_c)=0.0757369(1)$ & $F_{\rm bulk}(\beta_c)=0.0757375(1)$ \\
             & $\Theta_+=0.492(5)$              & $\Theta_+=0.622(5)$              & $\Theta_+=0.845(5)$             \\
             & $\cp=0.36(3)$                   & $\cp=-0.48(2)$                  & $\cp=-0.44(1)$             \\
$12$         & $\chi^2/DOF=8.0/12$            & $\chi^2/DOF=7.5/12$            & $\chi^2/DOF=13.2/12$       \\
             & $F_{\rm bulk}(\beta_c)=0.0757368(2)$ & $F_{\rm bulk}(\beta_c)=0.0757368(2)$ & $F_{\rm bulk}(\beta_c)=0.0757375(2)$ \\
             & $\Theta_+=0.495(10)$             & $\Theta_+=0.634(11)$             & $\Theta_+=0.84(1)$             \\
             & $\cp=0.40(9)$                   & $\cp=-0.39(7)$                  & $\cp=-0.44(5)$             \\
$16$         & $\chi^2/DOF=7.4/9$             & $\chi^2/DOF=6.5/9$             & $\chi^2/DOF=7.7/9$       \\
             & $F_{\rm bulk}(\beta_c)=0.0757368(2)$ & $F_{\rm bulk}(\beta_c)=0.0757368(2)$ & $F_{\rm bulk}(\beta_c)=0.0757372(2)$ \\
             & $\Theta_+=0.50(2)$               & $\Theta_+=0.63(2)$               & $\Theta_+=0.88(2)$             \\
             & $\cp=0.4(2)$                    & $\cp=-0.39(15)$                 & $\cp=-0.23(12)$             \\
\hline
\hline
\end{tabular} 
\caption{Fit of our {\MC} data at $T_c$ for the {\BC} of Figs.~\ref{bcplusopen} and \ref{bcstripes}, to \eref{I_critical2} with free parameters $F_{\rm bulk}(\beta_c)$, $\Theta_+(\kappa=s_+/L_z)$, and $\cp$. $L_{\rm min}$ is the smallest lattice size taken into account for the fit. $DOF$ denotes degrees of freedom. The quoted error bars of the fit parameters correspond to one standard deviation; see, e.g., Ref.~\cite{numrecipes} for a discussion of the method of minimum $\chi^2$ data fitting.}
\label{fit_critical}
\end{table*}

\begin{table*}
\begin{tabular}{l@{\hspace{2em}}c@{\hspace{4em}}c@{\hspace{4em}}c}
\hline
\hline
$L_{\rm min}$ & $\kappa=1$                     & $\kappa=2$                     & $\kappa=3$ \\
\hline
$8$          & $\chi^2/DOF=8.9/15$            & $\chi^2/DOF=12.7/15$           & $\chi^2/DOF=9.0/15$       \\
             & $F_{\rm bulk}(\beta_c)=0.0757370(1)$ & $F_{\rm bulk}(\beta_c)=0.0757366(1)$ & $F_{\rm bulk}(\beta_c)=0.0757369(1)$ \\
             & $\Theta_+=1.383(4)$              & $\Theta_+=1.875(5)$              & $\Theta_+=2.053(4)$             \\
             & $\cp=-0.264(8)$                 & $\cp=-0.138(8)$                 & $\cp=-0.097(5)$             \\
$12$         & $\chi^2/DOF=4.8/12$            & $\chi^2/DOF=11.0/12$           & $\chi^2/DOF=7.0/12$       \\
             & $F_{\rm bulk}(\beta_c)=0.0757369(2)$ & $F_{\rm bulk}(\beta_c)=0.0757367(2)$ & $F_{\rm bulk}(\beta_c)=0.0757369(2)$ \\
             & $\Theta_+=1.387(8)$              & $\Theta_+=1.869(8)$              & $\Theta_+=2.048(8)$             \\
             & $\cp=-0.25(2)$                  & $\cp=-0.15(2)$                  & $\cp=-0.11(2)$             \\
$16$         & $\chi^2/DOF=4.2/9$             & $\chi^2/DOF=7.1/9$             & $\chi^2/DOF=5.0/9$       \\
             & $F_{\rm bulk}(\beta_c)=0.0757369(2)$ & $F_{\rm bulk}(\beta_c)=0.0757368(2)$ & $F_{\rm bulk}(\beta_c)=0.0757369(2)$ \\
             & $\Theta_+=1.394(12)$             & $\Theta_+=1.86(1)$               & $\Theta_+=2.05(1)$             \\
             & $\cp=-0.22(5)$                  & $\cp=-0.18(4)$                  & $\cp=-0.09(3)$             \\
\hline
\hline
\end{tabular}
\caption{Same as Table~\protect\ref{fit_critical} for $\kappa=s_+/L_z=1$, $2$, $3$ and for the {\BC} of \fref{bcstripes}.}
\label{fit_critical2}
\end{table*}

\begin{table*}
\begin{tabular}{l@{\hspace{2em}}c@{\hspace{4em}}c@{\hspace{4em}}c}
\hline
\hline
$L_{\rm min}$ & $\kappa\rightarrow 0:\ (o,o)$                        & $\kappa=1/4$                   & $\kappa=1/2$ \\
\hline
$8$          & $\chi^2/DOF=6.9/12$           & $\chi^2/DOF=7.5/12$            & $\chi^2/DOF=13.7/12$       \\
             & $F_{\rm bulk}(\beta_c)=0.07573678(9)$ & $F_{\rm bulk}(\beta_c)=0.0757369(1)$ & $F_{\rm bulk}(\beta_c)=0.0757369(1)$ \\
             & $\Theta_o=-0.030(2)$             & $\Theta_o=-0.039(2)$             & $\Theta_o=-0.054(1)$             \\
             & $\cp=0.8(2)$                    & $\cp=0.02(9)$                   & $\cp=0.07(6)$             \\
$12$         & $\chi^2/DOF=3.7/9$            & $\chi^2/DOF=3.8/9$             & $\chi^2/DOF=11.0/9$       \\
             & $F_{\rm bulk}(\beta_c)=0.0757368(1)$ & $F_{\rm bulk}(\beta_c)=0.0757370(2)$ & $F_{\rm bulk}(\beta_c)=0.0757369(2)$ \\
             & $\Theta_o=-0.030(5)$             & $\Theta_o=-0.045(5)$             & $\Theta_o=-0.053(3)$             \\
             & $\cp=0.7(7)$                   & $\cp=0.5(4)$                    & $\cp=0.0(3)$             \\
$16$         & $\chi^2/DOF=3.2/6$            & $\chi^2/DOF=2.5/6$             & $\chi^2/DOF=7.5/6$       \\
             & $F_{\rm bulk}(\beta_c)=0.0757368(3)$ & $F_{\rm bulk}(\beta_c)=0.0757369(3)$ & $F_{\rm bulk}(\beta_c)=0.0757368(3)$ \\
             & $\Theta_o=-0.035(15)$             & $\Theta_o=-0.038(10)$            & $\Theta_o=-0.05(1)$             \\
             & $\cp=1.5 \pm 2.3$                 & $\cp=-0.2 \pm 1.2$                 & $\cp=-0.1(9)$             \\
\hline
\hline
\end{tabular}
\caption{Same as Table~\protect\ref{fit_critical} for the {\BC} of Figs.~\ref{bcopenopen} and \ref{bcopenstripes}.}
\label{fit_open_critical}
\end{table*}

\begin{table*}
\begin{tabular}{l@{\hspace{1.5em}}c@{\hspace{2em}}c@{\hspace{2em}}c@{\hspace{2em}}c}
\hline
\hline
$L_{\rm min}$ & $\kappa=3/4$                    & $\kappa=1$                     & $\kappa=2$ & $\kappa=3$ \\
\hline
$8$          & $\chi^2/DOF=9.9/12$             & $\chi^2/DOF=8.0/12$            & $\chi^2/DOF=13.1/12$ & $\chi^2/DOF=12.0/12$\\
             & $F_{\rm bulk}(\beta_c)=0.07573679(9)$ & $F_{\rm bulk}(\beta_c)=0.0757370(1)$ & $F_{\rm bulk}(\beta_c)=0.0757365(2)$ & $F_{\rm bulk}(\beta_c)=0.0757368(2)$ \\
             & $\Theta_o=-0.062(2)$              & $\Theta_o=-0.032(2)$             & $\Theta_o=0.185(4)$ & $\Theta_o=0.287(4)$ \\
             & $\cp=0.37(6)$                    & $\cp=1.3(2)$                    & $\cp=0.34(5)$ & $\cp=0.36(4)$ \\
$12$         & $\chi^2/DOF=7.4/9$              & $\chi^2/DOF=7.9/9$             & $\chi^2/DOF=8.9/9$ & $\chi^2/DOF=8.3/9$ \\
             & $F_{\rm bulk}(\beta_c)=0.0757367(1)$  & $F_{\rm bulk}(\beta_c)=0.0757370(2)$ & $F_{\rm bulk}(\beta_c)=0.0757369(3)$ & $F_{\rm bulk}(\beta_c)=0.0757366(3)$\\
             & $\Theta_o=-0.058(4)$              & $\Theta_o=-0.032(5)$             & $\Theta_o=0.173(9)$ & $\Theta_o=0.292(10)$ \\
             & $\cp=0.1(2)$                     & $\cp=1.2(7)$                    & $\cp=0.04(20)$ & $\cp=0.45(14)$ \\
$16$         & $\chi^2/DOF=4.4/6$              & $\chi^2/DOF=3.4/6$             & $\chi^2/DOF=5.6/6$ & $\chi^2/DOF=6.6/6$ \\
             & $F_{\rm bulk}(\beta_c)=0.0757369(3)$  & $F_{\rm bulk}(\beta_c)=0.0757367(2)$ & $F_{\rm bulk}(\beta_c)=0.0757361(6)$ & $F_{\rm bulk}(\beta_c)=0.0757363(6)$ \\
             & $\Theta_o=-0.07(1)$               & $\Theta_o=-0.021(8)$             & $\Theta_o=0.20(3)$ & $\Theta_o=0.30(2)$ \\
             & $\cp=0.9(8)$                     & $\cp=-0.8 \pm 1.7$                 & $\cp=0.9(6)$ & $\cp=0.65(35)$ \\
\hline
\hline
\end{tabular}
\caption{Same as Table~\protect\ref{fit_open_critical} for $\kappa=s_+/L_z=3/4$, $1$, $2$, $3$ and for the {\BC} of \fref{bcopenstripes}}
\label{fit_open_critical2}
\end{table*}

Inspection of the the fit results tells that we generally reach a good $\chi^2/DOF$  ratio and the results appear to be stable with respect to the choice of $L_{\rm min}$. ($DOF$ is the number of degrees of freedom, i.e., the number of statistically independent points minus the number of fit parameters.) While there is a clear dependence of the Casimir amplitude $\Theta$ on $\kappa$,  as expected the critical bulk free-energy density $F_{\rm bulk}(\beta_c)$ does not exhibit a dependence on $\kappa$. Furthermore, the latter is in agreement with the value $F_{\rm bulk}(\beta_c)=0.0757368(4)$ reported in Ref.~\cite{Hasenbusch-10c}. By conservatively judging the variation of the resulting $\Theta$ with respect to $L_{\rm min}$, from Tables \ref{fit_critical} and \ref{fit_critical2} we obtain the following estimates  for the {\BC} shown in Figs.~\ref{bcstripes} and \ref{bcplusopen}:
\begin{figure}
\begin{center}
\includegraphics[width=0.8\linewidth,keepaspectratio]{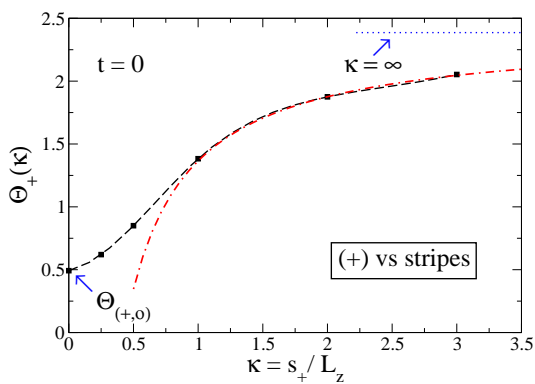}
\end{center}
\caption{(Color online) Critical Casimir force amplitude $\Theta_+(\kappa)=\theta_+(0,\kappa)$ (see Eqs.~(\protect\ref{casimir_fss_leading}) and (\protect\ref{theta_Delta})) at $T_c$  for the {\BC} of Figs.~\ref{bcstripes} and \ref{bcplusopen} and for $\kappa=S_+/L=0$, $1/4$, $1/2$, $1$, $2$, and $3$ as inferred from Tables \protect\ref{fit_critical} and \ref{fit_critical2} (see Eqs.~(\protect\ref{Delta0})-(\protect\ref{Delta3})). The amplitude at $\kappa=0$ is obtained for the $(+,o)$ {\BC} illustrated in Fig.~\protect\ref{bcplusopen}. The dashed line provides a smooth interpolation. The dashed-dotted line gives the estimate of the right-hand side of \eref{eq:limit-2}. These lines saturate at $\Theta_+(\kappa\rightarrow\infty)=\left(\Theta_{(+,+)}+\Theta_{(+,-)}\right)/2=2.386(5)$ \cite{PTD-10}, which is indicated by the dotted line. The omitted statistical error bars defined as one standard deviation and calculated with the standard Jackknife method (see, e.g., Ref.~\cite{AMM-book}) are comparable with the symbol size.}
\label{critvskappa}
\end{figure}
\begin{align}
\label{Delta0}
\text{$(+)$ vs stripes: \hspace{1em}}
&\Theta_+(\kappa=0)=\Theta_{(+,o)}= 0.492(5),\\
\label{Delta0.25}
&\Theta_+(\kappa=1/4)=0.62(1),\\
\label{Delta0.5}
&\Theta_+(\kappa=1/2)=0.85(1),\\
\label{Delta1}
&\Theta_+(\kappa=1)=1.383(4),\\
\label{Delta2}
&\Theta_+(\kappa=2)=1.875(6),\\
\label{Delta3}
&\Theta_+(\kappa=3)=2.053(5).
\end{align}
The subscript $+$ indicates the homogeneous $(+)$ {\BC} on one of the confining surfaces. These amplitudes are shown in \fref{critvskappa}. As expected, for decreasing values of $\kappa$ the critical Casimir amplitude $\Theta(\kappa)$ approaches the corresponding value for $(+,o)$ {\BC}. In particular, $\Theta_+(\kappa=1/4)$ is only $26\%$ larger than $\Theta_+(0)$.  In the opposite limit $\kappa\rightarrow\infty$, $\Theta_+(\kappa)$ approaches the critical Casimir amplitude for a single chemical step: $\Theta_+(\kappa\rightarrow\infty)=2.386(5)$ \cite{PTD-10}. In particular, $\Theta_+(\kappa=3)$ is only $14\%$ smaller than $\Theta_+(\kappa\rightarrow\infty)$.
Moreover, according to \eref{eq:limit-2}, the approach to the limit $\kappa\rightarrow\infty$ is determined by the contribution of the chemical steps. Using the results $\Theta_+(\kappa\rightarrow\infty)=2.386(5)$ and $E(\tau=0)=-2.04(3)$ of Ref.~\cite{PTD-10}, we can obtain the estimates $\Theta_+(\kappa=1/2)=0.35(3)$, $\Theta_+(\kappa=1)=1.37(2)$, $\Theta_+(\kappa=2)=1.876(9)$, and $\Theta_+(\kappa=3)=2.046(7)$. While we observe a large deviation between the estimate for $\kappa=1/2$ and the actual value reported in \eref{Delta0.5}, surprisingly the estimate of \eref{eq:limit-2} agrees rather well even for the relatively small value of $\kappa=1$. In \fref{critvskappa}, too, we compare our results with the estimate of the right-hand side of \eref{eq:limit-2}, finding a nice agreement for $\kappa\gtrsim 1$.
In the whole sampled region, $\Theta_+(\kappa)$ is a positive and monotonically increasing function of $\kappa$ so that the critical Casimir force at $T_c$ is always repulsive. The critical Casimir amplitude $\Theta_+(0)=\Theta_{(+,o)}$ for $(+,o)$ {\BC} can be compared with, e.g., the amplitude $\Theta_{(+,+)}$ resulting from homogeneous {\BC} $(+,+)$, for which the two confining surfaces exhibit the same adsorption preference. Within mean-field theory one has $\Theta_{(+,o)}/\Theta_{(+,+)}=-1/4$ \cite{Krech-97}. According to the MC results of Ref.~\cite{Hasenbusch-10c}, one has $\Theta_{(+,+)}=-0.820(15)$  so that the ratio between the two amplitudes is $\Theta_{(+,o)}/\Theta_{(+,+)}=-0.60(1)$.  Thus the fluctuations  produce a significant dependence of this ratio  on the spatial dimension. Accordingly, one concludes that in $d=3$ mean-field theory captures only the qualitative behavior of the critical Casimir force.  Our result for $\Theta_+(\kappa=0)=\Theta_{(+,o)}=0.492(5)$ is in agreement with the result $\Theta_{(+,o)}=0.497(3)$ of Ref.~\cite{Hasenbusch-11}, while it is not compatible with the earlier results \cite{Krech-97} $\Theta_{(+,o)}=0.33$ and $0.416$ obtained with the $\varepsilon$-expansion method and $0.375(14)$ obtained by {\MC} simulations \cite{Krech-97}.

\begin{figure}
\begin{center}
\vspace{1em}
\includegraphics[width=0.8\linewidth,keepaspectratio]{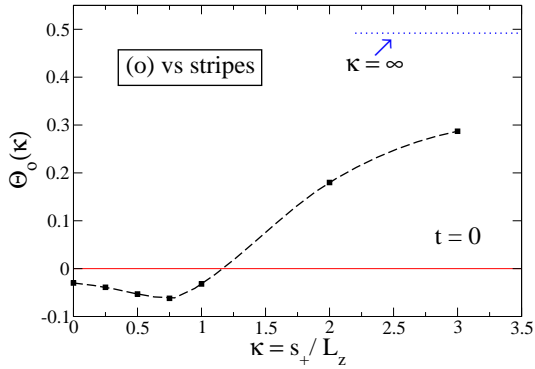}
\end{center}
\caption{(Color online) Critical Casimir force amplitude $\Theta_o(\kappa)=\theta_o(0,\kappa)$ [see Eqs.~(\protect\ref{casimir_fss_leading}) and (\protect\ref{theta_Delta})] at $T_c$ for the {\BC} of Figs.~\ref{bcopenstripes} and \ref{bcopenopen} and for $\kappa=S_+/L=0$, $1/4$, $1/2$, $3/4$, $1$, $2$, $3$, as inferred from Tables \protect\ref{fit_open_critical} and \ref{fit_open_critical2} [Eqs.~(\protect\ref{Deltao0})--(\protect\ref{Deltao3})]. The amplitude at $\kappa=0$ is obtained for the $(o,o)$ {\BC} illustrated in Fig.~\protect\ref{bcopenopen}. The dashed line provides a smooth interpolation. This line saturates at $\Theta(\kappa\rightarrow\infty)=\Theta_{(+,o)}=0.492(5)$ [\eref{Delta0}], which is indicated by the dotted line. The comparison with the thin full line tells that $\Theta_o(\kappa)$ changes sign at $\kappa\approx 1.2$. The omitted statistical error bars are comparable with the symbol size.}
\label{critopenvskappa}
\end{figure}

Inspecting the results reported in Tables \ref{fit_open_critical} and \ref{fit_open_critical2}, we obtain the following estimates for the {\BC} shown in Figs.~\ref{bcopenstripes} and \ref{bcopenopen}:
\begin{align}
\label{Deltao0}
\text{$(o)$ vs stripes: \hspace{1em}}
&\Theta_o(\kappa=0)=\Theta_{(o,o)}=-0.030(5)\\
\label{Deltao0.25}
&\Theta_o(\kappa=1/4)=-0.039(6),\\
\label{Deltao0.5}
&\Theta_o(\kappa=1/2)=-0.053(3),\\
\label{Deltao0.75}
&\Theta_o(\kappa=3/4)=-0.062(4),\\
\label{Deltao1}
&\Theta_o(\kappa=1)=-0.032(3),\\
\label{Deltao2}
&\Theta_o(\kappa=2)=0.18(1),\\
\label{Deltao3}
&\Theta_o(\kappa=3)=0.287(5),
\end{align}
where the subscript $o$ indicates the homogeneous Dirichlet {\BC} on one of the two confining surfaces. These amplitudes are shown in \fref{critopenvskappa}. As expected, for decreasing values of $\kappa$ the critical Casimir amplitude $\Theta_o(\kappa)$ approaches the corresponding value $\Theta_{(o,o)}$ for $(o,o)$ {\BC}, while in the opposite limit $\kappa\rightarrow\infty$ it approaches slowly the value $\Theta_{(+,o)}$ for $(+,o)$ {\BC}. Moreover, the critical Casimir amplitude changes sign: it is attractive for $\kappa=0$ and repulsive for $\kappa\rightarrow\infty$. Inspecting \fref{critopenvskappa}, we can estimate that $\Theta_o(\kappa)$ vanishes for $\kappa\approx 1.2$. Remarkably, different than $\Theta_+(\kappa)$ in \fref{critvskappa}, the critical Casimir amplitude $\Theta_o(\kappa)$ is not monotonic but exhibits a minimum close at $\kappa\approx 3/4$. Our result for $\Theta_o(\kappa=0)=\Theta_{(o,o)}=-0.030(5)$ is in agreement with the recent MC result $\Theta_o(\kappa=0)=\Theta_{(o,o)}=-0.028(16)$ of Ref.~\cite{VMD-11} and also with the earlier results \cite{Krech-97} $\Theta_o(0)=-0.0278$ and $-0.0328$ obtained with the $\varepsilon$-expansion method and $\Theta_o(0)=-0.023(4)$ obtained by MC simulations \cite{Krech-97}.

Finally, we can test the validity of \eref{cprime} by studying the behavior of the scaling corrections in the limit $\kappa\rightarrow 0$. To this end, we consider the {\BC} of \fref{bcstripes} and we take the limit of $\kappa\rightarrow 0$ at fixed $L_z$, i.e., $s_+\rightarrow 0$ in \eref{I_critical}. Assuming that $\Theta(\kappa)$ is analytic close to $\kappa=0$, we obtain
\begin{equation}
I\left(\beta_c, L_z, L_x, s_+\rightarrow 0\right) = F_{\rm bulk}(\beta_c) +\frac{\Theta_+(0)}{(L_z-1/2+c)^3}.
\label{I_critical3}
\end{equation}
A comparison of \eref{I_critical3} with \eref{I_critical2} gives $\cp(\kappa\rightarrow 0) = c$, a result which could also be obtained by taking the limit $\kappa\rightarrow 0$ in \eref{cprime}. On the other hand, in the limit $s_+\rightarrow 0$, the system effectively realizes the {\BC} shown in \fref{bcplusopen} but still in the presence of only $L_z-2$ fluctuating layers of spins (as for the {\BC} in \fref{bcstripes} with $s_+>0$). According to the convention fixed in Sec.~\ref{sec:model}, this corresponds to $(+,o)$ {\BC} for a film with $L_z-1$ layers and thickness $a(L_z-2)$,
\begin{equation}
\begin{split}
I\left(\beta_c, L_z, L_x, s_+\rightarrow 0\right) = I_{(+,o)}\left(\beta_c, L_z-1, L_x\right)\\
= F_{\rm bulk}(\beta_c) +\frac{\Theta_{(+,o)}}{(L_z-1-1/2+c_{(+,o)}')^3},
\end{split}
\label{I_critical4}
\end{equation}
where the subscript $(+,o)$ denotes explicitly the {\BC} of \fref{bcplusopen} with the convention of Sec~\ref{sec:model} and where we have used \eref{I_critical2}. By comparing \eref{I_critical3} with \eref{I_critical4} we finally obtain:
\begin{equation}
\lim_{\kappa\rightarrow 0}\cp(\kappa) = c = \cp_{(+,o)} - 1.
\label{c_from_cplusopen}
\end{equation}
We can extract $\cp_{(+,o)}=0.36(4)$ from the fit results of Table \ref{fit_critical} for the $(+,o)$ {\BC}. This result is in marginal agreement with the result $\cp_{(+,o)}=0.42(2)$ of Ref.~\cite{Hasenbusch-11} in which the same improved Blume-Capel Hamiltonian as the present one has been simulated. \footnote{Notice that, due to a different convention, the value $L_s=1.42(2)$ of the extrapolation length reported in Eq.~(58) of Ref.~\cite{Hasenbusch-11} is related to $\cp_{(+,o)}$ via $L_s=1+\cp_{(+,o)}$.} Using \eref{c_from_cplusopen} we obtain $c=\cp(\kappa\rightarrow 0)=-0.64(4)$. Inspecting the fit results of Tables \ref{fit_critical} and \ref{fit_critical2}, we observe that $\cp(\kappa)$ varies smoothly with $\kappa$ and indeed approaches the value of $c=-0.64(4)$ for $\kappa\rightarrow 0$. According to the results of Eqs.~(\ref{Delta0})--(\ref{Delta3}) and due to \fref{critvskappa}, the coefficient multiplying $(c-1/2)$ in \eref{cprime} is positive. This would imply that, due to $c-1/2<0$, $\cp(\kappa)<c$. However, within the current numerical precision such an inequality appears to be not satisfied by the fit results reported in Tables~\ref{fit_critical} and \ref{fit_critical2}. This suggests that the ansatz of \eref{casimir_fss} does not completely capture the scaling corrections for the striped {\BC}. One may need to modify in addition the second scaling argument of $\theta$ in \eref{casimir_fss}, for example by replacing $L$ with $L+aN$, with $N$ an integer number depending on the convention used to measure the film thickness or, more generally, by introducing a second nonuniversal length. A similar analysis of the scaling corrections for the {\BC} shown in \fref{bcopenstripes} is beyond the presently available numerical precision.

\section{The critical Casimir force scaling function}
\label{sec:off_criticality}
The determination of the critical Casimir force off criticality has been performed using essentially the algorithm introduced in Ref.~\cite{Hucht-07} and also used in Refs.~\cite{Hasenbusch-09b,Hasenbusch-09c,Hasenbusch-09d,Hasenbusch-10c,Hasenbusch-11}. By using the definition of the critical Casimir force given in \eref{casimir_def}, the definition of the reduced free-energy density given in \eref{Fdef}, and the definition of the reduced bulk free energy density given in \eref{Fbulkdef}, the critical Casimir force can be expressed as
\begin{equation}
\label{casimir_fromF}
\begin{split}
F_C\left(t, L=a\left(L_z-\frac{3}{2}\right), L_\parallel=aL_x,S_+=as_+\right)\\
= \Delta F(\beta, L_z, L_x, s_+) - F_{\rm bulk}(\beta),
\end{split}
\end{equation}
where
\begin{equation}
\label{deltaF}
\begin{split}
\Delta F(\beta, L_z, L_x, s_+) \equiv\ &L_zF(\beta, L_z, L_x, s_+)\\
&-(L_z-1)F(\beta, L_z-1, L_x, s_+).
\end{split}
\end{equation}
Analogous to \eref{I_casimir}, in \eref{casimir_fromF} the derivative in \eref{casimir_def} is implemented by a finite difference between the free energies of a film of thickness $L=a(L_z-1)$ and of a film of thickness $L-a=a(L_z-2)$, so that the resulting critical Casimir force corresponds to the intermediate thickness $a(L_z-3/2)$. This choice ensures that in the FSS limit no additional scaling corrections $\propto L_z^{-1}$ are generated \cite{PTD-10}. The reduced temperature $t$ is given by $t=(\beta_c-\beta)/\beta$, with $\beta_c=0.387721735(25)$ \cite{Hasenbusch-10}. As in \eref{I_casimir}, in \eref{casimir_fromF} corrections $\propto L_z^{-2}$ have been neglected.  We note that $\Delta F(\beta, L_z, L_x, s_+) \rightarrow F_{\rm bulk}(\beta)$ for $L_z, L_x\rightarrow\infty$,  which is in accordance with the vanishing of the critical Casimir force in the limit of large volume. Another useful relation follows from a comparison of Eqs.~(\ref{casimir_fromF}) and (\ref{I_casimir}):
\begin{equation}
\label{I_vs_deltaF}
\Delta F(\beta, L_z, L_x, s_+) = I\left(\beta, L_z, L_x, s_+\right).
\end{equation}
Instead of using the coupling parameter approach as in Sec.~\ref{sec:critical}, here we compute the free-energy differences by sampling the internal energy density $E(\beta, L_z, L_x, s_+)$ for various values of $\beta$ and for film thicknesses $a(L_z-1)$ and $a(L_z-2)$. Then $\Delta F(\beta, L_z, L_x, s_+)$ is computed by a numerical integration of \eref{F_from_E}.  For doing so, it is very useful to observe that it is not necessary to perform the integral  in full between $\beta'=0$ and $\beta'=\beta$ \cite{Hasenbusch-10c}. In fact, by inserting  a lower cutoff $\beta_0$ into the integral appearing in \eref{F_from_E} one can effectively compute the  difference between the critical Casimir force and the force at the inverse temperature $\beta_0$.  This implies that the critical Casimir force can be expressed as
\begin{equation}
\label{casimir_from_F_reduced}
\begin{split}
F_C\left(t, L=a\left(L_z-\frac{3}{2}\right), L_\parallel=aL_x, S_+=as_+\right)\\
= \Delta \widehat{F}(L_z, L_x, s_+;\beta,\beta_0)
- (F_{\rm bulk}(\beta)-F_{\rm bulk}(\beta_0))\\
+ F_C\left(t_0, L=a\left(L_z-\frac{3}{2}\right), L_\parallel=aL_x, S_+=as_+\right),
\end{split}
\end{equation}
with
\begin{equation}
\begin{split}
\Delta \widehat{F}(L_z, L_x, s_+;\beta,\beta_0) \equiv L_z\int_{\beta_0}^\beta d\beta' E(\beta',L_z,L_x,s_+)\\
-(L_z-1)\int_{\beta_0}^\beta d\beta' E(\beta',L_z-1,L_x, s_+),
\end{split}
\label{deltaF_reduced}
\end{equation}
and $t_0=(\beta_c-\beta_0)/\beta_0$ as the reduced temperature corresponding to the lower cutoff $\beta_0$. Since  for $L=a(L_z-1)\gg\xi$ the critical Casimir force vanishes $\propto\exp(-L/\xi)$, one can neglect the last term in \eref{casimir_from_F_reduced} if the correlation length $\xi$ at the lower cutoff $\beta_0$ is much smaller than $L=a(L_z-1)$. Moreover,  due to Eqs.~(\ref{I_vs_deltaF}) and (\ref{casimir_fromF}), the last term in \eref{casimir_from_F_reduced}  can be calculated independently with the coupling parameter approach  described in Sec.~\ref{sec:critical}.  This provides a precise control  of any approximation involving the cutoff $\beta_0$. We  did compute $F_C\left(t_0, L=a\left(L_z-\frac{3}{2}\right), L_\parallel=aL_x, S_+=as_+\right)$ within the  aforementioned coupling parameter approach and we have taken into account this term in \eref{casimir_from_F_reduced}  whenever it is relevant within the statistical precision. The numerical integrations in \eref{deltaF_reduced} have been  carried out according to Simpson's rule.  Certain technical details are reported in Appendix \ref{sec:mc}. Finally, the determination of the critical Casimir force  on the basis of \eref{casimir_from_F_reduced} requires the knowledge of the reduced  bulk free-energy density $F_{\rm bulk}(\beta)$ which is independent of the {\BC}.  We have determined  it via MC simulations of lattices size $L_z^3$ with $L_z=24$--$256$ and periodic {\BC}. In Appendix \ref{sec:bulk} we report  certain details of this computation, which is important for a successful determination of $F_C$.

 Along these lines we have computed the critical Casimir force for lattice thickness $L_z=8$, $12$, $16$, and $24$  with the {\BC} shown in Figs.~\ref{bcstripes} and \ref{bcplusopen} as well as for $\kappa=0$, $1/2$, $1$, $2$, and $3$. As in Sec.~\ref{sec:critical} we have considered  three aspect ratios for each value of $L_z$ and $\kappa$; accordingly, we have taken $\rho=1/8$, $1/12$, and $1/16$ for $\kappa\le 2$, as well as $\rho=1/12$, $1/18$, and $1/24$ for $\kappa=3$. We have checked that for these small values the data are independent of $\rho$ within the statistical accuracy. Therefore we  expect that our results  capture the limit $\rho\rightarrow 0$.

 In the present case, for $t\ne 0$ it is not easy to subtract the scaling corrections because according to \eref{casimir_fss} a part of the scaling corrections $\propto 1/L_z$ stem from the dependence on $L_z$ of the second scaling argument of $\theta$. This holds even if the scaling ansatz of \eref{casimir_fss} does not completely capture the $1/L_z$ scaling corrections. In fact, the nonuniversal length $\cp$, defined in \eref{cprime} and extracted from the fits reported in Tables~\ref{fit_critical} and \ref{fit_critical2}, shows a small but significant dependence on $\kappa$, which would be absent if scaling corrections were independent of $\kappa$. In Ref.~\cite{PTD-10} a similar problem was encountered in the {\MC} investigation of the critical Casimir force in the presence of an isolated chemical step. There the dependence of the force on the aspect ratio contributes to the scaling corrections. Since this dependence on $\rho$ was found to be linear, in that case it was possible to eliminate the scaling corrections via a first-order Taylor expansion of the critical Casimir force in $\rho$. As Figs.~\ref{critvskappa} and \ref{critopenvskappa} show, in the present case the critical Casimir force does not follow such a simple dependence on $\kappa$. Furthermore, the possible values of $\kappa$ which can be sampled by the MC simulations are constrained by the fact that the stripe width $s_+$ has to be an integer number. Due to these technical difficulties, here we implement an approximate scheme for the removal of the scaling corrections. For every value of $\kappa$ we extract the nonuniversal length $\cp$ from the fits of Tables~\ref{fit_critical} and \ref{fit_critical2}. Then we employ the substitution $L_z\rightarrow L_z+\cp$. Since such a substitution cannot completely eliminate the scaling corrections $\propto L_z^{-1}$, the resulting scaling function $\theta(\tau,\kappa)$ exhibits a residual scaling correction $\propto \psi(\tau,\kappa)/L_z$, where $\psi(\tau,\kappa)$ is a scaling function. By construction, we have $\psi(0,\kappa)=0$. Thus, since $\psi$ is a continuous function, there is an interval around $\tau=0$ in which the residual scaling corrections are negligible with respect to the numerical precision. Furthermore, for $\kappa\rightarrow 0$ and $\kappa\rightarrow\infty$ this method becomes exact and, thus, we have $\psi(\tau,\kappa\rightarrow 0) = \psi(\tau,\kappa\rightarrow\infty)=0$. Therefore, the interval of validity around $\tau=0$ is expected to increase as $\kappa$ is lowered towards $0$ or is increased toward $\infty$.

\begin{figure}
\vspace{3em}
\begin{center}
\includegraphics[width=0.85\linewidth,keepaspectratio]{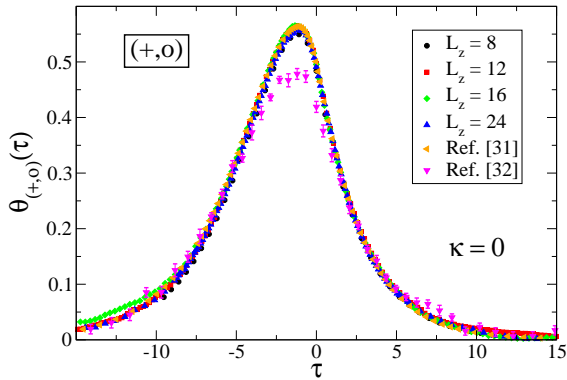}.

\end{center}
\caption{(Color online) The universal scaling function $\theta_{(+,o)}(\tau)$ of the critical Casimir force for the {\BC} $(+,o)$ shown in Fig.~\protect\ref{bcplusopen}, corresponding to the limit $\kappa=s_+/L_z\to0$  of the {\BC} shown in \fref{bcstripes}. Scaling corrections have been subtracted by using $\cp=0.36(4)$ (see the main text). We also compare our results with those of Ref.~\cite{Hasenbusch-11} for $L=16$ and of Ref.~\cite{VMD-11} for $L=20$. The omitted statistical error bars are, apart from $\tau\lesssim -10$, comparable with the symbol size.}
\label{casimir0}
\end{figure}

\begin{figure}
\vspace{3em}
\begin{center}
\includegraphics[width=0.8\linewidth,keepaspectratio]{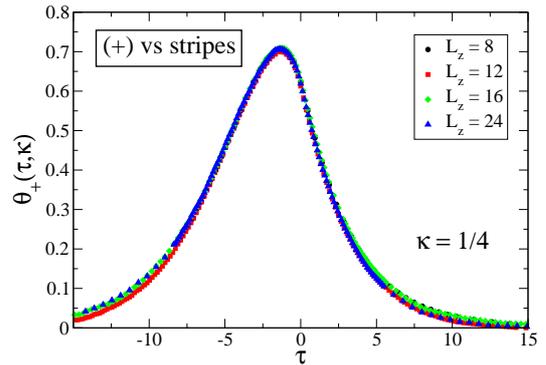}
\end{center}
\caption{(Color online) The universal scaling function $\theta_+(\tau,\kappa)$ of the critical Casimir force for the {\BC} shown in Fig.~\protect\ref{bcstripes} with $\kappa=S_+/L=1/4$  and $\cp=-0.48(2)$. The omitted statistical error bars are comparable with the symbol size.}
\label{casimir0.25}
\end{figure}

\begin{figure}
\vspace{3em}
\begin{center}
\includegraphics[width=0.8\linewidth,keepaspectratio]{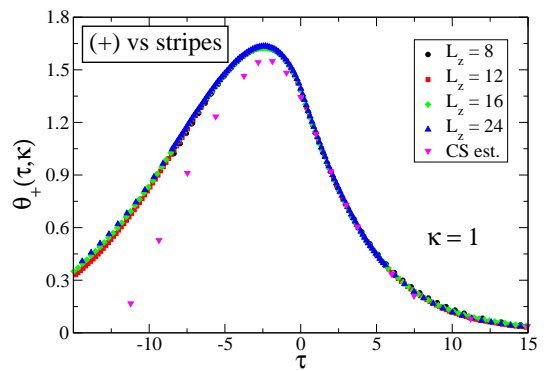}
\end{center}
\caption{(Color online) Same as Fig.~\protect\ref{casimir0.25} for $\kappa=1$  and $\cp=-0.26(1)$. The results are compared with the chemical-step estimate (CS est.) given in \eref{eq:limit-2}.}
\label{casimir1}
\end{figure}

\begin{figure}
\vspace{3em}
\begin{center}
\includegraphics[width=0.8\linewidth,keepaspectratio]{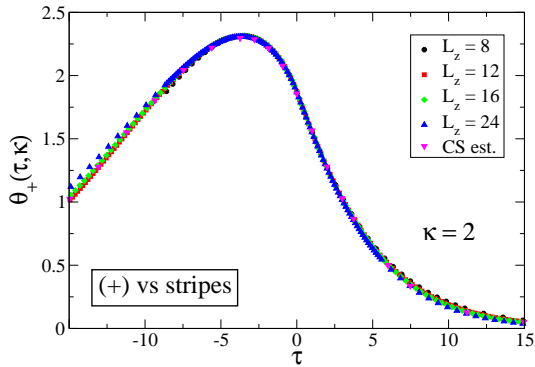}
\end{center}
\caption{(Color online) Same as Fig.~\protect\ref{casimir1} for $\kappa=2$  and $\cp=-0.14(1)$.}
\label{casimir2}
\end{figure}

\begin{figure}
\vspace{3em}
\begin{center}
\includegraphics[width=0.8\linewidth,keepaspectratio]{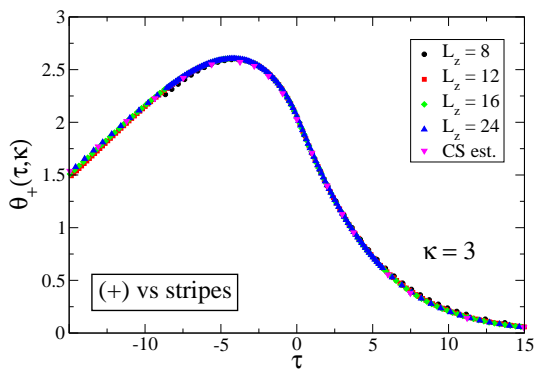}
\end{center}
\caption{(Color online) Same as Fig.~\protect\ref{casimir1} for $\kappa=3$  and $\cp=-0.10(1)$.}
\label{casimir3}
\end{figure}

In \fref{casimir0} we show our results for the {\BC} shown in \fref{bcplusopen}, corresponding to the limit $\kappa=s_+/L_z\rightarrow 0$ of the {\BC} shown in \fref{bcstripes}. In order to normalize the scaling variable $\tau$, one needs the value of the nonuniversal amplitude $\xi_{0l}^+$ of the correlation length $\xi$. From Ref.~\cite{Hasenbusch-10c} we  infer $\xi_{0l}^+=0.4145(4)$ in units of the lattice constant. As for the critical exponent $\nu$, we use the recent MC result $\nu=0.63002(10)$ of Ref.~\cite{Hasenbusch-10}. In \fref{casimir0} we also compare our results with those of Refs.~\cite{Hasenbusch-11} and \cite{VMD-11}. We observe a perfect agreement with the results of Ref.~\cite{Hasenbusch-11}, which in fact have been obtained by simulating precisely the same improved Blume-Capel model. The comparison with the results of Ref.~\cite{VMD-11} is less satisfactory and reveals a difference between the curves around the position of their maximum in the low-temperature phase, i.e., $\tau <0$. This difference may be due to the fact that the Ising model simulated in Ref.~\cite{VMD-11} suffers from larger scaling corrections than the improved model used here, which makes the extrapolation of the FSS limit more difficult. For the {\BC} illustrated in \fref{bcstripes}, in Figs.~\ref{casimir0.25}, \ref{casimir0.5}, \ref{casimir1}, \ref{casimir2}, and \ref{casimir3} we show our results for the scaling function $\theta_+(\tau, \kappa)$, for $\kappa=1/4$, $1/2$, $1$, $2$, and $3$, respectively.

\begin{figure}
\vspace{3em}
\begin{center}
\includegraphics[width=0.8\linewidth,keepaspectratio]{theta_stripes0.5_c-0.44_1.eps}
\end{center}
\caption{(Color online) Same as Fig.~\protect\ref{casimir0.25} for $\kappa=1/2$  and $\cp=-0.44(1)$.}
\label{casimir0.5}
\end{figure}

\begin{figure}
\vspace{3em}
\begin{center}
\includegraphics[width=0.8\linewidth,keepaspectratio]{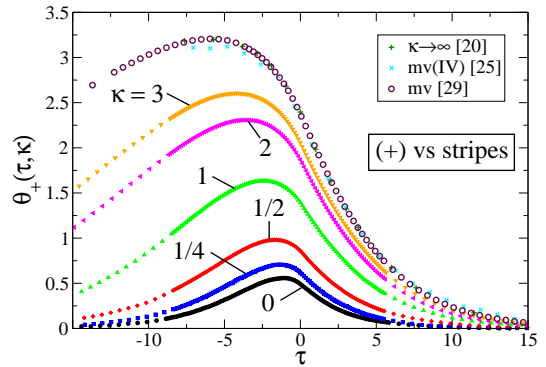}
\end{center}
\caption{(Color online) Comparison of the universal scaling function $\theta_+(\tau,\kappa)$ for $\kappa=0$, $1/4$, $1/2$, $1$, $2$, and $3$ as determined with $L=24$. We compare  the data also with the scaling function $\theta_+(\tau,\kappa\rightarrow\infty)$  in the limit of vanishing aspect ratio $\rho$, as obtained in Ref.~\cite{PTD-10} with $L=16$. The limit $\kappa\rightarrow\infty$ corresponds to the critical Casimir force between a homogeneous $(+)$ surface and a surface with an isolated chemical step which, for $\rho\rightarrow 0$, results in the mean value of the critical Casimir force for laterally homogeneous $(+,+)$ and $(+,-)$ {\BC} \cite{PTD-10}. We compare the results also with those latter mean values, which are either extracted from the so-called approximant IV of Ref.~\cite{VGMD-08} [mv (IV)] or which stem from the results of Ref.~\cite{Hasenbusch-10c} (mv).}
\label{casimirall}
\end{figure}

Inspection of Figs.~\ref{casimir0}--\ref{casimir3} reveals a satisfactory scaling collapse for the lattice sizes considered here. This supports the validity of the procedure described above to eliminate the scaling corrections.
In Figs.~\ref{casimir1}--\ref{casimir3} we also compare our results with the asymptotic estimate given in \eref{eq:limit-2}, which describes the approach to the limit $\kappa\rightarrow\infty$. For this purpose we have used the data of Ref.~\cite{Hasenbusch-10c} for computing the mean value $[\theta_{\pp}(\tau)+\theta_{\pmbc}(\tau)]/2$ and the results of Ref.~\cite{PTD-10} for the chemical-step contribution $E(\tau)$, as determined therein for thickness $L_z=12$. For $\kappa=1$ (\fref{casimir1}), the estimate of \eref{eq:limit-2} agrees well with our results for $\tau >0$, while for $\tau<0$ it shows a systematic deviation from $\theta(\tau,\kappa=1)$. For $\kappa\ge 2$ (Figs.~\ref{casimir2} and \ref{casimir3}), the chemical-step estimate given in \eref{eq:limit-2} agrees very well the {\MC} results throughout the critical region.
In \fref{casimirall} we show a comparison of the critical Casimir force for $\kappa=0$, $1/4$, $1/2$, $1$, $2$, and $3$, as obtained for $L_z=24$. We also compare the present results with the universal scaling function which describes the critical Casimir force for an isolated chemical step in the limit of vanishing aspect ratio, as determined in Ref.~\cite{PTD-10}.  This system corresponds to the limit $\kappa\rightarrow\infty$ and results in the mean value of the critical Casimir force for laterally homogeneous $(+,+)$ and $(+,-)$ {\BC}. In the whole range $0\le\kappa\le\infty$ the critical Casimir force is always repulsive.  This is expected because the stripe width for $(+)$ and for $(-)$ {\BC} are equal and the repulsive critical Casimir force for $(+,-)$ {\BC} is stronger than the attractive one for $(+,+)$ {\BC} \cite{VGMD-08}. In \fref{casimirall} we also show a comparison with the mean value of the critical Casimir force for the homogeneous $(+,+)$ and $(+,-)$ {\BC}, as obtained by MC simulations in Refs.~\cite{VGMD-08,Hasenbusch-10c}.

\begin{figure}
\begin{center}
\includegraphics[width=0.85\linewidth,keepaspectratio]{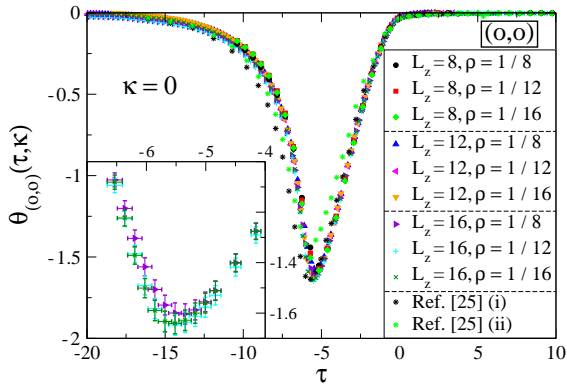}
\end{center}
\caption{(Color) Universal scaling function $\theta_{(o,o)}(\tau)$ of the critical Casimir force for the {\BC} $(o,o)$ shown in Fig.~\protect\ref{bcopenopen}, corresponding to the limit $\kappa=s_+/L_z\rightarrow 0$ of the {\BC} shown in \fref{bcopenstripes}. Scaling corrections have been subtracted by using $\cp=0.8(2)$ (see the main text). We compare our results with those of Ref.~\cite{VGMD-08} obtained from the approximants (i) and (ii) presented therein and for the film thickness $L=20$. The inset provides a magnification of the resulting curves close to the minimum of the force, for the largest available film thickness $L=16$ and for the three aspect ratios $\rho\equiv L_z/L_x$ considered here.}
\label{casimir_open0}
\end{figure}

\begin{figure}
\vspace{3em}
\begin{center}
\includegraphics[width=0.8\linewidth,keepaspectratio]{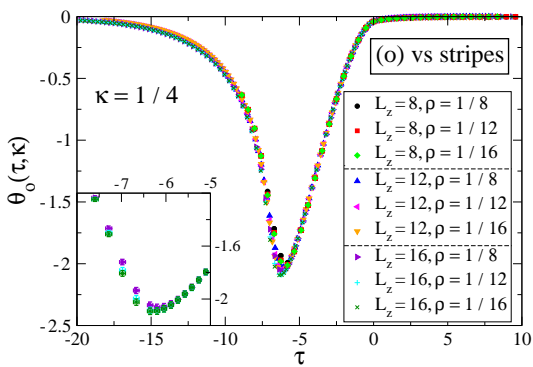}
\end{center}
\caption{(Color) Universal scaling function $\theta_o(\tau,\kappa)$ of the critical Casimir force for the {\BC} shown in Fig.~\protect\ref{bcopenstripes} with $\kappa=s_+/L_z=1/4$ and $\cp=0.02(9)$. The data points for $L=8$ and $\rho=1/8$, $1/12$ are hardly visible because they overlap with the other data sets. The inset provides a magnification of the resulting curves close to the minimum of the force, for the largest available film thickness $L=16$ and for the three aspect ratios $\rho\equiv L_z/L_x$ considered here.}
\label{casimir_open0.25}
\end{figure}

\begin{figure}
\vspace{3em}
\begin{center}
\includegraphics[width=0.8\linewidth,keepaspectratio]{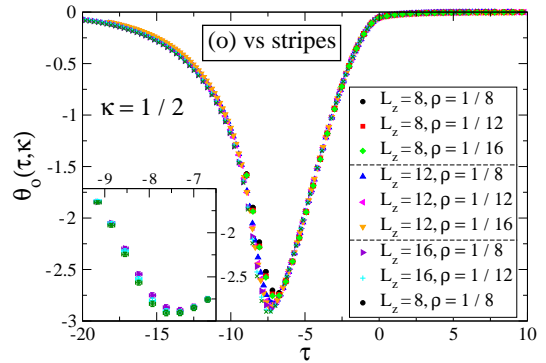}
\end{center}
\caption{(Color) Same as Fig.~\protect\ref{casimir_open0.25} for $\kappa=1/2$ and $\cp=0.05(8)$.}
\label{casimir_open0.5}
\end{figure}

\begin{figure}
\vspace{3em}
\begin{center}
\includegraphics[width=0.8\linewidth,keepaspectratio]{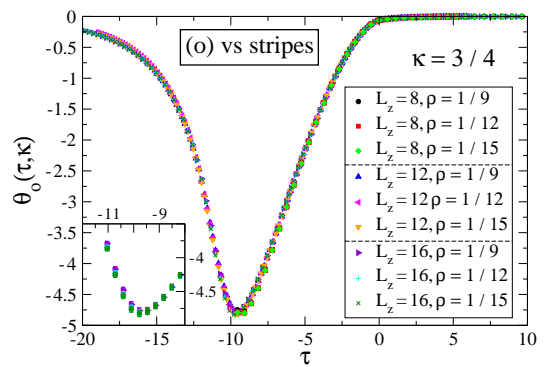}
\end{center}
\caption{(Color) Same as Fig.~\protect\ref{casimir_open0.25} for $\kappa=3/4$ and $\cp=0.37(7)$.}
\label{casimir_open0.75}
\end{figure}

\begin{figure}
\vspace{3em}
\begin{center}
\includegraphics[width=0.8\linewidth,keepaspectratio]{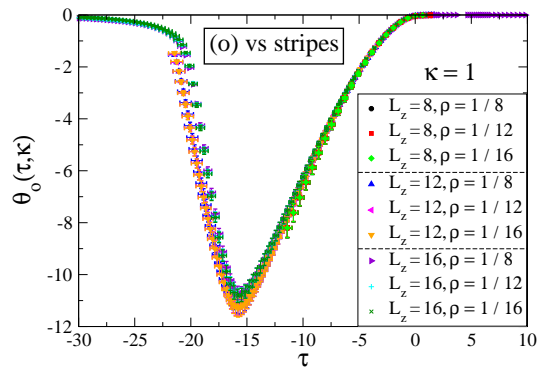}
\end{center}
\caption{(Color) Same as Fig.~\protect\ref{casimir_open0.25} for $\kappa=1$ and $\cp=1.3(2)$.}
\label{casimir_open1}
\end{figure}

\begin{figure}
\vspace{3em}
\begin{center}
\includegraphics[width=0.8\linewidth,keepaspectratio]{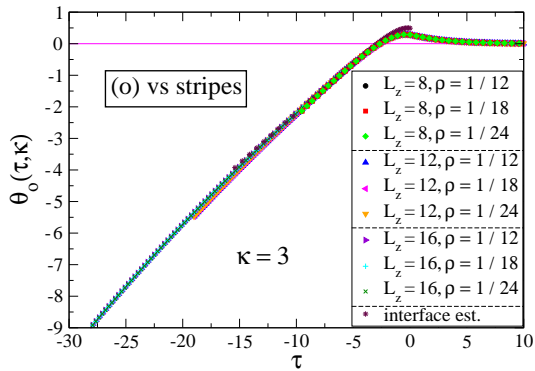}
\end{center}
\caption{(Color) Same as Fig.~\protect\ref{casimir_open0.25} for $\kappa=3$ and $\cp=0.36(9)$. We also compare our results with the interface estimate given by the right-hand side of \eref{eq:sum-1}. The scaling function changes sign at $\tau=\tau_0\simeq -2.7$.}
\label{casimir_open3}
\end{figure}

In \fref{casimir_open0} we show our results for the $(o,o)$ {\BC} shown in \fref{bcopenopen}, corresponding to the limit $\kappa=s_+/L_z\rightarrow 0$ of the {\BC} $(o)$ vs stripes shown in \fref{bcopenstripes}. We also compare our results with those of Ref.~\cite{VGMD-08} for the approximants (i) and (ii) presented therein. The approximant (i) agrees with our results for $\tau \gtrsim -6$, whereas the approximant (ii) displays a systematic deviation from our results. For $\tau \lesssim -6$ both approximants show a disagreement with our results. While the approximant (ii) displays a small but visible deviation from our results, the approximant (i) exhibits a larger, systematic deviation from our results. Such deviations may be due to the difficulty in extrapolating the FSS limit of the Ising model used in Ref.~\cite{VGMD-08}, which exhibits larger scaling corrections than the improved model of \eref{bc}. For the {\BC} illustrated in \fref{bcopenstripes}, in Figs.~\ref{casimir_open0.25}, \ref{casimir_open0.5}, \ref{casimir_open0.75}, \ref{casimir_open1}, and \ref{casimir_open3} we show our results for the scaling function $\theta_o(\tau, \kappa)$, for $\kappa=1/4$, $1/2$, $3/4$, $1$, and $3$, respectively.

The numerical determination of the critical Casimir forces in the presence of a Dirichlet {\BC} at one of the two confining surfaces has turned out to be much more involved than the computation for the {\BC} of Figs.~\ref{bcstripes} and \ref{bcplusopen}. First, at variance with the previous cases, we observed the onset of a dependence of the critical Casimir force on the aspect ratio $\rho=L_z/L_x$. As illustrated in the insets of Figs.~\ref{casimir_open0}--\ref{casimir_open3}, such a dependence on $\rho$ appears in a narrow interval of $\tau$ in the low-temperature phase. Although small, the differences between the calculated scaling functions $\theta_o(\tau,\kappa)$ for the three aspect ratios considered here is visible and larger than the statistical error bars. \footnote{We note that the error bars shown in Figs.~\ref{casimir_open0}--\ref{casimir_open3} are the sum of the statistical error bars originating from the MC sampling and the uncertainty in the determination of $\cp$, this last one being the dominant contribution to the error bars. The dependence on $\rho$ is more clearly seen in the raw MC data.} The observed dependence on $\rho$ implies the onset of a lateral correlation length, associated with an ordering process in the low-temperature phase. In order to understand this point, it is useful to consider the limit $\beta\rightarrow\infty$, i.e., the ground state of the model with the {\BC} illustrated in Figs.~\ref{bcopenstripes} and \ref{bcopenopen}. For the {\BC} shown in \fref{bcopenopen}, it is easy to see that the ground state is a spatially homogeneous state in which all spins take the same value. For the {\BC} shown in \fref{bcopenstripes}, besides the homogeneous state shown in \fref{homogeneousgs}, one can consider also a ``striped'' state, in which each spin in the film takes the value corresponding to the underlying stripe, so that the configuration of the system consists of columns of cross-sectional area $s_+\times L_x$ and height $L_z$. In \fref{stripesgs} we illustrate such a configuration. In view of the periodic {\BC} in the two lateral directions, the area $\cal A$ of the interface between $+$ and $-$ spins is given by
\begin{equation}
  \label{eq:transition-area}
\begin{aligned}
&{\cal A} = \frac{L_x^2}{2}, &\text{homogeneous state},\\
&{\cal A} = \frac{L_x}{s_+}L_zL_x = \frac{L_x^2}{\kappa}, &\text{striped state}.\\
\end{aligned}
\end{equation}
Thus, at low temperature, the system orders in a homogeneous state for $\kappa<2$ and in a striped state for $\kappa>2$. As a function of the parameter $\kappa$, the ground state undergoes a first-order transition at $\kappa=2$. Moreover, for $\kappa=2$, besides the homogeneous (see \fref{homogeneousgs}) and the striped (see \fref{stripesgs}) ground states, there are other states which have the same (minimal) energy: such states can be obtained by flipping the value of the spins in a single column in the striped state illustrated in \fref{stripesgs}. We note that the number of these additional ground states diverges in the thermodynamic limit. The emergence of these ground states at $\kappa=2$ gives rise to a sort of glassy behavior at low temperatures, which results in a considerable technical difficulty in simulating these systems. We leave this issue for future research.

This lateral ordering process at low temperatures corresponds to a phase transition which occurs in the film geometry characterized by the {\BC} described by Figs.~\ref{bcopenstripes} and \ref{bcopenopen}. This causes the dependence on the aspect ratio exhibited in Figs.~\ref{casimir_open0}--\ref{casimir_open3}. We note that, for the {\BC} corresponding to Figs.~\ref{bcstripes} and \ref{bcplusopen}, the striped state illustrated in \fref{stripesgs} is never a ground state. Moreover, without an external bulk field the presence of a surface field at the upper surface rounds the transition between the paramagnetic high-temperature phase and the homogeneous ground state to a simple crossover. This is in agreement with the independence of $\rho$ observed in Figs.~\ref{casimir0}--\ref{casimir3}. The appearance of a lateral correlation length breaks the scaling behavior discussed in Sec.~\ref{sec:fss}. On the other hand, inspection of Figs.~\ref{casimir_open0}--\ref{casimir_open3} reveals that the data for the two smallest aspect ratios agree within the statistical error. Therefore, since one expects a smooth dependence of the scaling function $\theta_o(\tau,\kappa)$ on $\rho$, in particular in the limit of $\rho\rightarrow 0$, we can regard our results for the smallest aspect ratio as a reliable extrapolation of the limit $\rho\rightarrow 0$.

Another difficulty in the numerical determination of the critical Casimir force for the {\BC} shown in Figs.~\ref{bcopenstripes} and \ref{bcopenopen} lies in the fact that the scaling function $\theta_o(\tau,\kappa)$ exhibits a minimum in the low-temperature phase which is shifted towards more negative values of $\tau$ upon increasing $\kappa$. Thus, in order to study this important feature of the scaling function, one has to generate MC data for temperatures lower than the ones needed for the {\BC} shown in Figs.~\ref{bcstripes} and \ref{bcplusopen}. Upon lowering the temperature the simulations become increasingly difficult because of the appearance of many metastable states associated with the aforementioned ground-state phase transition at $\kappa=2$.

Finally, in order to eliminate the leading scaling corrections, we have implemented the procedure outlined above. We note that for the {\BC} shown in \fref{bcopenstripes} such a method appears to be less reliable. While for $\kappa\le 1/2$ and $\kappa=3$ the overall scaling collapse is good, for $\kappa=3/4$ and for sufficiently negative values of $\tau$, there is a small but systematic deviation between the data for lattice size $L=12$ and $L=16$. The scaling collapse is even worse for $\kappa=1$; in this case a further complication seems to be that, apparently, in this case scaling corrections are stronger (see Table~\ref{fit_open_critical2}).

According to the discussion in Sec.~\ref{sec:model}, for the {\BC} shown in \fref{bcopenstripes} in the limit $\kappa\rightarrow\infty$ one expects to recover the {\BC} shown in \fref{bcplusopen}. Since for $\kappa=0$ the force is always attractive (see \fref{casimir_open0}) and for $\kappa\rightarrow\infty$ the force is repulsive (see \fref{casimir0}), at a certain intermediate value of $\kappa$ the force has to change sign. According to \fref{critopenvskappa}, at criticality this occurs at $\kappa=\kappa_0\approx 1.2$. Besides a change of sign of the force as a function of $\kappa$ there is also a change of sign as a function of $\tau$. This is nicely illustrated in \fref{casimir_open3}, where for $\kappa=3$ the force is found to be repulsive (respectively attractive) for $\tau\ge\tau_0$ (respectively $\tau\le\tau_0$), with $\tau_0\approx -2.7$. This implies that in the scaling regime and for a given temperature $T<T_c$, i.e., $t=(T-T_c)/T_c<0$ the force is repulsive (respectively attractive) for $L<L_0(t)$ [respectively $L>L_0(t)$], with $ L_0(t)=\xi_{0l}(\tau_0/t)^\nu$. Therefore $L=L_0(t)$ is a mechanically stable point of equilibrium for the critical Casimir force which can be sensitively tuned by varying the reduced temperature. This can be exploited for levitation purposes \cite{TKGHD-10}. In \fref{casimir_open3} we also compare our result with the interface estimate, i.e., the right-hand side of \eref{eq:sum-1}, which is expected to hold for $\kappa>2$ and $\tau \ll -1$. To this end, we employ the estimate of the universal amplitude ratio $R_\sigma=0.377(11)$ \cite{ZF-96}. The interface estimate is in nice agreement with our MC results for $\tau\lesssim -3.5$.

In principle, the determination of the full scaling function of the critical Casimir force at $\kappa=\kappa_0\approx 1.2$ would be of particular interest. According to the discussion in Sec.~\ref{sec:model}, due to $\kappa_o<2$ the scaling function $\theta_o(\tau,\kappa_0)$ is expected to develop a minimum for $\tau<0$ and to vanish for $\tau\rightarrow\pm\infty$. Therefore, if $\tau=0$ is the only zero of $\theta_o(\tau,\kappa_0)$, the function $\theta_o(\tau,\kappa_0)$ must have a positive maximum for $\tau>0$; in the presence of additional zeros beside the one at $\tau=0$, the scaling function $\theta_o(\tau,\kappa_0)$ may exhibit additional stationary points. Unfortunately, the study of such an interesting case is beyond the current technical capacities. On one hand, we note that for $\tau >0$ and within the available numerical precision the scaling function for the value of $\kappa$ closest to $\kappa_o$, i.e., $\theta_o(\tau, \kappa=1)$, is hardly distinguishable from $0$. Thus the possible stationary points of $\theta_o(\tau,\kappa_0)$ for $\tau >0$ and for $\tau<0$ close to $\tau=0$ are expected to be undetectable within the presently available precision. Moreover, the minimum in the low-temperature phase for $\kappa=\kappa_0$ is expected to be shifted towards a more negative value of $\tau$ with respect to the corresponding minimum for $\kappa=1$; this fact could lead to further technical difficulties, because lower temperatures have to be investigated in order to study the critical Casimir force close to this minimum.
On the other hand, it is even technically impossible to simulate the present lattice Hamiltonian for a generic value of $\kappa$. This is so because all lattice lengths $L_z$, $L_x$, and $s_+$ must be integer numbers. Even so, the need of studying several values of $L_z$ together with the limited computational resources, further constraints the (rational) values of $\kappa$ which can be analyzed.

\begin{figure}
\begin{center}
\includegraphics[width=0.8\linewidth,keepaspectratio]{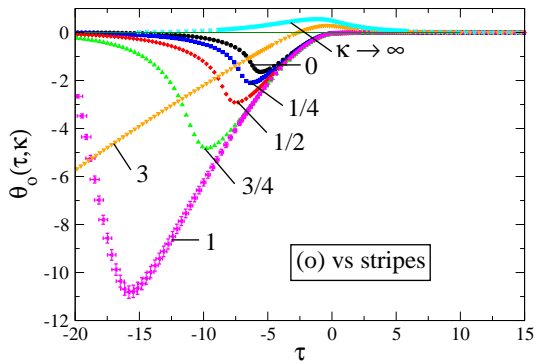}
\end{center}
\caption{(Color online) Comparison of the universal scaling function $\theta_o(\tau,\kappa)$ for $\kappa=0$, $1/4$, $1/2$, $3/4$, $1$, and $3$ for the {\BC} $(o)$ vs stripes shown in \fref{bcopenstripes}, as determined with $L=16$ and the smallest aspect ratio $\rho$ available. We compare these data also with the scaling function $\theta_o(\tau,\kappa\rightarrow\infty) =\theta_{(+,o)}(\tau)$, as obtained in \fref{casimir0} with $L=24$. For further discussions see the main text.}
\label{casimir_openall}
\end{figure}

In \fref{casimir_openall} we show a comparison of the scaling function $\theta_o(\tau,\kappa)$ of the critical Casimir force for the {\BC} shown in \fref{bcopenstripes} for $\kappa=0$, $1/4$, $1/2$, $3/4$, $1$, and $3$ as determined with $L=16$ and with the smallest aspect ratio $\rho$ available. We also compare these results with the Casimir scaling functions for the {\BC} $(+,o)$ shown in \fref{bcplusopen}, which corresponds to the limit $\kappa\rightarrow\infty$. Figure~\ref{casimir_openall} suggests that the approach of the limit $\kappa\rightarrow\infty$ is somehow singular. Apparently, for every finite value of $\kappa$, the force becomes attractive for sufficiently negative values of $\tau$ and exhibits a minimum which deepens and shifts to more negative values of $\tau$ as $\kappa$ is increased. Simultaneously, the zero of $\theta_o(\tau,\kappa)$ shifts towards lower values of $\tau$.

\section{Mean-field theory}
\label{sec:mft}
 Within the field-theoretic approach, bulk and surface critical phenomena of the Ising universality class 
are described by the standard Landau-Ginzburg-Wilson fixed-point Hamiltonian 
given by \cite{Binder-83,Diehl-86,Diehl-97}
\begin{multline} 
  \label{eq:hamiltonian}
   \mathcal{H}[\phi]=\int_V\,\upd^d r\,\left\{
        \frac{1}{2}(\nabla\phi)^2
       +\frac{\tilde\tau}{2}\phi^2
       +\frac{u}{4!}\phi^4
			 \right\}+\\
       \int_{\partial V}\upd^{(d-1)} r\left\{\frac{c(\vec{r})}{2}\phi^2-h_1(\vec{r})\phi\right\},
\end{multline} 
where $\phi(\vec{r})$ is the spatially varying order parameter describing the critical medium, which completely fills the
volume $V$ bounded by the boundaries $\partial V$ in $d$-dimensional space.
In \eref{eq:hamiltonian} $\tilde\tau\propto t$ and $u>0$ is a coupling constant providing stability for $t<0$;
$c(\vec{r})$ is the surface
enhancement, which, within mean-field theory, can be interpreted as an inverse extrapolation length of the order parameter field, 
and $h_1(\vec{r})$ is an (external) surface field acting on the order parameter at the boundaries.
Here, we consider surface fields and enhancements which can differ for the two confining surfaces
and which may also vary along one lateral direction of a single surface.
In the strong adsorption limit, i.e., $(\pm)$ {\BC}, corresponding to the so-called normal surface UC,
the surface behavior is described by the renormalization-group fixed-point values
$h_1\to\pm\infty$, and the order parameter diverges close to the surface: $\phi|_{\partial V}\to\pm\infty$.
The ordinary surface UC corresponds to the fixed point values $\{c=\infty,h_1=0\}$
and a vanishing order parameter $\phi|_{\partial V}=0$, i.e., Dirichlet $(o)$ {\BC}.
The film geometry considered here is bounded by surfaces at $z=0$ and at $z=L$ with either 
homogeneous $(+)$ or $(o)$ {\BC} or periodically alternating $(+)$/$(-)$ {\BC} of width $S_+=P/2$
along the lateral $x$ direction (see Figs.~\ref{bcstripes}--\ref{bcopenopen}).
\par
The Hamiltonian given in \eref{eq:hamiltonian} is minimized by 
the mean-field order parameter profile $m\equiv u^{1/2}\langle\phi\rangle$: $\updelta \mathcal{H}[\phi]/\updelta\phi|_{\phi=u^{-1/2}m}=0$.
Renormalization group arguments tell that mean-field theory (MFT)
provides the correct universal properties of critical phenomena for spatial dimensions
above the upper critical dimension $d\ge d_{\text{uc}}=4$ 
(up to logarithmic corrections in $d=d_{\text{uc}}$).
Mean-field theory provides the lowest-order contribution to universal properties within
an expansion in terms of $4-d=\varepsilon$.
Thus, universal properties in $d=4$ can be determined from MFT, up to two independent nonuniversal 
amplitudes appearing in the description of bulk critical phenomena (two-scale universality \cite{Binder-83,Diehl-86}):
the amplitude $B$ of the bulk order parameter $\langle\phi\rangle=\pm B|t|^\beta$ for $t<0$, 
where $\beta(d=4)=1/2$, and the amplitude $\xi_0^+$ of the correlation length [see \eref{xi_crit}, where $\nu(d=4)=1/2$].
Since here we are dealing only with vanishing or diverging values of $h_1$ and $c$, within MFT  
all quantities appearing in \eref{eq:hamiltonian} can be expressed in terms of 
these amplitudes: $\tilde\tau=t (\xi_0^+)^{-2}$ and $u=6(B\xi_0^+)^{-2}$.
Using the stress tensor method \cite{Krech-97} the mean-field universal scaling functions of the critical Casimir forces at the 
upper critical dimension $d_{\text{uc}}=4$ can be inferred directly from the MFT order parameter profiles 
up to an overall prefactor $\propto u^{-1}$.
\par
For the laterally homogeneous $(+,+)$, $(+,-)$, $(+,o)$, or $(o,o)$ {\BC} the MFT order parameter profiles across the film 
\cite{Krech-97,gambassi:2006} and the corresponding universal scaling functions of the critical Casimir force 
are known analytically \cite{Krech-97,oomft}.
Accordingly, the critical Casimir amplitude $\Theta_{\pp}=8K^4(1/\sqrt{2}) (B\xi_0^+)^2 \simeq -47.2682 (B\xi_0^+)^2$, 
where $K(k)$ is the complete elliptic integral of the first kind \cite{Krech-97}.
Note that, within MFT, the scaling functions $\theta_{\pmbc}(\tau)=-4\theta_{\pp}(-\tau/2)$ \cite{corr}, 
and $\theta_{(+,o)}(\tau)=\theta_{\pmbc}(4\tau)/16$ \cite{Krech-97} are directly related to each other, so that at $T_c$
$\Theta_{\pmbc}=-4\Theta_{\pp}$ and $\Theta_{\po}=-\Theta_{\pp}/4$.
In contrast to the case $d=3$, the MFT scaling function for $(o,o)$ {\BC} vanishes for $\tau\ge0$ [i.e., $\Theta_{\oo}(d=4)=0$] and
exhibits a cusplike singularity at its minimum at $\tau=-\pi^2$ below which $\theta_{\oo}(\tau<-\pi^2)=\theta_{\pp}(\tau)$ and
above which an analytic expression for $\theta_{\oo}$ has been derived in Ref.~\cite{oomft}.
\par
In order to obtain the spatially inhomogeneous MFT order parameter profile for the film geometry involving chemically striped
surfaces, we have minimized $\mathcal{H}[\phi]$ numerically using a quadratic finite element method.
Here, we extend previous investigations \cite{SSD-06} to negative values $t<0$ and to a broader range of geometrical parameters.
The corresponding scaling functions for the critical Casimir force are obtained via the stress tensor \cite{Krech-97}.
\par
The boundary condition for the diverging order parameter profile at those parts of the surface where there are $(+)$ or $(-)$ {\BC} can be implemented numerically
only approximately via a short-distance expansion of the corresponding profile for the semi-infinite systems \cite{Binder-83,Diehl-86}.
Thus, the MFT data presented below are subject to a numerical error which contains also the uncertainties due to the fineness of the numerical
mesh.
We estimate the numerical error for the data presented below to be less than $1\%$ or $\pm0.004\times|\Theta_{\pp}|$ if the latter is bigger.
\begin{figure}
  \begin{center}
    \includegraphics[width=0.9\linewidth,keepaspectratio]{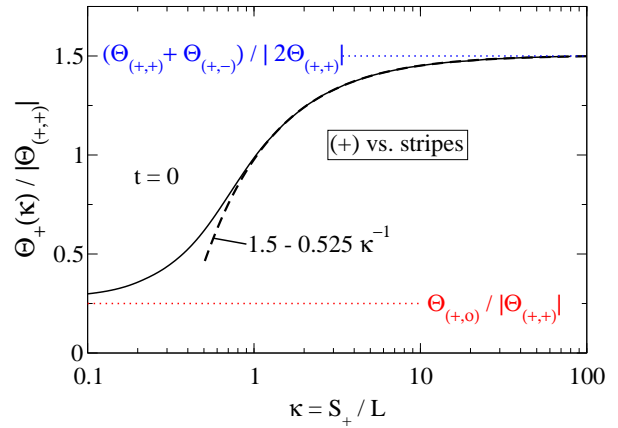}
  \end{center}
  \caption{(Color online)
  Reduced critical Casimir force amplitude $\Theta_{+}(\kappa)$ [\eref{theta_Delta}] in units of $|\Theta_{\pp}|$
  for the {\BC} shown in \fref{bcstripes} as obtained within mean-field theory. 
  For $\kappa\to0$ the Casimir amplitude approaches the value for $(+,o)$ {\BC} shown in Fig.~\protect\ref{bcplusopen}, i.e., $\Theta_{\po}/|\Theta_{\pp}|=\tfrac{1}{4}$,
  indicated by the lower red dotted line.
  For large stripes, $\Theta_{+}(\kappa\to\infty)/|\Theta_{\pp}|$ approaches the average value of the reduced 
  Casimir amplitudes for $(+,+)$ and $(+,-)$ {\BC}, i.e., $(\Theta_{\pp}+\Theta_{\pmbc})/|{2}\Theta_{\pp}|=\tfrac{3}{2}$ shown as upper 
  blue dotted line.
  For $\kappa\gg1$ the behavior of the Casimir amplitude $\Theta_{+}(\kappa)/|\Theta_{\pp}|$ approaches the function 
  $\tfrac{3}{2}-\tfrac{5}{4}\alpha_{\protect\raisemath{-2pt}{+}}\kappa^{-1}$ (see the black dashed line and the main text). 
  From a least-squares fit we have obtained $\alpha_{\protect\raisemath{-2pt}{+}}=0.420(4)$.
  Compare \fref{critvskappa}, where $\Theta_{(+,o)}/|\Theta_{\pp}|=0.60(1)$ and $\Theta_{+}(\kappa\to\infty)/|\Theta_{\pp}|\simeq2.91(5)$.
  }
  \label{fig:crit-mft}
\end{figure}
\vspace{-1em}
\subsection{Critical Casimir amplitude at $T_c$}
\label{sec:critical-amplitude-mft}
In \fref{fig:crit-mft} the amplitude of the critical Casimir force $\Theta_+(\kappa)=\theta_+(0,\kappa)$ 
(see Eqs.~(\protect\ref{casimir_fss_leading}) and (\protect\ref{theta_Delta})) 
for a striped surface opposite to a homogeneous surface with $(+)$ {\BC} is shown as obtained numerically
within MFT in units of $|\Theta_{\pp}|$.
We have been able to calculate the values of $\Theta_+(\kappa)$ numerically within the range $\kappa=0.1$ to $\kappa=80$.
As discussed above, for $\kappa\to0$ the Casimir amplitude approaches the value for $(+,o)$ {\BC} 
shown in Fig.~\protect\ref{bcplusopen}, i.e., $\Theta_{\po}$, so that for relatively narrow stripes
the chemically striped wall effectively mimics a wall with $(o)$ {\BC}.
On the other hand, for $\kappa\to\infty$ the Casimir amplitude approaches the average value of the Casimir 
amplitudes for $(+,+)$ and $(+,-)$ {\BC}, i.e., $\Theta_+(\kappa\to\infty)=(\Theta_{\pp}+\Theta_{\pmbc})/2
=-\tfrac{3}{2}\Theta_{\pp}$,
whereas $\Theta_+(\kappa)$ monotonically interpolates between these two limits.
\par
For $\kappa\gg1$, according to \eref{eq:limit-2}, we find for the critical Casimir amplitude
\begin{multline} 
  \label{eq:large-kappa}
  \Theta_+(\kappa\gg1)\\
  \simeq\Theta_{\po}+\left(\frac{\Theta_{\pp}+\Theta_{\pmbc}}{2}-\Theta_{\po}\right)\left(1-\frac{\alpha_{\raisemath{-2pt}{+}}}{\kappa}\right)\\
  =-\,\,\Theta_{\pp}\left(\frac{3}{2}-\frac{5}{4}\alpha_{\raisemath{-2pt}{+}}\kappa^{-1}\right),
\end{multline} 
where the proportionality constant $\alpha_{\raisemath{-2pt}{+}}$ is related to the scaling function $E(\tau)$ according to $E(0)=-\alpha_{\raisemath{-2pt}{+}}(\Theta_{\pp}+\Theta_{\pmbc})$ and  by using a least-squares fit it has been determined within MFT as $\alpha_{\raisemath{-2pt}{+}}=0.420(4)$. In three spatial dimensions, using the results $(\Theta_{\pp}+\Theta_{\pmbc})/2=2.386(5)$ and $E(0)=2.04(3)$ of Ref.~\cite{PTD-10}, we obtain $\alpha_{\raisemath{-2pt}{+}}=0.427(7)$, in nice agreement with the MFT result.
\par
\begin{figure}
  \begin{center}
    \includegraphics[width=0.85\linewidth,keepaspectratio]{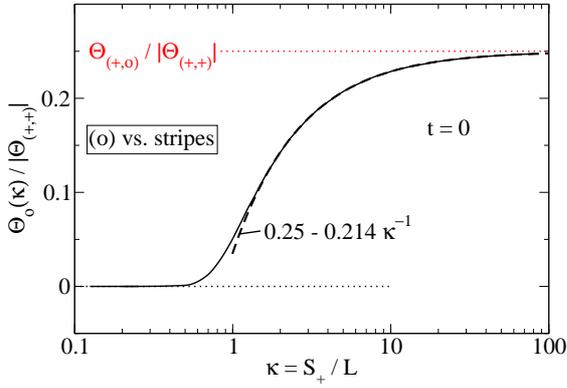}
  \end{center}
  \vspace{1em}
  \caption{(Color online)
  Reduced  Casimir amplitude $\Theta_o(\kappa)$ [\eref{theta_Delta}] in units of $|\Theta_{\pp}|$ 
  for the {\BC} shown in \fref{bcopenstripes} as obtained within MFT. 
  For $\kappa\to0$ the Casimir amplitude approaches monotonically from positive values the limiting value $\Theta_{\oo}=0$ shown by the lower 
  green dotted line.
  According to \eref{eq:limit-3}, for $\kappa\gg1$ the reduced Casimir amplitude $\Theta_o(\kappa)/|\Theta_{\pp}|$ approaches  
  $\tfrac{1}{4}(1-\alpha_o\kappa^{-1})$ shown as black dashed line.
  From a least-squares fit we have obtained, within MFT, $\alpha_o=0.857(9)$
  [\eref{eq:large-kappa-open}].
  For $\kappa\to\infty$, $\Theta_o(\kappa)/|\Theta_{\pp}|$ approaches the Casimir amplitude for $(+,o)$ {\BC}, i.e., 
  $\Theta_{\po}/|\Theta_{\pp}|=1/4$ shown as the upper red dotted line.
  Compare \fref{critopenvskappa}, where $\Theta_o(\kappa=0)/|\Theta_\pp|=0.037(6)$ and $\Theta_o(\kappa\to\infty)/|\Theta_\pp|=0.60(1)$.
  }
  \label{fig:crit-open-mft}
\end{figure}
Figure~\ref{fig:crit-open-mft} shows the reduced critical Casimir force amplitude $\Theta_o(\kappa)$ in units of $|\Theta_{\pp}|$ 
for the case of a striped surface opposite
to a surface with a homogeneous $(o)$ {\BC} (see Figs.~\ref{bcopenstripes} and \ref{bcopenopen}).
Similarly to \fref{fig:crit-mft}, $\Theta_o(\kappa)$ monotonically interpolates between the limiting values for $\kappa\to0$
and $\kappa\to\infty$, i.e., $\Theta_{\oo}/|\Theta_{\pp}|=0$ and $\Theta_{\po}/|\Theta_{\pp}|=1/4$, respectively.
For narrow stripes the amplitude $\Theta_o(\kappa\to0)$ approaches its limit already for larger values of $\kappa$ than
in the case of a homogeneous $(+)$ {\BC} shown in \fref{fig:crit-mft}.
This indicates that the strength of the tendency of a chemically striped surface to effectively mimic an
$(o)$ {\BC} in the limit $\kappa\to0$ also depends on the type of homogeneous {\BC}
at the opposing surface of the film.
According to \eref{eq:limit-3}, for $\kappa\gg1$ the dependence of the Casimir amplitude $\Theta_o(\kappa)$ on $\kappa$ approaches
the following form:
\begin{multline} 
  \label{eq:large-kappa-open}
  \Theta_o(\kappa\gg1)\simeq\Theta_{\oo}+\left(\Theta_{\po}-\Theta_{\oo}\right)\left(1-\frac{\alpha_o}{\kappa}\right)\\
  =-\frac{\Theta_{\pp}}{4}\left(1-\alpha_o\,\,\kappa^{-1}\right),
\end{multline} 
where we have determined $\alpha_o=0.857(9)$ via a least-squares fit.
\par
Whereas the behavior of the Casimir amplitude $\Theta_+(\kappa)$ for the case of a homogeneous $(+)$ {\BC} as calculated 
within MFT (\fref{fig:crit-mft}) is similar to the one 
obtained from {\MC} simulations (\fref{critvskappa}), the form of $\Theta_o(\kappa)$ for the case of a homogeneous $(o)$ {\BC} as obtained
within MFT (\fref{fig:crit-open-mft}) is qualitatively different from the one obtained from MC simulations (\fref{critopenvskappa}).
This will be addressed in more detail in Sec.~\ref{sec:comparison} below.
\subsection{Scaling function of the critical Casimir force}
\label{sec:scaling-function-mft}
\begin{figure}
  \begin{center}
    \includegraphics[width=0.84\linewidth,keepaspectratio]{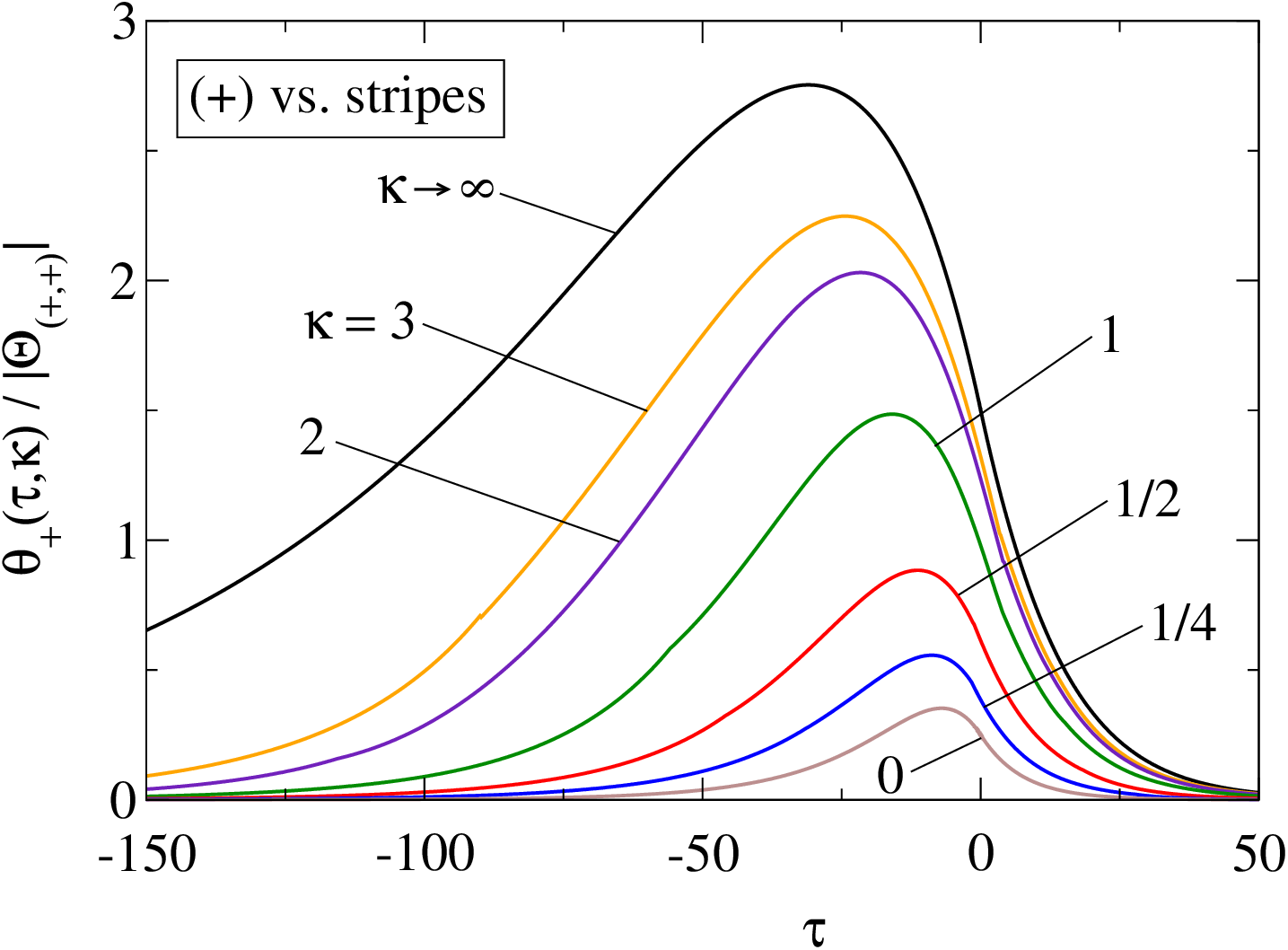}
  \end{center}
  \caption{(Color online)
  Reduced universal scaling function $\theta_+(\tau,\kappa)/|\Theta_{\pp}|$ [\eref{casimir_fss_leading}] 
  for a striped surface opposite to a surface with homogeneous $(+)$ {\BC} (\fref{bcstripes}),
  as determined numerically within MFT for various values of $\kappa$.
  For $\kappa\to0$ and $\kappa\to\infty$, the reduced scaling functions approach their limiting behaviors $\theta_{\po}(\tau)/|\Theta_{\pp}|$ [\eref{eq:limit-1}] 
  and $(\theta_{\pp}(\tau)+\theta_{\pmbc}(\tau))/|{2}\Theta_{\pp}|$ [\eref{eq:limit-2}], respectively.
  Compare \fref{casimirall} by taking into account that there, i.e., in $d=3$, $|\Theta_\pp|=0.820(15)$.
  }
  \label{fig:scaling-mft}
\end{figure}
The reduced scaling function $\theta_+(\tau,\kappa)/|\Theta_{\pp}|$ [\eref{casimir_fss_leading}] of the critical Casimir force between a chemically striped surface and a homogeneous
surface with $(+)$ {\BC} (\fref{bcstripes}) is shown in \fref{fig:scaling-mft} 
for $d=4$ (MFT) and for various values of $\kappa$.
For $\kappa\to0$, $\theta_+(\tau,\kappa)/|\Theta_{\pp}|$ approaches the scaling function $\theta_{\po}(\tau)/|\Theta_{\pp}|$,
i.e., the striped surface effectively mimics a surface with homogeneous $(o)$ {\BC}.
On the other hand, for $\kappa\to\infty$, the universal scaling function of the critical Casimir force approaches
the average of the scaling functions for $(+,+)$ and $(+,-)$ {\BC}, i.e., $ \theta_+(\tau,\kappa\to\infty)/|\Theta_{\pp}|
=(\theta_{\pp}(\tau)+\theta_{\pmbc}(\tau))/|{2}\Theta_{\pp}|$
[\eref{eq:limit-2}].
For intermediate values of $\kappa$, the scaling functions smoothly and monotonically interpolate between these limiting cases.
\par
\begin{figure}
  \begin{center}
    \includegraphics[width=0.8\linewidth,keepaspectratio]{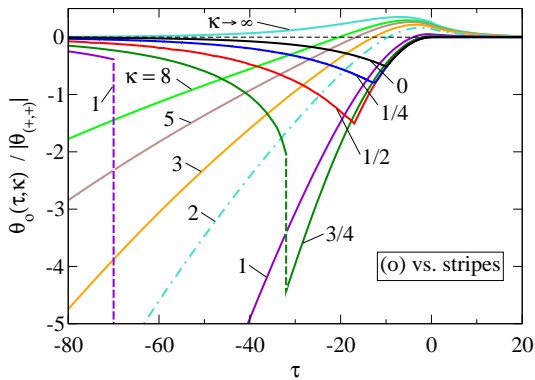}
  \end{center}
  \caption{(Color online)
  Reduced universal scaling function $\theta_o(\tau,\kappa)/|\Theta_{\pp}|$ of the critical Casimir force 
  for a striped surface opposite to a surface with homogeneous $(o)$ {\BC} (\fref{bcopenstripes}),
  as determined numerically within MFT for various values of $\kappa$. 
  We compare the data also with the reduced scaling functions $\theta_{\oo}(\tau)/|\Theta_{\pp}|$ and $\theta_{\po}(\tau)/|\Theta_{\pp}|$,
  which correspond to the limits $\kappa\to0$ [\eref{eq:limit-2}] and $\kappa\to\infty$ [\eref{eq:limit-3}], respectively.
  For $\kappa<2$ the numerically obtained MFT scaling functions suggest the occurrence of a cusplike singularity
  or a finite jump of $\theta_o(\tau,\kappa)$ at its minimum position $\tau_\text{min}$.
  (Due to the numerical difficulties in determining the thermodynamically stable configuration, both the positions and the depths of the minima of $\theta_o(\tau,\kappa<2)/|\Theta_{\pp}|$ are affected by an estimated numerical
  error of around $10\%$, which is one order of magnitude larger than for the remaining data.)
  For $\kappa>2$ the scaling functions diverge for $\tau\to-\infty$ [\eref{eq:sum-1}].
  Compare \fref{casimir_openall} by taking into account that there, i.e., in $d=3$, $|\Theta_\pp|=0.820(15)$.
  }
  \label{fig:scaling-mft-open}
\end{figure}
As discussed in Sec.~\ref{sec:off_criticality}, the behavior of the universal scaling scaling function $\theta_o(\tau,\kappa)$ for a striped surface 
opposite to a surface with homogeneous $(o)$ {\BC} (\fref{bcopenstripes}) is more complex than the one in the previous case.
Whereas for $\tau\ge0$ the scaling function $\theta_o(\tau,\kappa)$ smoothly interpolates between its limiting behaviors 
$\theta_{\oo}(\tau)$ for $\kappa=0$ and $\theta_{\po}(\tau)$ for $\kappa\to\infty$, for negative values of $\tau$ its dependence on $\kappa$ is nonmonotonic and involves 
a phase transition associated with the one at $\kappa=2$ between the ground states of the system (see \eref{eq:transition-area}).
For $\kappa<2$ the ground states are spatially homogeneous, which results in a vanishing value $\theta_o(\tau\to-\infty,\kappa<2)\to0$.
The numerically obtained MFT data shown in \fref{fig:scaling-mft-open} suggest that the minima of the scaling functions for $\kappa<2$ correspond to a cusplike
singularity or even a finite jump. (Recall that $\theta_o$ is the scaling function of the critical Casimir 
\emph{force}, which is the derivative of the Casimir interaction.)
However, due to the presence of metastable striped and homogeneous states the numerics even within MFT is so involved that the present data 
suffer from an error of the position of the minimum of around $10\%$.
Moreover, due to using the short-distance expansion in the numerical implementation of $(\pm)$ {\BC}, it is technically difficult
to distinguish these metastable states for $\kappa\simeq2$.
For $\kappa>2$ a striped ground state is stable, which involves a divergence of the scaling function for $\tau\to-\infty$ so that
for $\tau<0$ the transition to its limiting behavior $\theta_{\po}(\tau)>0$ for $\kappa\to\infty$ is somewhat singular.
Since at $T=T_c$, the critical Casimir amplitude $\Theta_o(\kappa)$ is non-negative for all values of $\kappa$ (see \fref{fig:crit-open-mft}; for
$\kappa\lesssim0.5$, $\Theta_o$ is vanishingly small), within MFT the scaling function $\theta_o(\tau,\kappa)$ changes sign for all values
of $\kappa$ at a certain value $\tau^*(\kappa)<0$.

\par
In the following we consider the contribution of the interface tension to the critical Casimir force for $\tau<0$ 
[see \eref{eq:sum-1}].
Near $T_c$ the interface tension varies as $\sigma=\sigma_0|t|^\mu$ where $\mu=(d-1)\nu$, so that $\mu=3/2$ within MFT
\cite{ZF-96};
$\sigma_0$ is the corresponding nonuniversal amplitude which forms the universal amplitude ratio
$\tfrac{1}{k_BT_c}\sigma_0(\xi_0^+)^{(d-1)}=R_\sigma$.
Within MFT $\sigma/(k_BT_c)=4\sqrt{2}u^{-1}(\xi_0^+)^{-(d-1)}|t|^{\mu}$ \cite{brezin:1984}  
so that $R_\sigma=\tfrac{2}{3}\sqrt{2}(B\xi_0^+)^2$ and $R_\sigma/|\Theta_{\pp}|\simeq0.020$.
For the homogeneous configuration with the interfaces parallel to the film (i.e., for $\kappa<2$), 
the interface energy does not contribute explicitly to the resulting force
because the area of these interfaces is not changed upon varying of the film thickness.
(Note, however, that the order parameter profile across these interfaces does depend on $L$.)
For the striped configurations, i.e., for $\kappa>2$, in which the interfaces are oriented perpendicular 
to the film, the interface tension dominates the resulting 
force for large negative $\tau$ (i.e., $L$ large), because approximately the interface along the $z$ direction 
has an area $L_\parallel^{d-2}L$ which is proportional to the film thickness $L$.
Thus, the free energy $\Gamma^i_s$ of such a \textit{s}ingle \textit{i}nterface is given by
\begin{equation} 
  \label{eq:gamma}
  \Gamma^i_s=L_\parallel^{d-2}L\sigma,
\end{equation} 
where $L_\parallel$ is the extension of the system along the invariant direction(s).
For a single such interface
this gives rise to a force along the normal direction,
\begin{equation} 
  \label{eq:fgamma}
    F_{\Gamma,s}^i=-\frac{\partial\Gamma^i_s}{\partial L}=-L_\parallel^{d-2}\sigma.
\end{equation} 
For the striped state there are $2\times L_\parallel/P=L_\parallel/S_+$ such interfaces so
that the total force per area $L_\parallel^{d-1}$ of the film and per $k_BT_c$ is
\begin{multline} 
  \label{eq:fg}
  \frac{F_{\Gamma,\textit{tot}}^i}{k_BT_c L_\parallel^{d-1}}= -\frac{1}{S_+}\frac{\sigma}{k_B T_c} = 
  -\frac{1}{L^d}\frac{1}{\kappa}L^{d-1}\frac{\sigma_0}{k_BT_c}|t|^{(d-1)\nu}\\
  =\frac{1}{L^d}\left(-\frac{1}{\kappa}R_\sigma|\tau|^\mu\right),
\end{multline}
so that its contribution $\theta_{o,\Gamma}(\tau,\kappa)$ to the universal scaling function of the critical Casimir force reads
[see \eref{casimir_fss_leading}]
\begin{equation} 
  \label{eq:fg2}
  \theta_{o,\Gamma}(\tau,\kappa)= - \frac{R_\sigma}{\kappa}|\tau|^{\mu},
\end{equation}
which is attractive and becomes as strong as $\Theta_{\pp}$ for $|\tau|^{\mu}/\kappa\gtrsim 50$ within MFT.
Accordingly, for the limit $\tau\ll-1$ and $\kappa>2$ the scaling function of the critical Casimir force approaches 
the expression given in \eref{eq:sum-1}, which corresponds to the sum of the homogeneous contribution and the contribution due to the interfaces
oriented perpendicular to the film surfaces.
\par
\begin{figure}
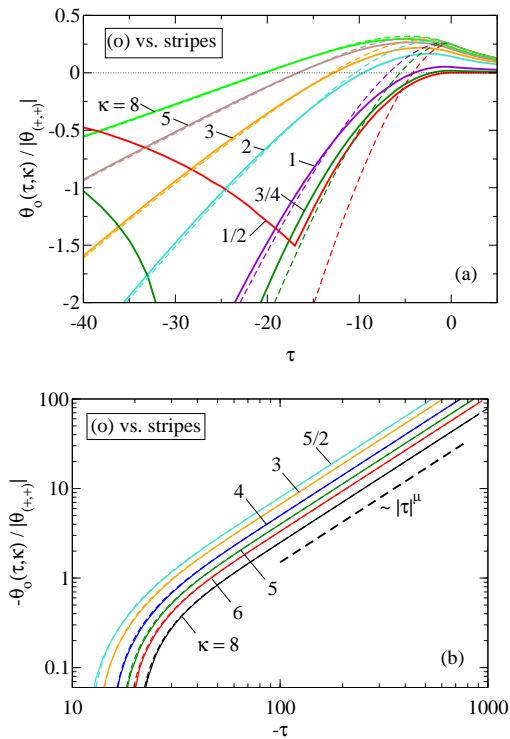

  \begin{center}
    \includegraphics[width=0.73\linewidth,keepaspectratio,clip=true]{mft-open-sum.eps}
  \end{center}
  \begin{center}
    \includegraphics[width=0.76\linewidth,keepaspectratio,clip=true]{logsumopen.eps}
  \end{center}
  \caption{(Color online)
  Reduced universal scaling function $\theta_o(\tau,\kappa)/|\Theta_{\pp}|$ of the critical Casimir force 
  for a striped surface opposite to a surface with homogeneous $(o)$ {\BC} (\fref{bcopenstripes}),
  as determined numerically within MFT (solid lines, same as \fref{fig:scaling-mft-open}).
  For $\tau\ll-1$ and $\kappa>2$ they agree well with the asymptotic expression given by the r.h.s. of \eref{eq:sum-1}
  shown as dashed lines (a).
  For $\kappa>2$ and large negative values of $\tau$, i.e.,  $\tau\ll-10$, the attractive interface contribution 
  $-R_\sigma \kappa^{-1}|\tau|^\mu/|\Theta_{\pp}|$ [\eref{eq:fg2}]
  dominates the the scaling function $\theta_o(\tau,\kappa)/|\Theta_{\pp}|$ (b). 
  }
  \label{fig:scaling-log-open}
\end{figure}
\par
Figure~\ref{fig:scaling-log-open} compares $\theta_o(\tau,\kappa)$ for a striped surface opposite to a surface 
with homogeneous $(o)$ {\BC} as determined numerically within MFT with the estimate of the corresponding
interface contribution as given in \eref{eq:sum-1}.
The dashed lines shown in \fref{fig:scaling-log-open} correspond to \eref{eq:sum-1}.
They are approached by the actual scaling functions shown as solid lines in \fref{fig:scaling-log-open}.
As expected, \eref{eq:sum-1} describes neither the behavior for $\kappa<2$ nor the one for small absolute values of $\tau$.
However, for $\tau\ll-1$ and $\kappa>2$, the scaling functions agree rather well with their asymptotic behavior
given in \eref{eq:sum-1}.

\section{Comparison between mean-field theory and Monte Carlo data}
\label{sec:comparison}
\subsection{Critical Casimir amplitude at $T_c$}
\label{sec:critical-amplitude-compare}
Differing from the MC data for $d=3$, the universal scaling functions of the critical Casimir force obtained within 
mean-field theory can be determined only up to an unknown constant amplitude.
In order to facilitate nonetheless a valuable comparison between them, which illustrates the dependence of the
scaling functions on the spatial dimension $d$, it is useful to normalize them by an overall amplitude
so that the unknown constant amplitude for the MFT results drops out.
In the previous section we normalized the various scaling functions by one and the same universal critical
Casimir amplitude $|\Theta_{\pp}|$.
Here, we propose an alternative normalization, which makes use only of that scaling function under consideration
and also normalizes the ratios between the corresponding critical Casimir amplitudes, which depend on $d$,
\begin{equation} 
  \label{eq:normalized}
  \hat{\Theta}(\kappa)\equiv
  \frac{\Theta(\kappa)-\Theta(\kappa\to0)}{\Theta(\kappa\to\infty)-\Theta(\kappa\to0)}
  \to\left\{
\begin{aligned}%
  0,&\qquad\kappa\to0,\\
  1,&\qquad\kappa\to\infty.
\end{aligned}\right.
\end{equation} 
\par
\begin{figure}
  \begin{center}
    \includegraphics[width=0.85\linewidth,keepaspectratio,clip=true]{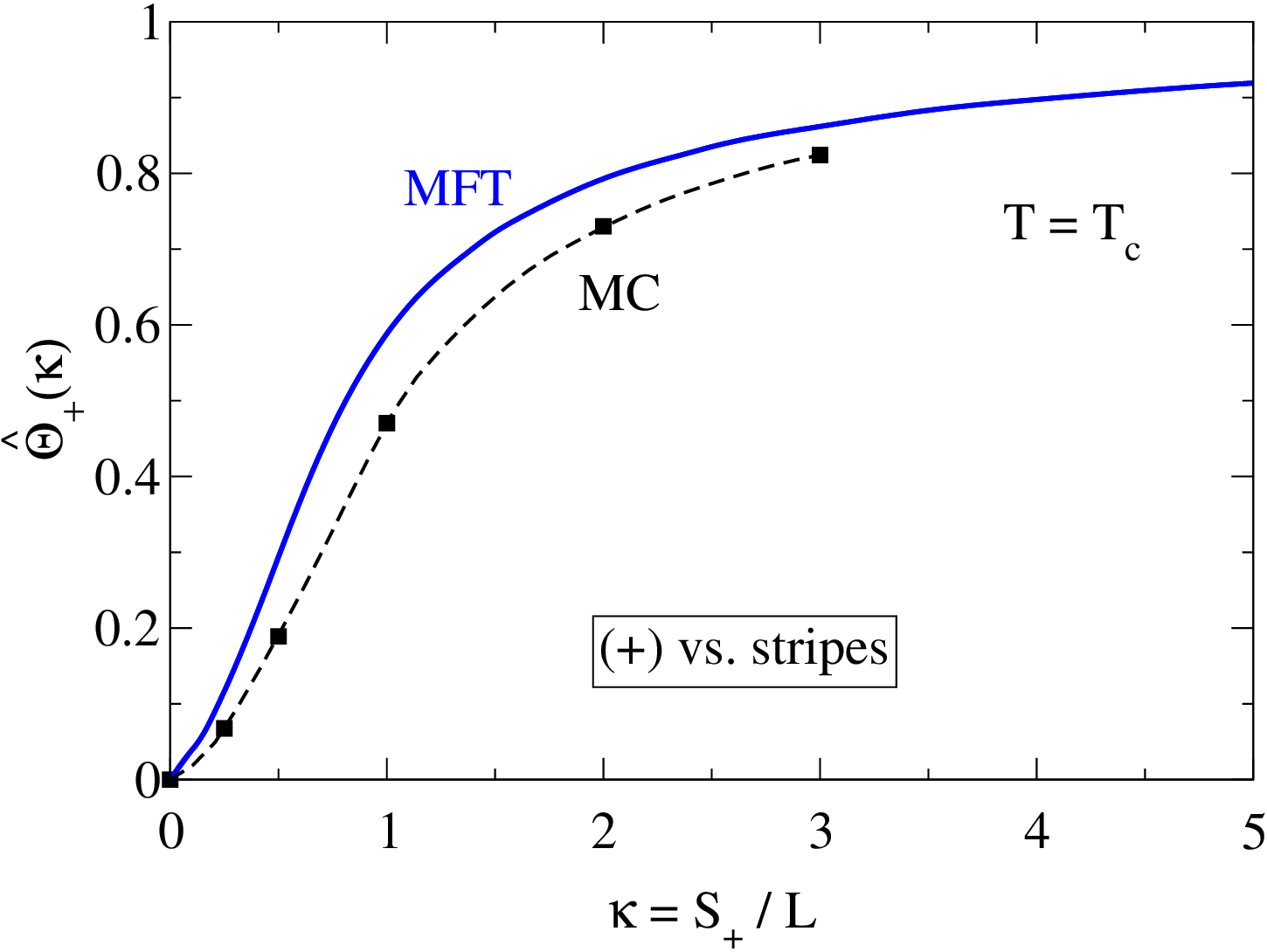}
  \end{center}
  \caption{(Color online)
  Comparison of the normalized critical \mbox{Casimir amplitude
  $\hat{\Theta}_+(\kappa)\hspace{-0.1em}=\hspace{-0.1em}[\Theta_+(\kappa)-\Theta_+(\kappa\to 0)]/[\Theta_+(\kappa\hspace{-0.15em}\to\hspace{-0.15em}\infty)$} $-\Theta_+(\kappa\to0)]$
  [\eref{eq:normalized}] for a homogeneous $(+)$ wall opposite to
  a striped wall (\fref{bcstripes}) as obtained from MC data (symbols; same as \fref{critvskappa}) and within MFT 
  (solid line; see \fref{fig:crit-mft}).
  }
  \label{fig:compare-crit}
\end{figure}
As discussed in the previous sections, the critical Casimir amplitude between a chemically striped wall
and a homogeneous wall with $(+)$ {\BC} interpolates between $\Theta_+(\kappa\to0)=\Theta_{\po}$ and
$\Theta_+(\kappa\to\infty)=(\Theta_{\pp}+\Theta_{\pmbc})/2$.
Figure~\ref{fig:compare-crit} shows the corresponding normalized critical Casimir amplitude $\hat{\Theta}_+(\kappa)$ [\eref{eq:normalized}]
as obtained from MC data (symbols) as well as obtained within MFT (full line).
As can be inferred from \fref{fig:compare-crit} the behavior of the normalized Casimir amplitude $\hat{\Theta}_+(\kappa)$
as a function of $\kappa$ as obtained from MFT ($d=4$) is rather similar to the one in $d=3$.
Thus, for this geometry the effects of the chemical patterning are captured even semiquantitatively by MFT.
\par
\begin{figure}
  \begin{center}
    \includegraphics[width=0.85\linewidth,keepaspectratio,clip=true]{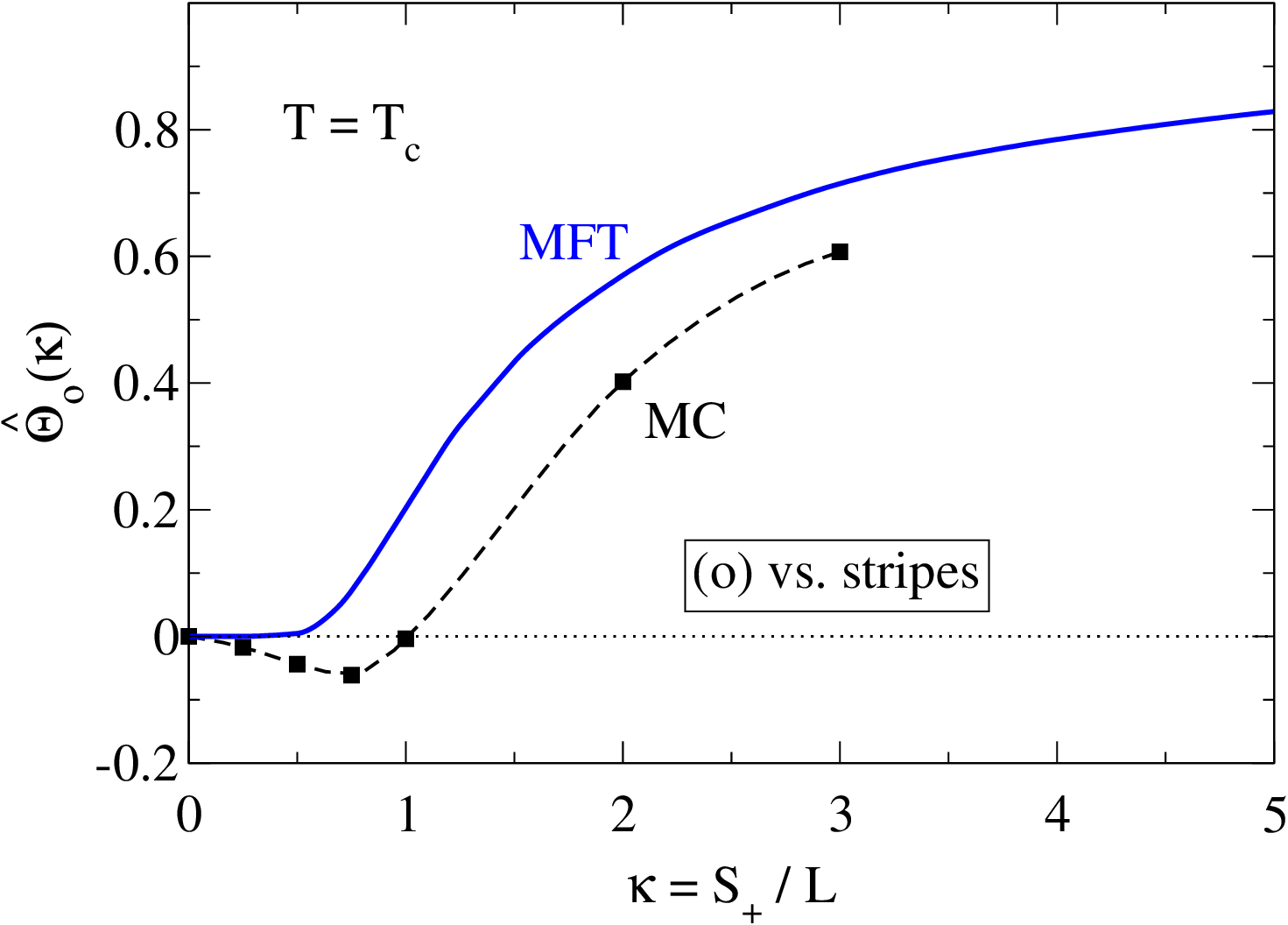}
  \end{center}
  \caption{(Color online)
  Normalized Casimir amplitude 
  $\hat{\Theta}_o(\kappa)=[\Theta_o(\kappa)-\Theta_o(\kappa\to 0)]/[\Theta_o(\kappa\to\infty)-\Theta_o(\kappa\to0)]$
  [\eref{eq:normalized}] for a homogeneous $(o)$ wall opposite to a 
  striped wall (\fref{bcopenstripes}) as obtained from MC data (symbols; same as \fref{critopenvskappa}) 
  and within MFT (solid line; see \fref{fig:crit-open-mft}).
  In contrast to the behavior shown in \fref{fig:compare-crit}, the MFT results differ qualitatively from the behavior in $d=3$. 
  In both cases MFT overestimates the strength of the force (here for $\kappa\gtrsim 0.75$).
  $\hat{\Theta}_o(\kappa\to\infty)$ attains its limiting value $1$ slower than $\hat{\Theta}_+(\kappa\to\infty)$.
  }
  \label{fig:compare-crit-open}
\end{figure}
In contrast, for the case of a homogeneous $(o)$ surface opposite to a striped one (\fref{bcopenstripes}), we find qualitative differences.
In \fref{fig:compare-crit-open} the normalized critical Casimir amplitude $\hat{\Theta}_o(\kappa)$  
[\eref{eq:normalized}], as obtained both in $d=3$ and within MFT, is shown, using the corresponding limits 
$\Theta_o(\kappa\to0)=\Theta_{\oo}$ and $\Theta_o(\kappa\to\infty)=\Theta_{\po}$.
Whereas the critical Casimir amplitude as obtained from {\MC} simulations shows a nonmonotonic behavior and changes
sign as a function of $\kappa$, the mean-field amplitudes are always positive and monotonically increasing as function of $\kappa$.
As expected, the absence of fluctuations within MFT affects the quantitative estimate of the Casimir amplitude more strongly
for the $(o)$ {\BC} than for the $(+)$ {\BC}.
\subsection{Scaling function of the critical Casimir force}
\label{sec:scaling-function-compare}
In order to compare also the temperature dependence of the scaling functions $\theta_{+/o}(\tau,\kappa)$ 
of the critical Casimir force in $d=3$ with their corresponding MFT estimates, it is useful to
not only normalize the amplitude of the latter but also
to rescale them along the $\tau$ axis by an overall factor.
Although this is an {\it ad hoc} procedure, it has turned out that a suitable combination of such rescaled
MFT results with only partly available MC data might be a successful method in order to obtain quantitatively
reliable approximations in an extended range of variables \cite{mohry:2012}.
\begin{figure}
  \begin{center}
    \includegraphics[width=0.85\linewidth,keepaspectratio,clip=true]{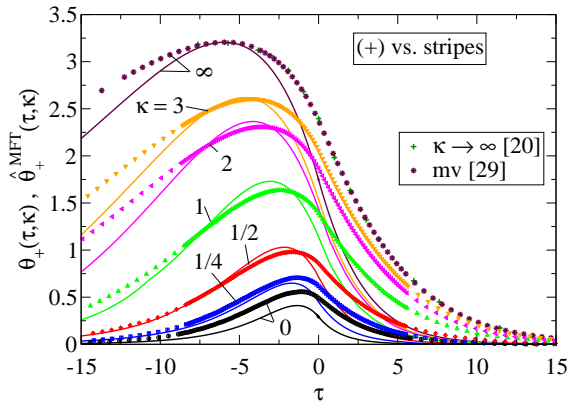}
  \end{center}
  \caption{(Color online)
  Comparison of the scaling functions $\theta_+(\tau,\kappa)$ for a wall with
    a homogeneous $(+)$ {\BC} opposite to a chemically striped wall (\fref{bcstripes})
    as obtained for $d=3$ and within MFT, i.e., for $d=4$.
    The symbols are the data obtained from the {\MC} simulations shown in \fref{casimirall}.
    The data obtained for $\kappa\to\infty$ \cite{PTD-10} agree with the \textit{m}ean \textit{v}alue of the
    data for $(+,+)$ and $(+,-)$ {\BC} of Ref.~\cite{Hasenbusch-10c}.
    The solid lines correspond to the MFT scaling functions $\hat{\theta}^{\text{MFT}}_+$ 
    shown in \fref{fig:scaling-mft} which have been rescaled according to \eref{eq:rescale}
    (see the main text and the caption of \fref{casimirall}).
    Upon construction, for $\kappa=\infty$ the positions and the heights of the maxima for $d=3$ and $d=4$ agree.
  }
  \label{fig:compare-all}
\end{figure}
\begin{figure}
  \begin{center}
    \includegraphics[width=0.85\linewidth,keepaspectratio,clip=true]{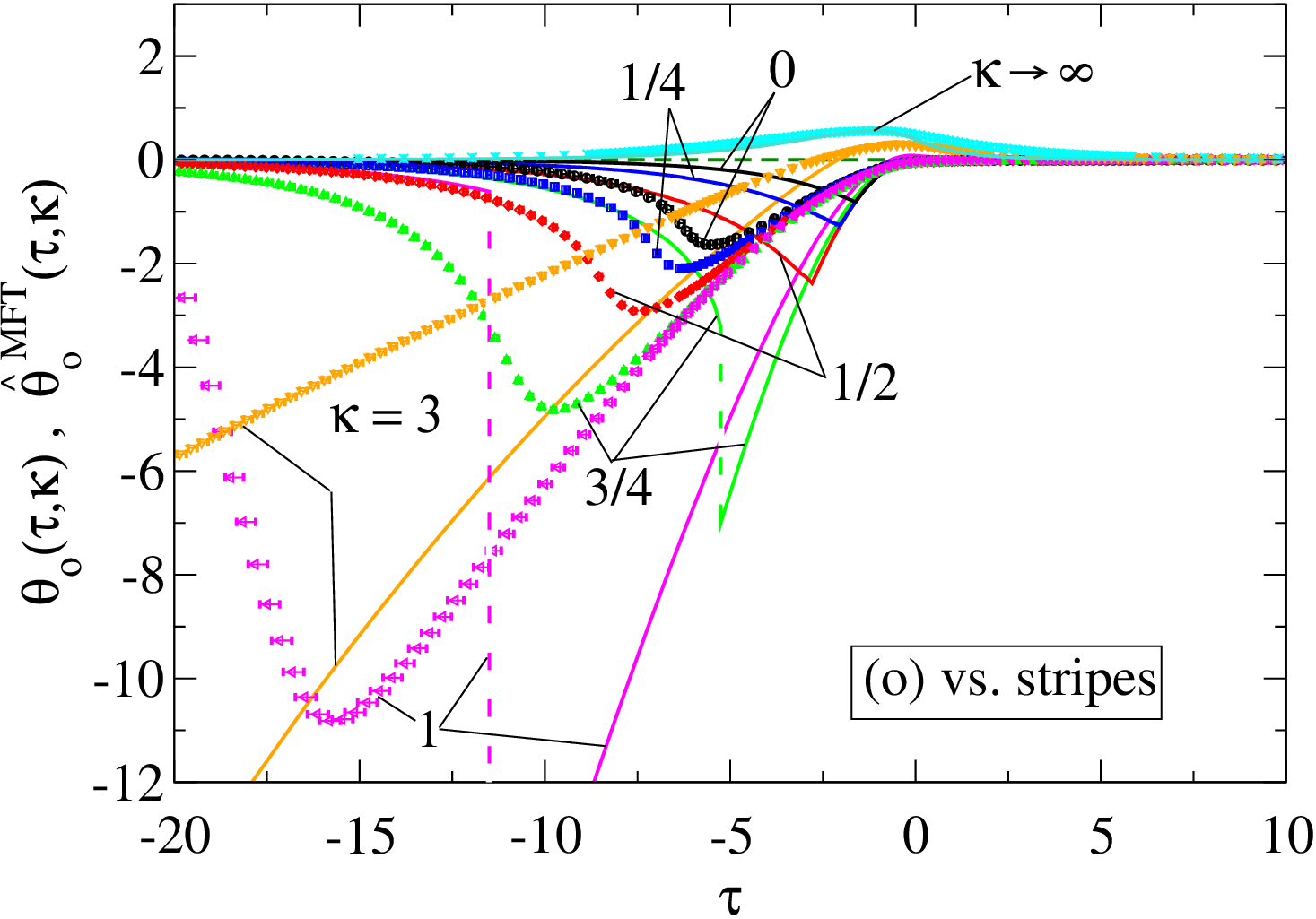}
  \end{center}
  \caption{(Color online)
    Comparison of the scaling functions $\theta_o(\tau,\kappa)$ for a homogeneous $(o)$ wall opposite to a 
    striped wall (\fref{bcopenstripes}).
    The symbols correspond to the {\MC} data ($d=3$) shown in \fref{casimir_openall}, whereas
    the solid lines correspond to the MFT scaling functions ($d=4$) shown in \fref{fig:scaling-mft-open} which
    have been rescaled according to \eref{eq:rescale}.
    In contrast to \fref{fig:compare-all}, the rescaled MFT scaling functions differ qualitatively
    from the corresponding ones in $d=3$.
    Upon construction, for $\kappa=\infty$ the positions and the heights of the maxima for $d=3$ and $d=4$ agree.
  }
  \label{fig:compare-all-open}
\end{figure}
In the following we use a simple normalization of the MFT scaling functions
$\theta^\text{MFT}_{+/o}(\tau,\kappa)$.
In Figs.~\ref{fig:compare-all} and \ref{fig:compare-all-open} the mean-field scaling functions
are rescaled linearly according to
\begin{multline}
  \label{eq:rescale}
  \hat{\theta}^\text{MFT}_{+/o}(\tau,\kappa)\\\equiv
  \frac{\theta_{+/o}(\tau_\text{max,$+/o$},\kappa\to\infty)}{\theta_{+/o}^{\text{MFT}}(\tau_\text{max,$+/o$}^{\text{MFT}},\kappa\to\infty)}
  \theta^\text{MFT}_{+/o}\left(\frac{\tau_{\text{max,$+/o$}}^{\text{MFT}}}{\tau_{\text{max,$+/o$}}}\tau,\kappa\right)
\end{multline}
so that for $\kappa\to\infty$ the positions and the values of the maxima of the 
rescaled scaling functions $\hat{\theta}^{\text{MFT}}_{+/o}$ agree
with those of the MC data.
In \eref{eq:rescale} $\tau_\text{max,$+/o$}$ and $\tau_\text{max,$+/o$}^\text{MFT}$ correspond to the position of 
the maximum of the scaling functions for $\kappa\to\infty$ in $d=3$ and $d=4$, respectively.
For the case of a homogeneous $(+)$ wall opposite to a striped wall
we can infer from the data of Ref.~\cite{Hasenbusch-10c} the rough estimates 
$\tau_\text{max,$+$}\simeq-6.0$ and $\theta_+(\tau_\text{max,$+$},\kappa\to\infty)\simeq3.21$ in $d=3$
(see the caption of \fref{casimirall} and Refs.~\cite{PTD-10,Hasenbusch-10c}) 
and $\tau^\text{MFT}_\text{max,$+$}\simeq -31.960$ and $\theta^\text{MFT}_+(\tau^{\text{MFT}}_\text{max,$+$},\kappa\to\infty)\simeq 2.7531|\Theta_{\pp}|$ in $d=4$
(by taking the mean value of the scaling functions for $(+,+)$ and $(+,-)$ {\BC} from Ref.~\cite{Krech-97}; see 
\fref{fig:scaling-mft}).
For a homogeneous $(o)$ wall opposite to a striped wall one has
$\tau_\text{max,$o$}=-1.174(10)$ and $\theta_o(\tau_\text{max,$o$},\kappa\to\infty)=0.564(3)$ in $d=3$
(see Ref.~\cite{Hasenbusch-11} which agrees with the result shown in \fref{casimir0}) 
and $\tau^\text{MFT}_\text{max,$o$}\simeq- 7.0275$ and $\theta^\text{MFT}_o(\tau^{\text{MFT}}_\text{max,$o$},\kappa\to\infty)\simeq 0.35280|\Theta_{\pp}|$ in $d=4$
as obtained from Ref.~\cite{Krech-97}.
\par
Figure~\ref{fig:compare-all} shows the comparison of the scaling functions of the critical Casimir force
for a homogeneous $(+)$ wall opposite to a striped wall (see \fref{bcstripes}).
All MFT curves have been rescaled by the same factors according to \eref{eq:rescale} so that the position and the height of the 
maximum of the MFT curve for $\kappa\to\infty$ agrees with the one obtained from the MC simulations in $d=3$.
As can be inferred from \fref{fig:compare-all}, the rescaled MFT behaviors as a function of $\tau$ show a 
qualitative agreement with the corresponding MC results even for finite values of $\kappa$.
\par
In \fref{fig:compare-all-open} we compare the scaling functions of the critical Casimir force
for a homogeneous wall with $(o)$ {\BC} opposite to a striped one (see \fref{bcopenstripes}).
The MFT scaling functions have been rescaled according to \eref{eq:rescale}.
In contrast to the case shown in \fref{fig:compare-all}, these rescaled MFT scaling functions for the $(o)$ case shown
in \fref{fig:compare-all-open} differ qualitatively from the corresponding behavior in $d=3$.
Whereas for $\kappa<2$ the MFT results suggest that the minima of the scaling functions exhibit
a cusplike singularity or a finite jump, the scaling functions $\theta_o(\tau,\kappa)$ in $d=3$
are analytic at their minima.
These differences are analogous to the ones obtained for homogeneous $(o,o)$ {\BC} at both surfaces \cite{gambassi:2006,oomft}.

\section{Summary and Outlook}
\label{sec:summary}
Within the Ising universality class we have studied the critical Casimir force for 
a film of thickness $L$ by using Monte Carlo (\MC) simulations in $d=3$ spatial dimensions and  by using mean-field theory. 
Along the lateral directions we have employed periodic boundary bonditions,  whereas along the normal direction  at the 
two confining surfaces fixed {\BC} have been imposed.
We have considered two cases: a homogeneous wall with $(+)$ {\BC} opposite to a wall patterned with alternating chemical stripes of equal width
$S_+=S_-$ with $(+)$/$(-)$ {\BC} (\fref{bcstripes}) and a homogeneous wall corresponding to $(o)$ {\BC} opposite to a striped
wall (\fref{bcopenstripes}).
In the limit of very narrow stripes, i.e.,  $\kappa\equiv S_+/L\to0$, the striped wall effectively mimics the behavior of Dirichlet $(o)$ {\BC},
so that for $\kappa\to0$ the system reduces to the homogeneous cases with $(+,o)$ or $(o,o)$ {\BC}, respectively (see 
Figs.~\ref{bcplusopen} and \ref{bcopenopen}).
In the opposite limit $\kappa\to\infty$, i.e., very broad stripes,  
in the first case (+; \fref{bcstripes}) the critical Casimir force equals the mean value
of the corresponding forces for films with homogeneous $(+,+)$ and $(+,-)$ boundary conditions at both surfaces, respectively.
On the other hand, in the second case (o; \fref{bcopenstripes}), deep in the two-phase regime, the corresponding limit is singular.
\par
We have investigated this system by combining {\MC} simulations and numerical integration as well as by carrying out numerically the corresponding
MFT calculation.
We have employed an improved lattice model, for which the leading scaling corrections are suppressed. 
We have obtained the following main results.

\begin{enumerate}

\item 
In the finite-size scaling limit the critical Casimir force $F_C=L^{-d}\theta(\tau,\kappa)$ per area and in units of $k_BT$ is described [\eref{casimir_fss_leading}] by a universal scaling function $\theta(\tau,\kappa)$, with the scaling variables $\tau\equiv t(L/\xi_0^+)^{1/\nu}$ and $\kappa\equiv S_+/L$. Here $t\equiv (T-T_c)/T_c$ is the reduced temperature, $\xi_0^+$ is the nonuniversal amplitude of the correlation length $\xi(t\rightarrow 0^{+})=\xi_0^+|t|^{-\nu}$, and $S_+$ is the width of the stripes on the lower surface. In the limit $\kappa\rightarrow 0$ the patterned surface attains an effective Dirichlet {\BC} [\eref{eq:limit-1}]. Within the range of aspect ratios $ \rho=L/L_\parallel$ (Figs.~\ref{bcstripes}--\ref{bcopenopen}) considered here, the MC data do not display a detectable dependence on $\rho$. Therefore we regard our results as the ones corresponding to the extrapolation to the film limit $\rho\rightarrow 0$. 

\item 
  In the limit of broad stripes, i.e., $\kappa\gg1$, the effects of the chemical steps separating the stripes vanish as $\propto\kappa^{-1}$ [Eqs.~\eqref{eq:limit-2} and \eqref{eq:limit-3}]. Thus, the total critical Casimir force effectively approaches the sum of the forces between the individual stripes and the opposing wall. Accordingly, the assumption of additivity of the forces (which underlies the Derjaguin or proximity force approximation) generally holds for $\kappa\to\infty$. However, in the case of a homogeneous wall with $(o)$ {\BC} opposite to a chemically striped wall, for $\kappa>2$ and $\tau\ll-1$, due to the formation of interfaces perpendicular to the film surfaces, the scaling function of the force varies as $\propto\kappa^{-1}|\tau|^\mu\propto L^d/S_+$ [for a fixed temperature $t<0$; \eref{eq:sum-1}], so that $F_C$ does not decay for $L\to\infty$ as long as $L<S_+/2$. Accordingly, for $\tau\rightarrow -\infty$, in the subsequent limit $\kappa\to\infty$ force additivity breaks down. The two limits $\kappa\rightarrow\infty$ and $\tau\rightarrow -\infty$ do not commute. 

\item 
By using {\MC} simulations for $d=3$, we have determined the critical Casimir amplitude at $T_c$ for various values of $\kappa$, in the case of the {\BC} illustrated in Figs.~\ref{bcstripes} and \ref{bcopenstripes} as well as in the limit $\kappa\rightarrow 0$, which corresponds to the {\BC} shown in Figs.~\ref{bcplusopen} and \ref{bcopenopen}. The results are reported in Eqs.~(\ref{Delta0})--(\ref{Delta1}) for the case of \fref{bcstripes} and in Eqs.~\eqref{Deltao0}--\eqref{Deltao3} for the case of \fref{bcopenstripes}. Whereas in the first case involving a homogeneous $(+)$ wall, the critical Casimir force is always repulsive (\fref{critvskappa}), in the case of a homogeneous $(o)$ wall the critical Casimir amplitude is nonmonotonic and changes sign as a function of $\kappa$ (\fref{critopenvskappa}).

\item Concerning $T\neq T_c$ we have determined the critical Casimir scaling functions $\theta_{+/o}(\tau,\kappa)$ in $d=3$ for various values of $\kappa$, as well as in the limit $\kappa\rightarrow 0$. In Figs.~\ref{casimir0}--\ref{casimir3} and \ref{casimir_open0}--\ref{casimir_open3} we show the scaling functions $\theta_+(\tau,\kappa)$ and $\theta_o(\tau,\kappa)$, respectively, as determined for various film thicknesses. In \fref{casimirall} we compare the universal scaling function $\theta_+(\tau,\kappa)$ of the critical Casimir force between a homogeneous wall with $(+)$ {\BC} and a striped wall (\fref{bcstripes}) for various values of $\kappa$, as determined from systems with the largest film thickness considered here, i.e., $L_z=24$, where $L_z=L/a+1$ and $a$ is the MC lattice constant. We also compare our results with the universal scaling function for the geometry consisting of a single chemical step (in the limit of vanishing aspect ratio studied in Ref.~\cite{PTD-10}) which corresponds to the limit $\kappa\rightarrow \infty$. Moreover, using the results of Ref.~\cite{PTD-10}, we have computed the asymptotic estimate for $\theta(\tau,\kappa)$ given in \eref{eq:limit-2}, which describes the approach to the limit $\kappa\rightarrow\infty$. We observe that this estimate agrees very well with our {\MC} results for $\kappa\ge 2$, as well as for $\kappa=1$ and $\tau >0$. In this case, within the entire range $0\le\kappa\le\infty$ the critical Casimir force is always repulsive. 

\item
  In contrast, for the case of a homogeneous $(o)$ wall opposite to a striped one (\fref{bcopenstripes}), the scaling function of the critical Casimir force exhibits a rather different behavior. As shown in \fref{casimir_openall}, the critical Casimir force varies nonmonotonically and changes sign as a function of $\kappa$ as well as a function of $\tau$. Moreover, for $\tau<0$ and for finite values of $\kappa$ the force may become much stronger than the ones for its limiting homogeneous cases $(o,o)$ and $(+,o)$ attained for $\kappa\to0$ and $\kappa\to\infty$, respectively, which are also shown in \fref{casimir_openall}. At $\kappa=2$ the system exhibits a transition of ground states from homogeneous states for $\kappa<2$ to vertically striped states for $\kappa>2$. Whereas the scaling functions of the critical Casimir force for the homogeneous states exhibit a minimum at finite $\tau<0$ and vanish for $\tau\to-\infty$, for $\kappa>2$ the scaling functions diverge for $\tau\to-\infty$ as $\propto|\tau|^\mu$ in accordance with \eref{eq:sum-1}. This is confirmed by the {\MC} results for $\kappa=3$ as shown in \fref{casimir_open3}. Thus, the scaling functions of the critical Casimir force as obtained for this case---belonging to the \emph{Ising} bulk universality class---do not vanish for $\tau\to-\infty$. So far this peculiar feature is only known for the critical Casimir force acting in films belonging to the \emph{XY} bulk universality class and thus exhibiting Goldstone modes \cite{DK-04,Hucht-07,VGMD-07,VGMD-08,Hasenbusch-09b,Hasenbusch-09c,Hasenbusch-09d,oomft,dohm:2013}. 

\item 
  In Sec.~\ref{sec:mft}, within MFT we have calculated the corresponding scaling functions for the critical Casimir force for the two cases sketched in \fref{bcstripes} and \fref{bcopenstripes}. The results for the suitably reduced critical Casimir amplitudes are shown in Figs.~\ref{fig:crit-mft} and \ref{fig:crit-open-mft} as a function of $\kappa$ within a wide range of values. For $\kappa\gg1$ the numerical MFT results agree with the asymptotic behaviors of the scaling functions $\Theta_+$ and $\Theta_o$ given in \eref{eq:limit-2} and \eref{eq:limit-3}, respectively, according to which they approach their corresponding limits for $\kappa\to\infty$ as $\kappa^{-1}$. The suitably reduced universal scaling functions for $\tau\ne0$, as obtained within MFT, are shown in \fref{fig:scaling-mft} for various values of $\kappa$ in the case of a homogeneous surface with $(+)$ {\BC} opposite to a striped surface. They interpolate smoothly between their limiting cases and always correspond to a repulsive critical Casimir force. 

\item
In the case of a homogeneous surface with $(o)$ {\BC} opposite to a striped surface the reduced MFT scaling functions are presented in \fref{fig:scaling-mft-open}. They show a rich dependence on $\kappa$. For $\kappa<2$ and $\tau<0$ the scaling function exhibits a minimum, and our numerical data suggest a cusplike singularity or a finite jump of the scaling function at its minimum. For $\kappa>2$ the scaling functions diverge for $\tau\to-\infty$ and the MFT scaling functions agree to large extent with the interface estimate given by \eref{eq:sum-1} (\fref{fig:scaling-log-open}). 

\item
The comparison of the suitably normalized Casimir amplitudes as obtained from {\MC} simulations in $d=3$ with the corresponding MFT ones reveals a good agreement for a homogeneous $(+)$ surface opposite to a striped surface (\fref{fig:compare-crit}) but qualitative differences for the corresponding $(o)$ case (\fref{fig:compare-crit-open}). Whereas in the latter case the data for $d=3$ show a nonmonotonic behavior and a change of sign as function of $\kappa$, in $d=4$ the MFT Casimir amplitudes are always positive. 

\item
Similarly, as shown in \fref{fig:compare-all}, the behaviors of the full scaling functions $\theta_{+/o}(\tau,\kappa)$ as obtained from simulations and within MFT plus a suitable rescaling (\eref{eq:rescale}) agree qualitatively to large extent for a homogeneous $(+)$ wall opposite to a striped one. On the other hand, for a homogeneous $(o)$ surface opposite to a striped surface the MFT scaling functions show, even after rescaling, qualitative differences to the ones obtained via {\MC} simulations (\fref{fig:compare-all-open}). However, within MFT as well as in $d=3$, in the latter $(o)$ case (\fref{bcopenstripes}) we always observe a change of sign of the critical Casimir force from negative values at $\tau\ll-1$ to positive values for $\tau>0$. In $d=3$ this occurs for $\kappa\gtrsim 3$ and within MFT for all values of $\kappa\gtrsim0.5$. At a fixed reduced temperature $t$, this zero at $\tau=\tau_0$ corresponds to a \emph{stable} distance $L_0(t)=\xi_0^+(\tau_0/t)^\nu$ at which the upper plate levitates above the lower plate due to critical Casimir forces alone. The levitation height $L_0(t)$ varies very sensitively as function of the reduced temperature $t=(T-T_c)/T_c$. 

\item 
The computation of the critical Casimir force requires to subtract the bulk free-energy density  from the free-energy density of the film. This bulk quantity is independent of the {\BC}. For $d=3$ we have determined it using a combination of MC simulations and numerical integration (see Appendox \ref{sec:bulk}). 

\end{enumerate}

The present study is relevant for the critical behavior of films belonging to the Ising universality class and in the presence of a chemically structured substrate. This can be experimentally realized by considering complete wetting films of binary liquid mixtures near their critical end points of demixing and by exposing their vapor phases to a chemically structured substrate \cite{cwetting,KD-92b}. The critical Casimir forces can be inferred by monitoring the thicknesses of the wetting films. This realizes the $(+)$ BC versus a striped surface. The surface fields describe the preferences of the two species for the confining interfaces of the wetting films. 

Another realization consists of studying directly the force acting on a colloidal particle immersed in a critical binary liquid mixture and exposed to a chemically structured substrate, as has been done in Ref.~\cite{SZHHB-08}. In this case the \emph{normal} critical Casimir force is approximately the one for the film geometry investigated here, provided the radius of the colloidal particle is sufficiently large relative to its distance from the wall. However, near $T_c$ for such a system an additional \emph{lateral} critical Casimir force sets in. In Ref.~\cite{TKGHD-10} the critical Casimir force for a sphere in front of a chemically structured substrate has been studied by means of mean-field theory as well as in $d=3$ by using the Derjaguin approximation. In this study, it was found that for suitable geometric features of the stripes on the substrate and in the presence of homogeneous {\BC} on the spherical colloid levitation is possible even for $\tau>0$, i.e., in the homogeneous phase of the solvent. Although these experimental studies \cite{SZHHB-08,TZGVHBD-11} are closely related to the setup studied here, a re-evaluation of the existing data is not sufficient in order to compare them with the present theoretical predictions. On one hand, the authors of Refs.~\cite{SZHHB-08,TZGVHBD-11} have measured only the \emph{lateral} forces acting on the colloidal particles, and not the \emph{normal} ones studied here. On the other hand, in order to effectively mimic the film geometry studied here, the radius of the colloidal particles should be much larger than the stripe widths $S_+$ and $S_-$, whereas the length scales realized in those experimental studies are of the same size \cite{SZHHB-08,TZGVHBD-11}.

In view of recent {\MC} results for the critical Casimir force of a sphere in front of a homogeneous wall \cite{Hasenbusch-12b}, it would be very interesting to extend this study by considering a sphere in front of a chemically structured wall. Besides analyzing directly the latter experimental setup, this would also provide the possibility of elucidating the range of validity of the Derjaguin approximation, which is commonly employed for curved geometries \cite{TKGHD-09,TKGHD-10}.

Here we have determined the critical Casimir force in the presence of a chemically structured substrate by using MC simulations in spatial dimension $d=3$, and by using mean-field theory, which holds in $d=4$. In order to complement this spatial dependence and to further probe the relevance of fluctuations, it would be interesting to investigate the corresponding system in $d=2$, where some exact results  are available \cite{AM-10} and conformal invariance allows one to determine exactly certain critical properties.

The present study also lends itself to further extensions. Here we have considered stripes with $(+)$ and $(-)$ {\BC} of equal widths. A natural extension of the present study would consist of calculating the critical Casimir force as a function of the ratio of the widths of the $(+)$ and $(-)$ stripes. Moreover, by considering two striped surfaces, one can also investigate the corresponding lateral critical Casimir force. So far the case of two striped surfaces has been investigated by mean-field theory for the film geometry \cite{SSD-06}; the issue of the lateral force has been analyzed by mean-field theory and in $d=3$ within the Derjaguin approximation for the sphere-wall geometry \cite{TKGHD-10}. 

Finally, as mentioned in Secs.~\ref{sec:off_criticality} and \ref{sec:mft}, for the {\BC} shown in \fref{bcopenstripes} and for $\kappa=2$, the system displays a rich glassy behavior at low temperatures. This deserves further investigation.
\section*{Acknowlegdements}

We are grateful to Svyatoslav Kondrat and Thomas F. Mohry for useful discussions and thank
Martin Hasenbusch and Oleg Vasilyev for providing MC data for the homogeneous cases.

\appendix
\section{Monte Carlo simulations}
\label{sec:mc}

In this appendix we report certain technical details of the {\MC} simulations we have performed.
As explained in Sec.~\ref{sec:critical}, the evaluation of the Casimir force at $T_c$  has been carried out in two steps. First, we  have determined the thermal average $\<{\cal H}_2-{\cal H}_1\>_\lambda$ which appears in \eref{I_def}. This is obtained by a standard {\MC} simulation for the ensemble characterized by the crossover Hamiltonian ${\cal H}_\lambda$ defined in \eref{crossover_H}. We  have implemented a combination of the standard Metropolis and Wolff cluster algorithms. Each MC step consists of $1$ Metropolis sweep over the entire lattice in lexicographic order and $L_z$ Wolff single-cluster flips; $L_z$ denotes the total number of lattice layers, including the surfaces of fixed spins, so that there are $L_z-2$ layers of fluctuating spins in the case of the {\BC}  shown in \fref{bcstripes}, $L_z-1$ layers in the case of {\BC} shown in Figs.~\ref{bcopenstripes} and \ref{bcplusopen}, and $L_z$ layers in the case of the $(o,o)$ {\BC} illustrated in \fref{bcopenopen}. As random number generator we have used the double precision SIMD-oriented Fast Mersenne Twister (dSFMT) \cite{Mersenne}.  Important details of the simulations performed at the critical temperature are reported in Tables \ref{mc_criticality_stripes}--\ref{mc_criticality_open_stripes3}.  Additional details  concerning the implementation of the simulating algorithm can be found in Ref.~\cite{PTD-10}.

\begin{table*}
\begin{tabular}{c@{\hspace{2em}}ccc@{\hspace{4em}}c@{\hspace{2em}}ccc@{\hspace{4em}}c@{\hspace{2em}}ccc}
\hline
\hline
\multicolumn{4}{c}{$\kappa\rightarrow 0:(+,o)$} & \multicolumn{4}{c}{$\kappa=1/4$} & \multicolumn{4}{c}{$\kappa=1/2$}\\
$L_z$ & $\rho$ & $N_{\rm steps}/10^3$ & $N_{\rm therm}/10^3$ & $L_z$ & $\rho$ & $N_{\rm steps}/10^3$ & $N_{\rm therm}/10^3$ & $L_z$ & $\rho$ & $N_{\rm steps}/10^3$ & $N_{\rm therm}/10^3$\\
\hline
$24$ & $1/8$ & $1200$ & $200$ & $24$ & $1/8$ & $500$ & $100$ & $24$     & $1/8$ & $400$ & $80$ \\
$24$         & $1/12$ & $800$ & $100$ & $24$         & $1/12$ & $200$ & $40$ & $24$ & $1/12$ & $160$ & $32$ \\
$24$ & $1/16$ & $600$ & $100$ & $24$ & $1/16$ & $100$ & $20$ & $24$     & $1/16$ & $100$ & $20$ \\
$32$ & $1/8$ & $3000$ & $500$ & $32$ & $1/8$ & $1500$ & $300$ & $32$     & $1/8$ & $1200$ & $240$ \\
$32$         & $1/12$ & $1400$ & $200$& $32$         & $1/12$ & $700$ & $140$ & $32$ & $1/12$ & $500$ & $100$ \\
$32$ & $1/16$ & $800$ & $100$ & $32$ & $1/16$ & $350$ & $70$ & $32$     & $1/16$ & $250$ & $50$ \\
$48$ & $1/8$ & $1500$ & $200$ & $48$ & $1/8$ & $1500$ & $200$ & $48$     & $1/8$ & $1500$ & $200$ \\
$48$         & $1/12$ & $700$ & $100$ & $48$         & $1/12$ & $700$ & $100$ & $48$ & $1/12$ & $700$ & $100$ \\
$48$ & $1/16$ & $350$ & $50$  & $48$ & $1/16$ & $350$ & $50$ & $48$     & $1/16$ & $350$ & $50$ \\
\hline
\hline
\end{tabular}
\caption{The total  number $N_{\rm steps}$ of MC steps and the  number $N_{\rm therm}$  of MC steps discarded in order to achieve thermalization  as used to determine the critical Casimir amplitudes for  film thicknesses $L_z\ge 24$, for aspect ratios $\rho=L_z/L_x\le 1/8$, and for the {\BC} shown in Figs.~\protect\ref{bcstripes} and \ref{bcplusopen}. Every MC step consists  of 1 Metropolis sweep over the entire lattice and $L_z$ Wolff single-cluster flips.  Additional details  concerning the simulation algorithm can be found in Ref.~\protect\cite{PTD-10}.}
\label{mc_criticality_stripes}
\end{table*}

\begin{table*}
\begin{tabular}{c@{\hspace{2em}}ccc@{\hspace{4em}}c@{\hspace{2em}}ccc@{\hspace{4em}}c@{\hspace{2em}}ccc}
\hline
\hline
\multicolumn{4}{c}{$\kappa=1$} & \multicolumn{4}{c}{$\kappa=2$} & \multicolumn{4}{c}{$\kappa=3$}\\
$L_z$ & $\rho$ & $N_{\rm steps}/10^3$ & $N_{\rm therm}/10^3$ & $L_z$ & $\rho$ & $N_{\rm steps}/10^3$ & $N_{\rm therm}/10^3$ & $L_z$ & $\rho$ & $N_{\rm steps}/10^3$ & $N_{\rm therm}/10^3$\\
\hline
$24$     & $1/8$ & $1600$ & $320$ & $24$     & $1/8$ & $1600$ & $320$ & $24$     & $1/12$ & $800$ & $160$ \\
$24$ & $1/12$ & $1100$ & $220$ & $24$ & $1/12$ & $1100$ & $220$ & $24$ & $1/18$ & $550$ & $110$ \\
$24$     & $1/16$ & $800$ & $160$ & $24$     & $1/16$ & $800$ & $160$ & $24$     & $1/24$ & $400$ & $80$ \\
$32$     & $1/8$ & $2600$ & $520$ & $32$     & $1/8$ & $2600$ & $520$ & $32$     & $1/12$ & $1300$ & $260$ \\
$32$ & $1/12$ & $1700$ & $200$ & $32$ & $1/12$ & $1700$ & $340$ & $32$ & $1/18$ & $850$ & $170$ \\
$32$     & $1/16$ & $1300$ & $250$ & $32$     & $1/16$ & $1300$ & $200$ & $32$     & $1/24$ & $650$ & $130$ \\
$48$     & $1/8$ & $1500$ & $200$ & $48$     & $1/8$ & $1500$ & $300$ & $48$     & $1/12$ & $750$ & $150$ \\
$48$ & $1/12$ & $700$ & $100$ & $48$ & $1/12$ & $700$ & $140$ & $48$ & $1/18$ & $350$ & $70$ \\
$48$     & $1/16$ & $350$ & $50$ & $48$     & $1/16$ & $350$ & $50$ & $48$     & $1/24$ & $170$ & $34$ \\
\hline
\hline
\end{tabular}
\caption{Same as Table~\ref{mc_criticality_stripes} for $\kappa=1$, $2$, $3$.}
\label{mc_criticality_stripes2}
\end{table*}

\begin{table*}
\begin{tabular}{c@{\hspace{2em}}ccc@{\hspace{4em}}c@{\hspace{2em}}ccc@{\hspace{4em}}c@{\hspace{2em}}ccc}
\hline
\hline
\multicolumn{4}{c}{$\kappa\rightarrow 0:(o,o)$} & \multicolumn{4}{c}{$\kappa=1/4$} & \multicolumn{4}{c}{$\kappa=1/2$}\\
$L_z$ & $\rho$ & $N_{\rm steps}/10^3$ & $N_{\rm therm}/10^3$ & $L_z$ & $\rho$ & $N_{\rm steps}/10^3$ & $N_{\rm therm}/10^3$ & $L_z$ & $\rho$ & $N_{\rm steps}/10^3$ & $N_{\rm therm}/10^3$\\
\hline
$24$ & $1/8$ & $12000$ & $1200$ & $24$ & $1/8$ & $12000$ & $1200$ & $24$ & $1/8$ & $12000$ & $1200$ \\
$24$ & $1/12$ & $8000$ & $800$ & $24$ & $1/12$ & $8000$ & $800$ & $24$ & $1/12$ & $8000$ & $800$ \\
$24$ & $1/16$ & $6000$ & $600$ & $24$ & $1/16$ & $6000$ & $600$ & $24$ & $1/16$ & $6000$ & $600$ \\
$32$ & $1/8$ & $32000$ & $600$ & $32$ & $1/8$ & $6000$ & $600$ & $32$ & $1/8$ & $6000$ & $600$ \\
$32$ & $1/12$ & $16000$ & $300$ & $32$ & $1/12$ & $3000$ & $300$ & $32$ & $1/12$ & $3000$ & $300$ \\
$32$ & $1/16$ & $8000$ & $150$ & $32$ & $1/16$ & $1500$ & $150$ & $32$ & $1/16$ & $1500$ & $150$ \\
\hline
\hline
\end{tabular}
\caption{Same as Table~\ref{mc_criticality_stripes} for the {\BC} of Figs.~\ref{bcopenstripes} and \ref{bcopenopen}.}
\label{mc_criticality_open_stripes}
\end{table*}

\begin{table*}
\begin{tabular}{c@{\hspace{2em}}ccc@{\hspace{4em}}c@{\hspace{2em}}ccc}
\hline
\hline
\multicolumn{4}{c}{$\kappa=3/4$} & \multicolumn{4}{c}{$\kappa=1$} \\
$L_z$ & $\rho$ & $N_{\rm steps}/10^3$ & $N_{\rm therm}/10^3$ & $L_z$ & $\rho$ & $N_{\rm steps}/10^3$ & $N_{\rm therm}/10^3$ \\
\hline
$24$     & $1/9$ & $12000$ & $120$ & $24$     & $1/8$ & $12000$ & $1200$ \\
$24$ & $1/12$ & $8000$ & $800$ & $24$ & $1/12$ & $8000$ & $800$ \\
$24$     & $1/15$ & $6000$ & $600$ & $24$     & $1/16$ & $6000$ & $600$ \\
$32$     & $1/9$ & $6000$ & $600$ & $32$     & $1/8$ & $6000$ & $600$ \\
$32$ & $1/12$ & $3000$ & $300$ & $32$ & $1/12$ & $3000$ & $300$ \\
$32$     & $1/15$ & $1500$ & $150$ &$32$      & $1/16$ & $1500$ & $150$ \\
\hline
\hline
\end{tabular}
\caption{Same as Table~\ref{mc_criticality_open_stripes} for $\kappa=3/4$ and $1$.}
\label{mc_criticality_open_stripes2}
\end{table*}

\begin{table*}
\begin{tabular}{c@{\hspace{2em}}ccc@{\hspace{4em}}c@{\hspace{2em}}ccc}
\hline
\hline
\multicolumn{4}{c}{$\kappa=2$} & \multicolumn{4}{c}{$\kappa=3$} \\
$L_z$ & $\rho$ & $N_{\rm steps}/10^3$ & $N_{\rm therm}/10^3$ & $L_z$ & $\rho$ & $N_{\rm steps}/10^3$ & $N_{\rm therm}/10^3$ \\
\hline
$24$     & $1/24$ & $600$ & $100$ & $24$     & $1/24$ & $600$ & $60$ \\
$24$ & $1/36$ & $400$ & $80$ & $24$ & $1/36$ & $400$ & $60$ \\
$24$     & $1/48$ & $300$ & $60$ & $24$     & $1/48$ & $300$ & $60$ \\
$32$     & $1/24$ & $150$ & $30$ & $32$     & $1/24$ & $150$ & $20$ \\
$32$ & $1/36$ & $70$ & $15$ & $32$ & $1/36$ & $70$ & $10$ \\
$32$     & $1/48$ & $40$ & $8$ & $32$     & $1/48$ & $40$ & $8$ \\
\hline
\hline
\end{tabular}
\caption{Same as Table~\ref{mc_criticality_open_stripes} for $\kappa=2$ and $3$.}
\label{mc_criticality_open_stripes3}
\end{table*}

As explained in Sec.~\ref{sec:off_criticality}, the determination of the scaling function for the critical Casimir force has been obtained by sampling the reduced energy densities $E(\beta',L_z,L_x,s_+)$  and $E(\beta',L_z-1,L_x,s_+)$ [see \eref{deltaF_reduced}] followed by carrying out numerically the integration in \eref{deltaF_reduced} by using Simpson's rule. An upper bound of the systematic error due to the discretization of the integrals can be  determined by sampling the fourth derivative of the integrand: by computing $\partial^4E(\beta,L_z,L_x,s_+)/(\partial\beta^4)$ we have checked that such  a systematic error is always negligible  compared to the statistical errors. (Since  for $L_z\rightarrow\infty$ the quantity $\partial^4E(\beta,L_z,L_x,s_+)/(\partial\beta^4)$ diverges at the critical point, the number of sampled points has to increase with $L_z$.) In Table \ref{MC_offcritical} we report important details  concerning these simulations associated with \eref{deltaF_reduced}.  For each film thickness and {\BC} we have considered the same three aspect ratios $\rho$ for determining the scaling functions as the ones used for determining the critical Casimir amplitude (see Tables~\ref{mc_criticality_stripes}--\ref{mc_criticality_open_stripes3}), except for the {\BC} shown in \fref{bcopenstripes} and $\kappa=3$ (see Table \ref{mc_criticality_open_stripes3} and \fref{casimir_open3}). For the {\BC} shown in Figs.~\ref{bcstripes} and \ref{bcplusopen}, we have verified that the sampled reduced energy densities are {\it de facto} independent of $\rho$. Therefore our results capture reliably the limit $\rho\rightarrow 0$; we have averaged them over the three aspect ratios considered. Concerning the {\BC} shown in Figs.~\ref{bcopenstripes} and \ref{bcopenopen}, as discussed in Sec.~\ref{sec:off_criticality}, the data exhibit a weak dependence on the aspect ratio at low temperatures and we have considered the three aspect ratios separately, i.e., without taking this average.

Finally, we mention that with the above described simulation algorithm and for the {\BC} shown in Fig~\ref{bcopenstripes}, we occasionally observed the appearance of metastable states at low temperatures, which cause the thermalization of the run to be rather long. We have found that this problem can be healed by starting the simulations with an ordered states.

\begin{table}
\begin{tabular}{l@{\hspace{2em}}c@{\hspace{2em}}c@{\hspace{2em}}c}
\hline
\hline
$L_z$ & $\beta_0$ & $\beta_{\rm max}$ & $\Delta\beta$ \\
\hline
$8$ & $0.327721735$ & $0.427721735$ & $0.0005$ \\
$12$ & $0.327721735$ & $0.427721735$ & $0.0001$ \\
$16$ & $0.327721735$ & $0.427721735$ & $0.0005$ \\
$24$ & $0.377721735$ & $0.397721735$ & $0.00002$ \\
\hline
\hline
\end{tabular}
\caption{The  lowest ($\beta_0$) and the  highest ($\beta_{\rm max}$) inverse temperatures used for the computation of the scaling functions associated with the free-energy differences  via \eref{deltaF_reduced}. The integrals have been computed numerically using Simpson's rule,  with the reported intervals $\Delta\beta$ between two consecutive points. For each film thickness we have considered the same three aspect ratios as the ones used for determining the critical Casimir amplitude (see Tables~\ref{mc_criticality_stripes}--\ref{mc_criticality_open_stripes3}).}
\label{MC_offcritical}
\end{table}

\section{Determination of the bulk free-energy density}
\label{sec:bulk}
Here we report  certain details concerning the determination of the bulk free-energy density which is needed  for calculating the critical Casimir force (see Eqs.~(\ref{casimir_fromF}) and (\ref{casimir_from_F_reduced})). For this purpose we have simulated the improved Blume-Capel model described by \eref{bc} for a simple cubic lattice with periodic {\BC} in all directions and lattice sizes $L_z=24$--$256$. For this system we  have determined the reduced energy density $E(\beta,L_z)$ and the reduced free-energy density $F(\beta,L_z)$ as defined in Eqs.~(\ref{Edef}) and (\ref{Fdef}). For the sake of simplicity, here we omit the dependence on $L_x$ and $s_+$ because the lattice considered here has the same size in all directions and it does not have any surface. Since  the aim is to extract the thermodynamic limit of these quantities from finite-size  results, we recall the expected behavior of the  corresponding finite-size parts. For $T\neq T_c$ and $L_z\gg \xi$, $E(\beta,L_z)$ approaches its infinite-volume limit $E_{\rm bulk}(\beta)$ as
\begin{equation}
\label{largeL_E}
\delta E(\beta,L_z) \equiv E(\beta,L_z) - E_{\rm bulk}(\beta) \sim \left(L_z/\xi\right)^{k+1}e^{-L_z/\xi},
\end{equation}
where $k$ is  an integer. Conversely, in the region where $\xi \ \approx\ L_z$,  one has ($\alpha=2-3\nu$)
\begin{equation}
\label{closetc_E}
\delta E(\beta,L_z) = t^{1-\alpha}\tilde{h}_E\left(L_z/\xi\right)=\frac{1}{L_z^{3-1/\nu}} h_E\left(L_z/\xi\right),
\end{equation}
where the scaling function $h_E(x)$ is universal up to a prefactor and $h_E(x)=O(1)$ for $\xi \ \approx\ L_z$. The reduced free-energy density $F(\beta,L_z)$ can be obtained by integrating $E(\beta,L_z)$  according to \eref{F_from_E}. It follows that, for $T>T_c$ and $L_z\gg\xi$, $F(\beta,L_z)$ approaches its infinite-volume limit $F_{\rm bulk}(\beta)$ as
\begin{equation}
\label{largeL_F_ht}
\delta F(\beta,L_z)\equiv  F(\beta,L_z) - F_{\rm bulk}(\beta) \sim \left(L_z/\xi\right)^ke^{-L_z/\xi}.
\end{equation}
In deriving \eref{largeL_F_ht}, we have used the fact that for $T>T_c$ the condition $L_z\gg\xi$ is satisfied throughout the interval of integration on the right-hand side of \eref{F_from_E}. This is not the case if $T < T_c$. For $T < T_c$, $L_z\gg \xi$, and by using \eref{F_from_E}, the finite-size correction $\delta F(\beta,L_z)$ can be expressed as
\begin{equation}
\begin{split}
\delta F(\beta,L_z) \equiv  F(\beta,L_z) - F_{\rm bulk}(\beta) \\
= \delta F(\beta\rightarrow\infty,L_z) + \int_\infty^\beta d\beta' \delta E(\beta',L_z).
\end{split}
\label{largeL_F_lt_partial}
\end{equation}
In the second term of the right-hand side of \eref{largeL_F_lt_partial} one has $L_z\gg\xi$ throughout the integration interval. Thus, by using \eref{largeL_E}, the integral on the right-hand side of \eref{largeL_F_lt_partial} varies as $\left(L_z/\xi\right)^ke^{-L_z/\xi}$. The finite-size correction $\delta F(\beta\rightarrow\infty,L_z)$ can be inferred from computing $F(\beta,L_z)$ for $\beta\rightarrow\infty$  and for a finite size $L_z$. For $\beta\rightarrow\infty$, the Gibbs measure is dominated by the twofold degenerate ground state, consisting of a configuration in which all spins are fixed to $+1$ or to $-1$. By using the definition of $F(\beta,L_z)$ given in \eref{Fdef}, one has
\begin{equation}
\begin{split}
F(\beta\rightarrow\infty,L_z) = \frac{1}{L_z^3}\ln \left[\frac{2e^{\left(3\beta-D\right)L_z^3}}{\left(1+2e^{-D}\right)^{L_z^3}}\right]\\
=\frac{\ln 2}{L_z^3} + \ln \left(\frac{e^{3\beta-D}}{1+2e^{-D}}\right),
\end{split}
\label{largeL_F_lt_partial2}
\end{equation}
where $D$ is the coupling constant appearing in the second term of the Hamiltonian  given in \eref{bc}. By taking the $L_z\rightarrow\infty$ in \eref{largeL_F_lt_partial2}, we identify the second term on the right-hand side of \eref{largeL_F_lt_partial2} as the infinite-volume limit $F_{\rm bulk}(\beta)$. \footnote{We note that $F_{\rm bulk}(\beta) \rightarrow\infty$ for $\beta\rightarrow\infty$. This is because $F_{\rm bulk}(\beta)$ is the bulk free energy per volume and in units of $-k_BT$. The free energy per volume $-F_{\rm bulk}(\beta)/\beta$ has instead a finite limit for $\beta\rightarrow\infty$.} Thus, we infer $\delta F(\beta\rightarrow\infty,L_z) = (\ln 2)/(L_z^3)$. Thus, for $T < T_c$ and  $L_z\gg \xi$, $F(\beta,L_z)$ approaches its infinite-volume limit $F_{\rm bulk}(\beta)$ as
\begin{equation}
\label{largeL_F_lt}
\delta F(\beta,L_z) \sim \left(L_z/\xi\right)^ke^{-L_z/\xi} + \frac{\ln 2}{L_z^3}.
\end{equation}
From Eqs.~(\ref{F_from_E})  and (\ref{largeL_F_lt}) one finds that
\begin{equation}
\label{ln2}
\int_0^\infty \delta E(\beta',L_z) d\beta' = \frac{\ln 2}{L_z^3},
\end{equation}
where the support of the integrand is actually  confined to the region where $L_z\ \approx\ \xi$. In  this region the finite-size correction of the reduced free-energy density  is given by
\begin{equation}
\label{closetc_F}
\delta F(\beta,L_z) =\frac{1}{L_z^3} h_F\left(L_z/\xi\right),
\end{equation}
where, as in \eref{closetc_E}, the universal scaling function $ h_F(x)=O(1)$ for $\xi\ \approx\ L_z$. A comparison of the finite-size corrections for the reduced energy density given in Eqs.~(\ref{largeL_E}) and (\ref{closetc_E}) and those for the reduced free-energy density  in Eqs.~(\ref{largeL_F_ht}), (\ref{largeL_F_lt}),  and (\ref{ln2}) shows that $F(\beta,L_z)$  converges faster to limit for $L_z\rightarrow\infty$ than $E(\beta,L_z)$. The only exception to this rule  occurs in the low-temperature phase, $T<T_c$ and $L_z\gg\xi$, where the reduced free-energy density exhibits an additional finite-size correction $\ln 2/L^3$ [see \eref{ln2}]. However, because this correction term is known exactly, one can  eliminate it by subtracting it explicitly.

\begin{table}
\begin{tabular}{l@{\hspace{2em}}c@{\hspace{2em}}c@{\hspace{2em}}c}
\hline
\hline
$L$   & $\beta_{\rm min}$ & $\beta_{\rm max}$ & $\Delta\beta$ \\
\hline
$24$  & $0.327721735$ & $0.427721735$ & $0.0002$ \\
$32$  & $0.347721735$ & $0.427721735$ & $0.0001$ \\
$48$  & $0.367721735$ & $0.407721735$ & $0.0001$ \\
$64$  & $0.377721735$ & $0.397721735$ & $0.0001$ \\
$96$  & $0.380521735$ & $0.395721735$ & $0.00005$ \\
$128$ & $0.381521735$ & $0.394521735$ & $0.00005$ \\
$192$ & $0.384121735$ & $0.393321735$ & $0.00002$ \\
$256$ & $0.385321735$ & $0.391321735$ & $0.00001$ \\
\hline
\hline
\end{tabular}
\caption{The interval of integration $[\beta_{\rm min},\beta_{\rm max}]$ for each lattice size $L$ used in the determination of the bulk free-energy density. We  have implemented Simpson's rule  with the reported distances  $\Delta\beta$ between two consecutive points.}
\label{mcbulk}
\end{table}

In order to compute the bulk free-energy density, we proceed as follows. At a given lattice size $L_z$, we compute the reduced energy density $E(\beta,L_z)$ in  an interval $[\beta_{\rm min},\beta_{\rm max}]$ around the inverse critical temperature $\beta_c=0.387721735(25)$ \cite{Hasenbusch-10}. In order to  minimize the error bars we have implemented the control-variates scheme introduced in Ref.~\cite{FMM-09}. Control variates are observables which have a vanishing mean value and therefore can be added to any observable without changing its mean value; control variates provide also an additional check of the MC simulations. In the second step $F(\beta,L_z)-F(\beta_{\rm min},L_z)$ is calculated by numerically integrating \eref{F_from_E}. For this purpose we have used Simpson's rule. The resulting quantity $F(\beta,L_z)-F(\beta_{\rm min},L_Z)$ suffers from two types of errors: a statistical error originating from the statistical error bars of the integrand $E(\beta,L_z)$ and a systematic error due to the chosen quadrature. In the present case and as mentioned above, the maximum systematic error in Simpson's rule can be computed by estimating the fourth derivative of $E(\beta,L_z)$. We have always checked that such an error is at least one order of magnitude smaller than the statistical error, so that it can be safely neglected and the statistical error bar is a correct measure of the uncertainty of the reduced free-energy density. The integration of $E(\beta,L_z)$ leads to the value of $F(\beta,L_z)-F(\beta_{\rm min},L_z)$ for several inverse temperatures $\beta\in [\beta_{\rm min},\beta_{\rm max}]$. For those values of $\beta<\beta_c$ for which $L_z\gg\xi$, we regard our results for finite $L_z$ to be the ones for infinite $L_z$ if the statistical error bars are  smaller than the finite-size correction. To this end, we  have checked that $E(\beta,L_z)$ is, within the numerical accuracy, independent of $L_z$ by comparing the values obtained for two consecutive lattice sizes. As discussed above, $E(\beta,L_z)$ is expected to converge to the thermodynamic limit slower than $F(\beta,L_z)$. Roughly speaking, with the present numerical accuracy, the finite-size scaling corrections are negligible for $L_z/\xi\le 20$. For $T<T_c$ we use the more conservative bound $L_z/\xi\le 35$--$40$, and we explicitly subtract the additional finite-size term $(\ln 2)/L_z^3$ which appears in \eref{ln2}. We note that the nonuniversal amplitude $\xi_{0l}^-$ of the correlation length below $T_c$ is roughly half of $\xi_{0l}^+$: $\xi_{0l}^+/\xi_{0l}^-=1.957(7)$ \protect\cite{CPRV-02}. At any given lattice size $L_z$, this procedure results in the estimate of the bulk free-energy density for a subset $[\beta_{\rm min}, \beta_{\rm inf}]\cup [\beta_{\rm sup}, \beta_{\rm max}]$ of the integration interval $[\beta_{\rm min},\beta_{\rm max}]$, with $\xi(\beta_{\rm inf})\ \approx\ L_z/20$ and $\xi(\beta_{\rm sup})\ \approx\ L_z/40$. Thus, for $\beta \le \beta_{\rm inf}$ and $\beta\ge \beta_{\rm sup}$, $F(\beta,L_z)-F(\beta_{\rm min},L_z)$ agrees within error bars with $F_{\rm bulk}(\beta)-F_{\rm bulk}(\beta_{\rm min})$, while $[\beta_{\rm inf},\beta_{\rm sup}]$ is the interval in which the finite-size correction $\delta F(\beta,L_z)$ is not negligible. In the next step we have applied the above procedure for a larger lattice size $L_z'>L_z$ and the smaller integration interval $[\beta'_{\rm min}=\beta_{\rm inf},\beta'_{\rm max}=\beta_{\rm sup}]$. This results in the quantity $F(\beta,L_z')-F(\beta'_{\rm min},L_z')$ to which we add $ F(\beta'_{\rm min},L_z)-F(\beta_{\rm min},L_z)\simeq F_{\rm bulk}(\beta'_{\rm min})-F_{\rm bulk}(\beta_{\rm min})$ as determined from the lattice size $L_z$, so that we finally obtain the desired quantity $F(\beta,L_z')-F_{\rm bulk}(\beta_{\rm min})$.  As before, this results in the estimate of the bulk free energy for $\beta \in [\beta_{\rm min}, \beta'_{\rm inf}]\cup [\beta'_{\rm sup}, \beta_{\rm max}]$, with $\beta'_{\rm inf}>\beta_{\rm inf} $ and $\beta'_{\rm sup}<\beta_{\rm sup}$. By iterating the procedure with increasing values of $L_z$, we progressively narrow the interval around $\beta_c$ where $\xi\ \approx\ L_z$ and finite-size scaling corrections are not negligible. In Table \ref{mcbulk} we report the interval used for each lattice size considered here. The final statistical error bars for $F_{\rm bulk}(\beta)$ are generally between $4\times 10^{-8}$ and $10^{-7}$.  Even for the largest lattice size $L_z=256$ we have considered, there exists of course an interval around $\beta_c$ for which the condition $L_z\gg\xi$ cannot be satisfied. In such  a region the finite-size scaling corrections are given by \eref{closetc_F}. In order to ensure that the residual finite-size correction is less than the statistical error bars, we have checked that the results for $L_z=192$ and $L_z=256$ differ at most by one error bar. As an additional check, using \eref{closetc_F} and the results of Ref.~\cite{Dohm-11}, we can infer that the finite-size correction term is at most $\approx\ 0.7/(256^3)=4\times 10^{-8}$. For the same interval the statistical error bar is between $8\times 10^{-8}$ and $10^{-7}$. Thus we conclude that our determination of the bulk free-energy density is reliable within the statistical error bars.


\begin{thebibliography}{99}
\bibitem{corr}
There is a misprint in Eq.~(27) of Ref.~\cite{VGMD-08},
  where the argument of the function on the right-hand side should be $-x/2$ instead of $-2x$.
  Additionally, in footnote [49] of Ref.~\cite{VGMD-08}, the estimate of the position of the maximum
  of the scaling function for a film with $(+,-)$ BC should read $\simeq-28.1$ instead of 
  $\simeq28.1$.
\end{thebibliography}
\end{document}